\documentclass[superscriptaddress,
aip,
 amsmath,amssymb,
 reprint,%
]{revtex4-1}

\usepackage{graphicx}
\usepackage{caption,subcaption}
\usepackage{dcolumn}
\usepackage{bm}
\usepackage{color}
\usepackage{ulem}
\usepackage{mathtools}
\usepackage{rotating}
\usepackage{tikz} 
\newcommand{\bnabla}{\boldsymbol{\nabla} } 
\newcommand{\bcdot}{\boldsymbol{\cdot} } 
\usepackage{array}
\newcolumntype{P}[1]{>{\centering\arraybackslash}p{#1}}
\newcolumntype{M}[1]{>{\centering\arraybackslash}m{#1}}
\DeclareMathOperator{\Tr}{Tr}
\begin{document}


\title{Global Organization of Three-Dimensional, Volume-Preserving
  Flows: \\ Constraints, Degenerate Points, and Lagrangian Structure}


\author{Bharath Ravu}
\email{brtreddy@iitb.ac.in}\affiliation{IITB-Monash Research Academy, Indian Institute of
  Technology Bombay, Powai, Mumbai-400076,India.}

\author{Guy Metcalfe}%
\email{gmetcalfe@swin.edu.au} \affiliation{\mbox{School of
    Engineering, Swinburne University of Technology,
    Hawthorn, VIC 3122, Australia}}

\author{Murray Rudman}
\email{murray.rudman@monash.edu}\affiliation{\mbox{Department of Mechanical and Aerospace Engineering, Monash University, Clayton, VIC 3800, Australia}}

\author{Daniel R. Lester}
\email{daniel.lester@rmit.edu.au}\affiliation{\mbox{School of Engineering, RMIT University, Melbourne, VIC 3000, Australia}}

\author{Devang V. Khakhar}
\email{khakhar@iitb.ac.in}\affiliation{%
Department of Chemical Engineering, Indian Institute of Technology Bombay, Powai, Mumbai-400076, India
}%

\date{\today}
\begin{abstract}
  Global organization of 3-dimensional (3D) Lagrangian chaotic
  transport is difficult to infer without extensive computation.  For
  3D time-periodic flows with one invariant we show how constraints on
  deformation that arise from volume-preservation and periodic lines result
  in resonant degenerate points that periodically have zero net
  deformation.  These points organize all Lagrangian transport in such
  flows through coordination of lower-order and higher-order periodic lines and
  prefigure unique transport structures that arise after perturbation
  and breaking of the invariant.  Degenerate points of periodic lines
  and the extended 3D structures associated with them are easily
  identified through the trace of the deformation tensor calculated
  along periodic lines.  These results reveal the importance of degenerate
  points in understanding transport in one-invariant fluid flows.

\end{abstract}

\maketitle

\begin{quotation} \bf
  Global organization of 3-dimensional (3D) Lagrangian chaotic
  transport is difficult to infer without extensive computation.  For
  3D time-periodic flows with one invariant we show how constraints on
  deformation from volume-preservation and periodic lines cause
  resonant degenerate points that periodically have zero net
  deformation.  These points organize all Lagrangian transport in the
  flow through coordination of lower-order and higher-order periodic lines.  We also
  show how to easily identify degenerate points of periodic lines and
  the extended 3D structures that go with them.
\end{quotation}


\section{Introduction}
\label{sec:Intro}


Design and control of Lagrangian fluid particle trajectories in
three-dimensional (3D) incompressible flow fields (conservative
dynamical systems) is a key step toward engineering efficient mixing
and transport properties for fluid devices.  This is particularly
important for intrinsically low Reynolds number applications, such as
microfluidic devices, processing highly viscous or delicate materials,
and many others \cite{Aref_frontiers_2017}.  Due to considerations of
simplicity or ease of manufacture, devices often have spatial
symmetries and time periodicity.  In the presence of such symmetries,
the dynamical system of the 3D incompressible flow has invariants or
conserved quantities \cite{Mezic_symmetry_1994,Haller_symmetry_1998}.
What organizes the global Lagrangian transport structure in such a
flow?  The answer is not completely known \cite{Wiggins_3D_2010}, but
fluid deformation around periodic lines plays an important role.  Here
we show how in the Stokes limit constraints imposed by
volume-preservation and by a lack of deformation in the tangent
direction to periodic lines organize global 3D transport.

We consider a 3D incompressible time-periodic flow with one invariant
that consists of nested topological spheroids, where the effective
radial coordinate (or action coordinate) is conserved.  Nested
spheroids, which we will call shells, conform to Bajer's general
formulation of a 3D flow that is nearly everywhere Hamiltonian except
in the neighbourhood of stagnation points, where the true 3D character
of the flow manifests \cite{Bajer_streamlines_1994}.  Only
two-dimensional (2D) flows can be truly Hamiltonian, i.e.\/ have the
Lagrangian motion derivable from a stream---or Hamiltonian---function
as
\begin{equation} 
  \frac{dx}{dt} = -\frac{\partial H}{\partial y} \qquad
  \frac{dy}{dt} =  \frac{\partial H}{\partial x},
\end{equation}
where $(x,y)$ are 2D Cartesian coordinates.  However in 3D flows with
one invariant, motion is restricted to surfaces on which the conserved
coordinate is a constant, so that we expect 2D Hamiltonian motion on
each shell \cite{Aref_frontiers_2017}.


In addition to space-filling shells, the fluid space is shot through
with periodic lines (stagnation lines of the flow map) that must either have both ends attached to solid boundaries, or that form
closed loops (see appendix~\ref{app:no_iso_periodic_points}), or that have one or both ends extend to infinity
(in an open flow).  Periodic lines are a 3D analogue of periodic
points in 2D periodic flows: after $n$ periods, a period-$n$ line is a
connected collection of points that all return to their original
locations.  Typically the lower-order period lines, e.g.\/ period-1 line (P1 line),
are most important for understanding the organization of transport.
The closed flow we examine here has the particularly simple situation
of having a single P1 line for all parameter values of the flow.
Moreover, due to the nature of the flow forcing, the single P1 line is
confined to a symmetry plane, making finding the line and analyzing
its properties straight-forward.

Analogous to 2D periodic points, 3D periodic lines have a
locally elliptic or hyperbolic character in that fluid trajectories in the neighbourhood
of elliptic line segments rotate around the
periodic line, whereas those in the neighbourhood of hyperbolic line segments have a direction of maximal contraction towards the
line and a direction of maximal stretching away from the line.
Elliptic and hyperbolic segments of a periodic line are separated by
degenerate points, where the local deformation is shear-like.  

This 2D local motion in the neighbourhood of 3D periodic lines comes
about because there is no deformation in the tangent direction of the
line.  If there were tangent deformation, points of the periodic line
would not periodically return to their initial locations.  As the
spheroidal shells are space-filling, continuity ensures that where a periodic
line touches a shell, the line shares its character---elliptic,
hyperbolic, or degenerate---with the local Hamiltonian motion on the shell
\cite{Gomez_invariant_2002}.  For example, where an elliptic segment (e.g. see figure~\ref{fig:prd1_line_sgmnt_beta_16_PS})
pierces a shell, a Poincar{\'e} section on that shell shows an
elliptic island centered at the piercing.

No deformation along periodic lines and volume conservation turn out
to be stringent---and key---constraints to organization of the flow.  In
3D the eigenvalues describing local deformation can be written in
terms of the three matrix invariants of the deformation tensor.  With
two constraints there is only one independent parameter that determines the
character of any periodic line.  Using these constraints we derive the
main results of this paper:
\begin{enumerate}
\item Degenerate points of periodic lines are analogous to resonances
  of classical planar bifurcation theory.
\item Resonances form extended 3D structures coordinating sets of
  lower-order and higher-order periodic lines.
\item While some of the 3D structures we observe could have been
  inferred from center manifold theory by expansion in tangent planes
  about critical points at shell piercings, other observed structures
  could not have been so inferred.  The latter are truly global in
  nature. 
\item There are a finite number of important resonances and associated
  3D structures.  
\item Because of (1) and (4) one-invariant 3D flows can be completley
  understood hierarchically by easily finding the
  resonant degenerate points and their extended transport structures
  first on P1 lines, then on P2 lines, etc., until as much detail of
  the total Lagrangian transport picture is uncovered as is required
  for any particular purpose.
\end{enumerate}
Degenerate resonance points and associated structures are generic
features of 3D one-invariant flows that previously have been only
partially described.  While these structures are interesting in
themselves, in a sequel publication we will show how this knowledge
forms a basis for investigating and understanding how breaking the
final invariant in these flows leads to fully 3D transport.

\section{Periodically Reoriented Enclosed Flow}
\label{sec:hemisphere_system}

In this study we consider a periodically reoriented lid-driven cavity flow as an example of a volume preserving, time-periodic 3D flow with one invariant.
Lid-driven cavity flows are one of the
most frequently studied problems in computational fluid dynamics
\cite{Shankar_driven_2000}; although, there are fewer studies of 3D
cavities. The cavity we use is a hemisphere, and we are aware of no
previous study of lid-driven flow in this geometry.  Importantly, this
flow possesses multiple symmetries that makes its Lagrangian properties easy to
identify, compute, and study; it is also amenable to experiment.

Lid-driven cavity flow without reorientation admits two families of
invariant surfaces (of which manifest as topological cylinders
oriented perpendicular to the direction of lid motion) that intersect
to form the closed particle trajectories shown in
figure~\ref{fig:ldhs_schematic}. The 2D material surfaces associated
with these invariants play a strong role in organizing the 3D
Lagrangian transport structure into the 1D closed orbits shown in
figure~\ref{fig:ldhs_schematic}. What does the periodic reorientation
do?  Consider the polar decomposition \cite{truesdell_mechanics_2004}
applied to the flow deformation tensor
$\boldsymbol{F} = \boldsymbol{R} \boldsymbol{\Phi}$, where
$\boldsymbol{R}$ is a rotation and $\boldsymbol{\Phi}$ is a stretching
tensor.  Instead of describing the deformation at a material point, we
can, through an abuse of notation, imagine the decomposition as a
prescription of how to generate deformation throughout a material by
imposing a sequence of flows followed by rotations.  In general, each
time interval $t_n$ could have a different flow map and rotation
\cite{Lester_global_2008}, leading to a total flow map
$\boldsymbol{\Psi} = \prod_n \boldsymbol{R}_n \boldsymbol{\Phi}_n$;
however, the more common practice, which we follow here, is to fix
$\boldsymbol{R}$ and $\boldsymbol{\Phi}$, leading to
$\boldsymbol{\Psi} = (\boldsymbol{R} \boldsymbol{\Phi})^n$ being a
function of the time that the flow $\boldsymbol{\Phi}$ operates
between reorientations and the angle of the rotation.  In 2D there is
only one reorientation angle, and periodic reorientation has been used
to successfully design industrial mixing devices
\cite{Speetjens_inline_2014}, biomedical diagnostic devices
\cite{Petkovic_hendra_2017}, and subsurface contaminant remediation
processes \cite{Trefry_subsurface_2012,Cho_field_2019}.  In 3D there
could be in general two angles of reorientation.  However, in the
driven-lid hemisphere only one axis of rotation makes sense, the
rotation about the hemisphere radius perpendicular to the lid, and the
driven-lid hemisphere is an example of the simplest extension from 2D
to 3D of a periodically reoriented flow: a geometry with a 3D flow but
only one reorientation angle.

In the rest of this section we introduce the base flow, discuss the
map that corresponds to its periodic reorientation, identify
symmetries in the reoriented flow, demonstrate its Hamiltonian nature
and briefly introduce the computational methods used to investigate
it; full computational details are given in Appendix~\ref{sec:method}.

\subsection{Periodically Reoriented Hemisphere Flow}

\subsubsection{Lid-Driven Hemispherical Cavity Flow}

\begin{figure}
\centering
\includegraphics[width=\columnwidth]{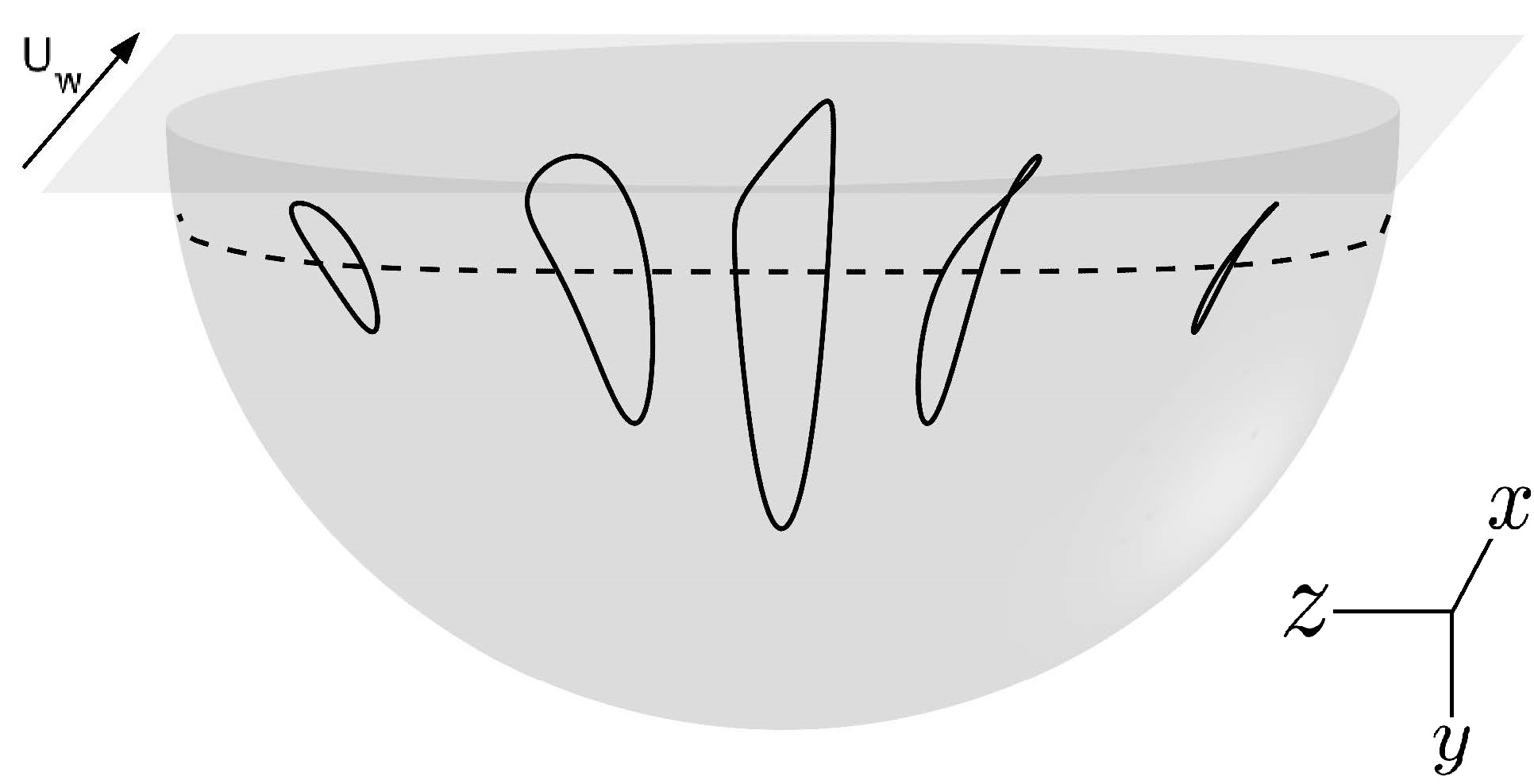}
\caption{Schematic of the base flow of the periodically reoriented lid
  driven hemisphere flow.  The lid moves in the $x$ direction with
  speed $U_w$.  Selected streamlines of the base flow are continuous;
  the stagnation line is dashed.}
\label{fig:ldhs_schematic}
\end{figure}

 
The Periodically Reoriented Hemisphere Flow (PRHF) studied here is a
Stokes flow generated in a hemispherical cavity in which the $x-z$
equatorial plane is driven across the cavity (Figure~\ref{fig:ldhs_schematic}). 
Distance is normalised by the hemisphere radius, $R$, giving a scaled radius of 1.  
The flow domain is thus defined in spherical polar coordinates by 
$r \in [0,1]$, $\theta \in [0,\pi] $ and $\phi \in [0,\pi] $ where $\theta$ is the
polar angle and $\phi$ is the azimuthal angle.  We define the
base flow as the flow generated with uniform lid motion in the
$\hat{\boldsymbol{x}}$ direction ($\phi=0$) with steady speed $U_w$.  
The lid velocity is normalised by $U_w$,  giving a scaled lid velocity of 1. 

 The base flow consists of one roll with
maximum speed at the sliding lid and a stagnation line through the
roll centre that is attached to the hemisphere wall (the dashed line
in Figure~\ref{fig:ldhs_schematic}) at coordinates $
(0, 0.1624, 0.9867)$ and $(0, 0.1624, −0.9867)$.  The base flow is constant in time and in the Stokes limit has two symmetries---a fore--aft
symmetry (reversal-reflection across the $z-y$ plane) and a
left--right symmetry (reflection across the $x-y$ plane).  With two
symmetries all fluid trajectories are closed 1D streamlines
(Figure~\ref{fig:ldhs_schematic}).

The equations of motion for the fluid inside the domain are the steady
Stokes equation, which gives the momentum balance between viscous and
pressure forces as
\begin{equation}
\label{eq:Stokes}
\mu \nabla^2 \boldsymbol{U} = -\bnabla p,
\end{equation}
and the conservation of mass or incompressibility condition is
\begin{equation}
\label{eq:incompressibility}
\bnabla \bcdot \boldsymbol{U} = 0,
\end{equation}
where $\boldsymbol{U}$ is the velocity field, $p$ is the pressure and
$\mu$ is the viscosity.  Although the domain is hemispherical, we
calculate the velocity field in cylindrical polar coordinates
$(\rho,\vartheta,z)$ using the spectral-element Fourier code {\it
  semtex}~\cite{Blackburn2004759}.  The velocity field is then
interpolated spectrally onto a uniform grid in spherical polar
coordinates $(r,\theta,\phi)$ with $n_r=100$, $n_\theta = 100$ and
$n_\phi = 320$.  The interpolated cylindrical polar velocity field on
this grid is then transformed to spherical polar velocity components
$(v_r,v_\theta,v_\phi)$.  A continuous analytically divergence-free
velocity field is created from this grid velocity data using the
method of Ravu~{\it et al.}~\cite{Ravu:2016jcp}.  Finally, MATLAB's
ode45 (with relative error $5 \times 10^{-14}$ and absolute error
$10^{-15}$) is used to integrate fluid particle trajectories using the
divergence-free velocity field defined analytically at all points
inside the hemisphere, and observed Lagrangian coherent structures
have been tested to be robust with respect to numerical integration
error.  The velocity discontinuity at the rim of the sliding lid and
hemisphere wall requires special consideration for the numerics; this
and additional computational details are in Appendix~\ref{sec:method}.



\subsubsection{Periodic Reorientation}  

\begin{figure}
 \centering
 \includegraphics[width=0.95\columnwidth]{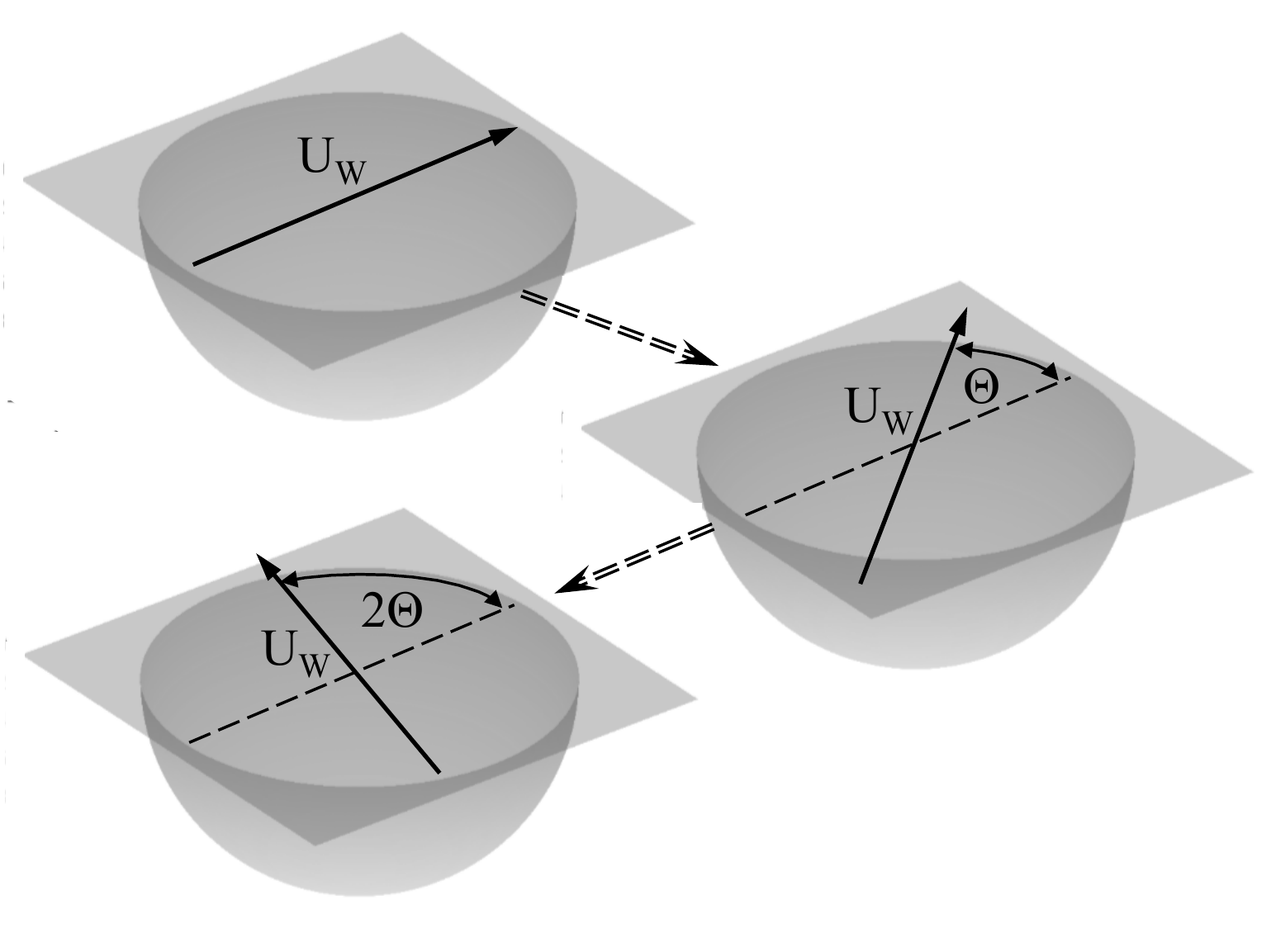}
 \caption{Schematic showing two steps of reorientation of the base
   flow by angle $\Theta$.}
 \label{fig:Reorient} 
\end{figure}

A time-dependant flow is created from the base flow by periodically
changing the direction of lid motion, n.b.\/ \emph{not} through
rotation of the lid.  The reorientation angle is defined as the angle
$\Theta$ between the directions of successive lid motions, and is one
of the key control parameters of the system.  The angle of lid motion
increases monotonically at each reorientation, with
Figure~\ref{fig:Reorient} showing the first two such reorientations.
In the periodically reoriented flow, the time between reorientations
is denoted by $\tau$.  Time is non-dimensionalized by $R/U_w$ giving a
non-dimensional period of $\beta = U_w\tau/R$.  The parameter $\beta$
is the second key control parameter of the reoriented system.

In the reoriented system, the velocity field satisfies
\begin{equation}
\boldsymbol{v}(\boldsymbol{x},t+\beta)= R_{\Theta}(\boldsymbol{v}(\boldsymbol{x},t)),
\end{equation}
or
\begin{equation}
\boldsymbol{v}(\boldsymbol{x},t)=R_{(m\Theta)}(\boldsymbol{u}(\boldsymbol{x})), \quad m = \lfloor{\frac{t}{\beta}}\rfloor
\end{equation}
for all $t>0$, with $\lfloor{.}\rfloor$ the floor function, $\boldsymbol{u}$ the base flow velocity field and
$R_{\Theta}$ the rotation operator for an anti-clockwise rotation
about the $\hat{y}$ axis.

An equivalent--and computationally more convenient--way of defining
the time-periodic flow is to move the lid in the $x-$direction
for a period $\beta$ followed by rotating the entire flow domain by an angle
$-\Theta$. In other words, we work with a frame that is invariant with respect to the lid direction. In this way the time periodic flow map is written
\begin{equation}
\boldsymbol{\Psi}_{(\beta, \Theta)}=R_{-\Theta}\boldsymbol{\Phi}_{\beta}, 
\label{eq:Periodic_gflow_map}
\end{equation}
where $\boldsymbol{\Phi}_{\beta}$ is the base flow map for lid displacement
$\beta$ and $R_{-\Theta}$ is rotation of the hemisphere in a clockwise
direction.   All analysis will be undertaken with this as the basic
building block of the time-periodic flow. We often drop the subscript $(\beta,\Theta)$ from equation~\ref{eq:Periodic_gflow_map} and simply write the time-periodic map as $\bm{\Psi}$, however $(\beta,\Theta)$ will always be implied. To reduce the complexity of higher dimensions, we always work with trajectories that are stroboscopic with respect to the reorientation period $\beta$. In this way, a trajectory contains fluid particle positions that are equally spaced in time for period $\beta$. They are also called temporal Poincar\'{e} sections and are obtained via
\begin{equation}
\label{eq:strobo_map}
\bm{x}_{k+1}=\bm{\Psi}(\bm{x}_k), \quad \bm{x}_k=\bm{x}(k\beta) ,
\end{equation}
where $\bm{x}_k$ is a particle position after $k$ periods (i.e. time=$k\beta$).


In the Stokes limit trajectories of the PRHF are confined to
spheroidal surfaces.  While in Appendix~\ref{app:invariants} we give a
formal proof of both the existence of a single invariant and that the
motion on each spheroidal shell is Hamiltonian, the simplest way to
see these results is to deform the hemisphere flow (formally via
conformal transform) to that contained within a sphere.  The
reorientation step in the flow map is then a solid body rotation about
the $\hat{y}$ axis.  The lid-driven shear is a non-uniform rotation
about the $\hat{x}$ axis.  Composed rotations do not change the
radius: the radius of every trajectory is invariant to the flow map
(\ref{eq:Periodic_gflow_map}), and every spherical surface at constant
radius contains a family of trajectories confined to that surface.
Reversing the conceptual conformal transform does not change the
topology, it merely deforms the nested spherical surfaces into nested
spheroids.  Also in Appendix~\ref{app:invariants} we generalize this
result to all lid-driven cavities by reducing the equations of motion
to Hamiltonian-Poisson form.


\subsection{Symmetries}
\label{sec:symmetry}

Symmetries of the base flow $\boldsymbol{u}(\boldsymbol{x})$ are
intimately tied to symmetries of Lagrangian trajectories and play an
important role in the organization of the global 3D Lagrangian
transport structure
\cite{Speetjens_topological_2006,Metcalfe_chaos_2010,Speetjens_inline_2014}.
The base flow velocity field defines the trajectory of a fluid particle with
initial condition $\boldsymbol{x}(0)=\boldsymbol{x}_0$ via the
kinematic equation
\begin{equation}
\frac{d\boldsymbol{x}}{dt}=\boldsymbol{u}(\boldsymbol{x}).
\label{eq:kinematic3d}
\end{equation}
After a time $t$, the particle will have moved to $x(t)$ and this
position can be written as the continuous flow map
\begin{equation}
\boldsymbol{x}(t)=\boldsymbol{\Phi}_t(\boldsymbol{x}_0).
\end{equation}

We introduce two maps $S_x$ and $S_z$ that reflect a particle position about $x=0$ and $z=0$: 
\begin{align}
\quad S_x &: (x,y,z) \rightarrow (-x,y,z) \\
\quad S_z &: (x,y,z) \rightarrow (x,y,-z)
\end{align}
Fore-aft symmetry of the base flow ({\it i.e.}~in the $x$--direction)
results in a time-reversal symmetry,
\begin{equation}
\label{eq:time_rev_sym_baseflow}
\boldsymbol{\Phi}_t = S_x {\boldsymbol{\Phi}_t}^{-1} S_x, 
\end{equation}
Left-right symmetry of the base flow ({\it i.e.}~in the
$z$--direction) results in a reflection symmetry,
\begin{equation}
\label{eq:reflec_sym_baseflow}
\boldsymbol{\Phi}_t = S_z {\boldsymbol{\Phi}_t} S_z, \end{equation}
So far, we discussed the symmetries of the base flow. We will discuss the symmetry of the reoriented flow in the following.

\begin{figure}
\centering 
\includegraphics[width=0.5\columnwidth]{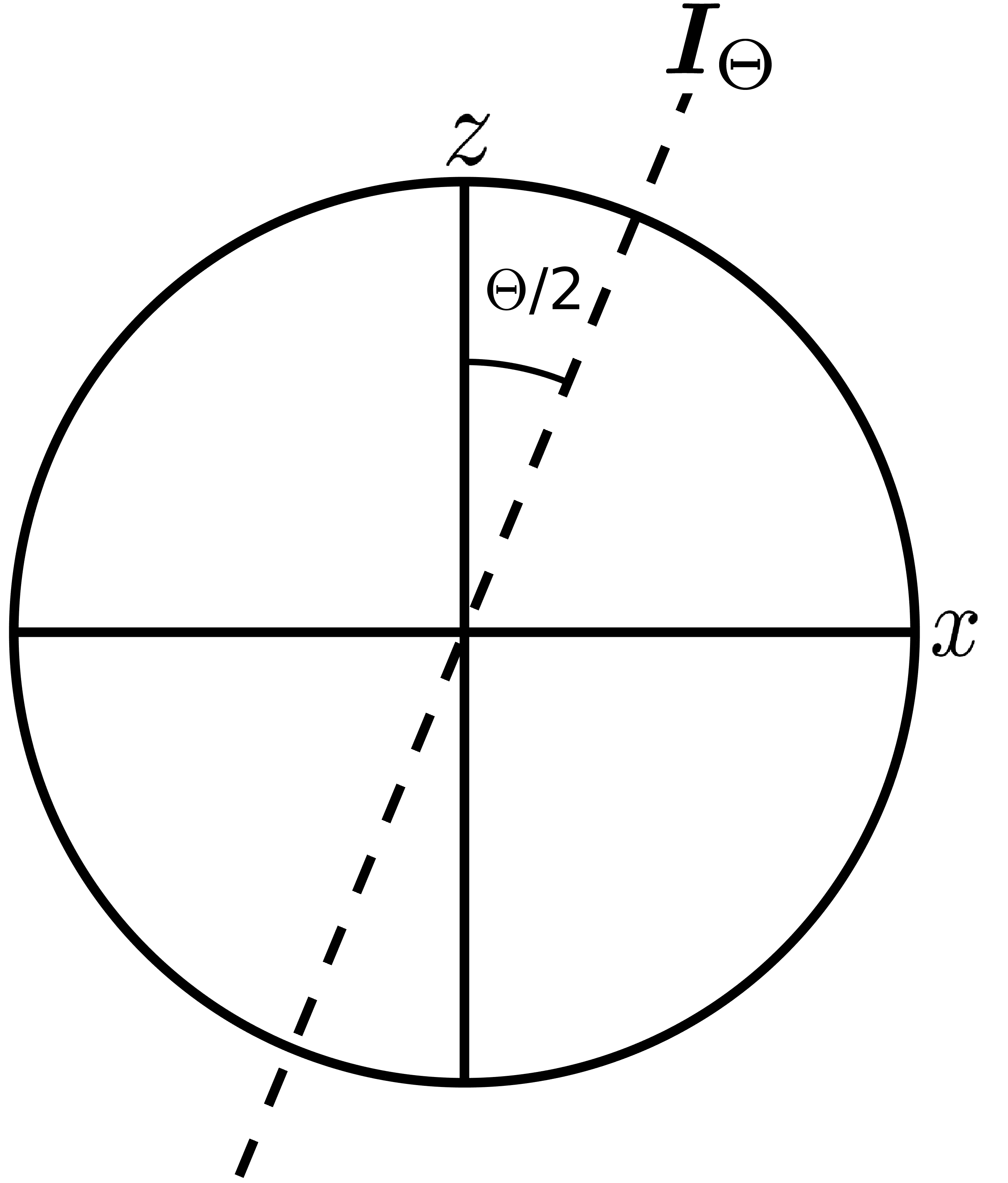}
\caption{ Symmetry plane for $\Theta=\pi/4$. }
\label{fig:sym_plane_angle}
\end{figure}

The time-periodic map $\boldsymbol{\Psi}$ has a time reversal
symmetry that can be seen by substituting
equation~\ref{eq:time_rev_sym_baseflow} (with $t=\beta$) into
equation~\ref{eq:Periodic_gflow_map}:
\begin{align*}
\boldsymbol{\Psi}  & = R_{-\Theta} \boldsymbol{\Phi}_{\beta} \\
& = R_{-\Theta}  S_x {\boldsymbol{\Phi}_{\beta}}^{-1} S_x \\
   & = R_{-\Theta}  S_x {\boldsymbol{\Phi}_{\beta}}^{-1} R^{-1}_{-\Theta} R_{-\Theta} S_x \\
   & = R_{-\Theta}  S_x {(R_{-\Theta}\boldsymbol{\Phi}_{\beta})}^{-1} R_{-\Theta}  S_x \\
   & = R_{-\Theta}  S_x  \boldsymbol{\Psi}^{-1} R_{-\Theta}  S_x .
 \end{align*}
Defining 
\begin{equation}
\label{eq:sym_plane}
R_{-\Theta}  S_x = S_{\Theta}, 
\end{equation}
we then write $\boldsymbol{\Psi}$ as
\begin{equation}
\boldsymbol{\Psi} = S_{\Theta} \boldsymbol{\Psi}^{-1} S_{\Theta}.  
\label{eq:PerTimeRev}
\end{equation}
This is the reversal-reflection symmetry of the map
$\boldsymbol{\Psi} $, where from equation~\ref{eq:sym_plane} $S_\Theta$ is the
map that reflects a particle about the plane $\theta=\pi/2-\Theta/2$
(figure~\ref{fig:sym_plane_angle}).  Defining the set of all the
points on the plane $\theta=\pi/2-\Theta/2$ as
$\boldsymbol{I}_\Theta$, it immediately follows that
\begin{equation}
\boldsymbol{I}_\Theta=S_{\Theta}\boldsymbol{I}_\Theta.
\end{equation} 
All periodically reoriented flows have this symmetry, another example of which is the driven-lid cylinder, which has been extensively
studied
\cite{Malyuga_stokes_2002,Speetjens_stokes_2004,Speetjens_inertia_2006,Speetjens_merger_2006,Pouransari_inertia_2010}
.

The importance of the symmetry (equation~\ref{eq:PerTimeRev}) is
twofold.  First, P1 points (and by extension P1 lines) must lie on the
symmetry plane $\boldsymbol{I}_\Theta$.  We prove this in
Appendix~\ref{app:symmetry}.  Second, we only need to consider Lagrangian transport on one side of the symmetry plane
$\boldsymbol{I}_{\Theta}$ because they will be mirrored across this
plane.

The symmetry manifests as an invariant and the flow becomes an action-angle-angle system in canonical coordinates \cite{Mezic_symmetry_1994, Haller_symmetry_1998}. In action-angle-angle systems, isolated periodic points can not exist, only periodic lines with
the ends attached to the boundary or closed periodic lines exist (see 
appendix~\ref{app:no_iso_periodic_points}).

\subsection{Lagrangian Skeleton}

In one invariant flows, as isolated periodic points cannot occur (see
appendix~\ref{app:no_iso_periodic_points}), periodic lines and their
manifolds constitute all of the Lagrangian structure, and periodic
lines may be called the Lagrangian skeleton of such flows.  As
generally, lower order periodic lines are more important in organising
Lagrangian transport than higher-order periodic lines, P1 lines, being
the most important of all, become the base of the Lagrangian skeleton
of the PRHF system.

In the Stokes limit the PRHF has a single P1 line. In the limit
$\beta \rightarrow 0$ there is no flow, only rotation, and all
trajectories are circular about the $\hat{y}$ axis and the line
between 0 and 1 on the $\hat{y}$ axis is the sole P1 line.  As $\beta$
increases above zero this distinguished P1 line bends but remains
anchored to the three non-trivial period-1 points (of which two are
fixed points) for all values of $\beta$ and $\Theta$.  What are these
three points?  One is the centroid of the foliation of shells, which
is a fixed point of the map for all $(\beta,\Theta)$; such a central
fixed point is generic for interior flows comprised of spheroidal
shells.  Another fixed point of the flow map is the point at the base
of the hemisphere intersected by the $y$--axis.  The third is an
attachment point in the symmetry plane {\it on} the sliding lid that
moves from the center point of the lid at $\beta = 0$ to the rim at
$\beta = 2 \sin(\Theta/2)$, where it remains for
$\beta>2 \sin(\Theta/2)$ due to the velocity discontinuity at this
point. The path of the lid attachment point as $\beta$ increases is
elaborated in Appendix~\ref{sec:lid_attach_point}.

\begin{figure}
\centering 
\begin{tabular}{cc}
\includegraphics[width=0.9\columnwidth]{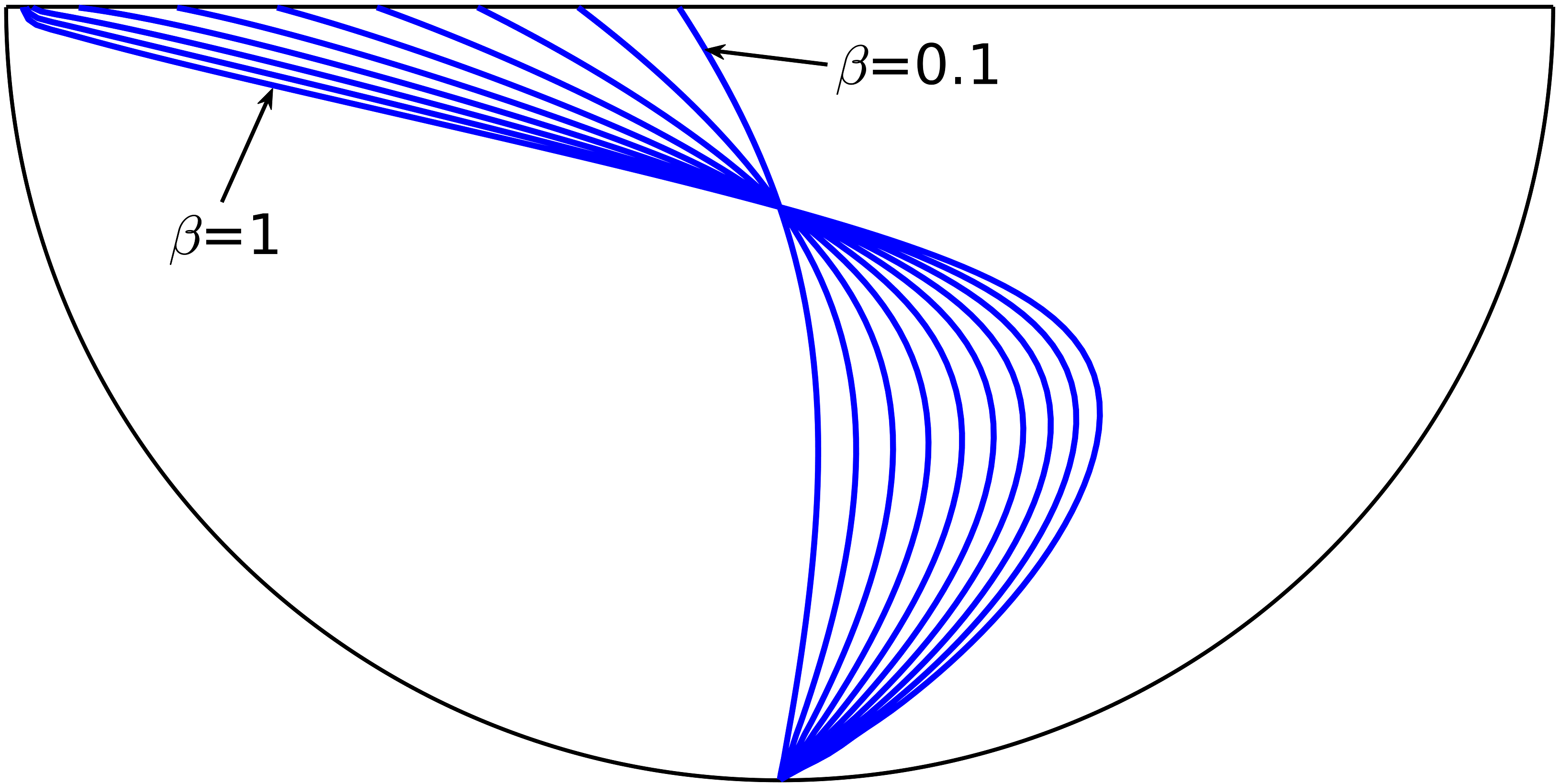} & \hspace*{-1cm} $\beta \in [0.1,1]$  \\
\includegraphics[width=0.9\columnwidth]{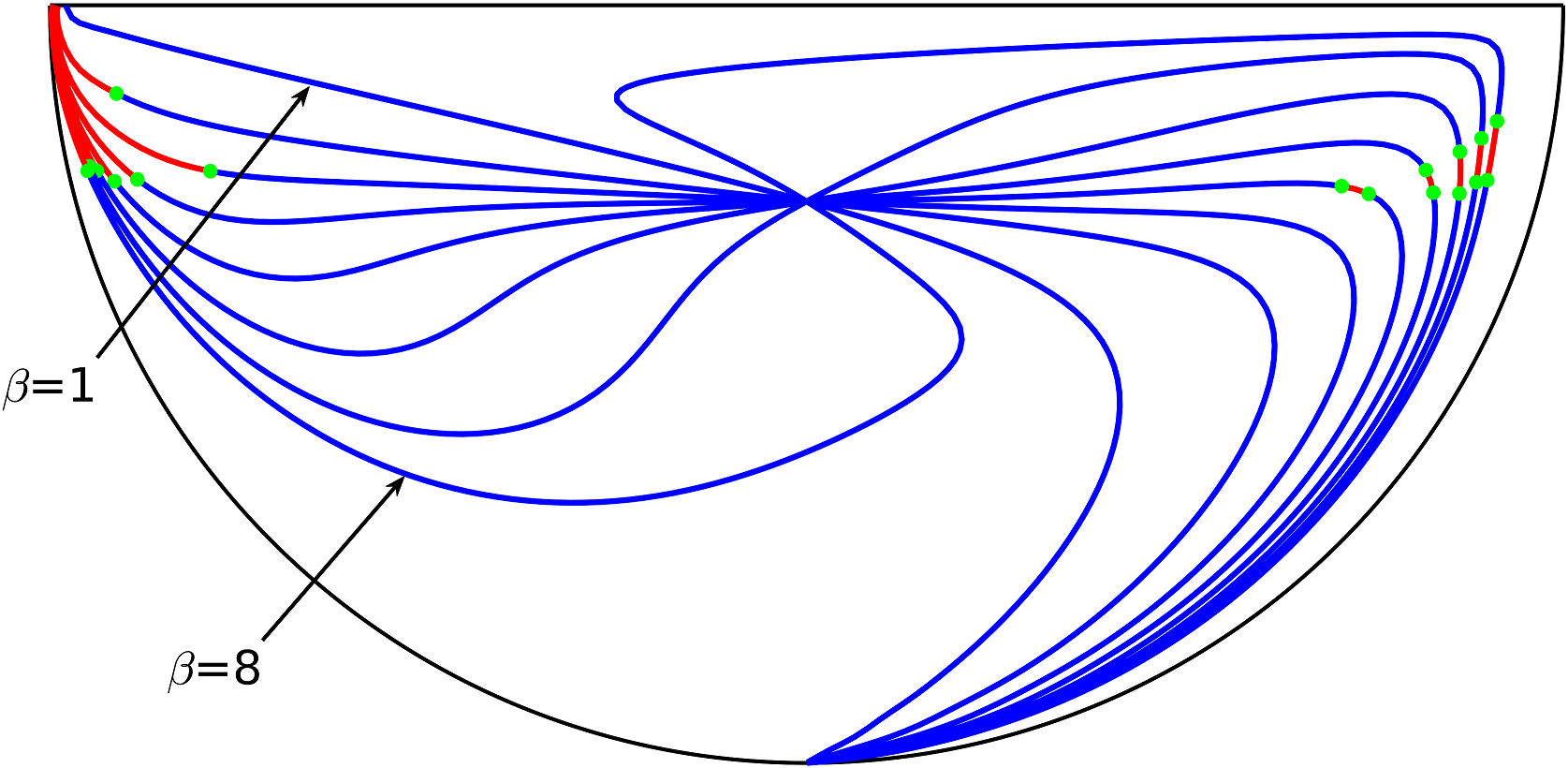} & 
\hspace*{-1cm} $\beta \in [1,8]$ \\
\includegraphics[width=0.9\columnwidth]{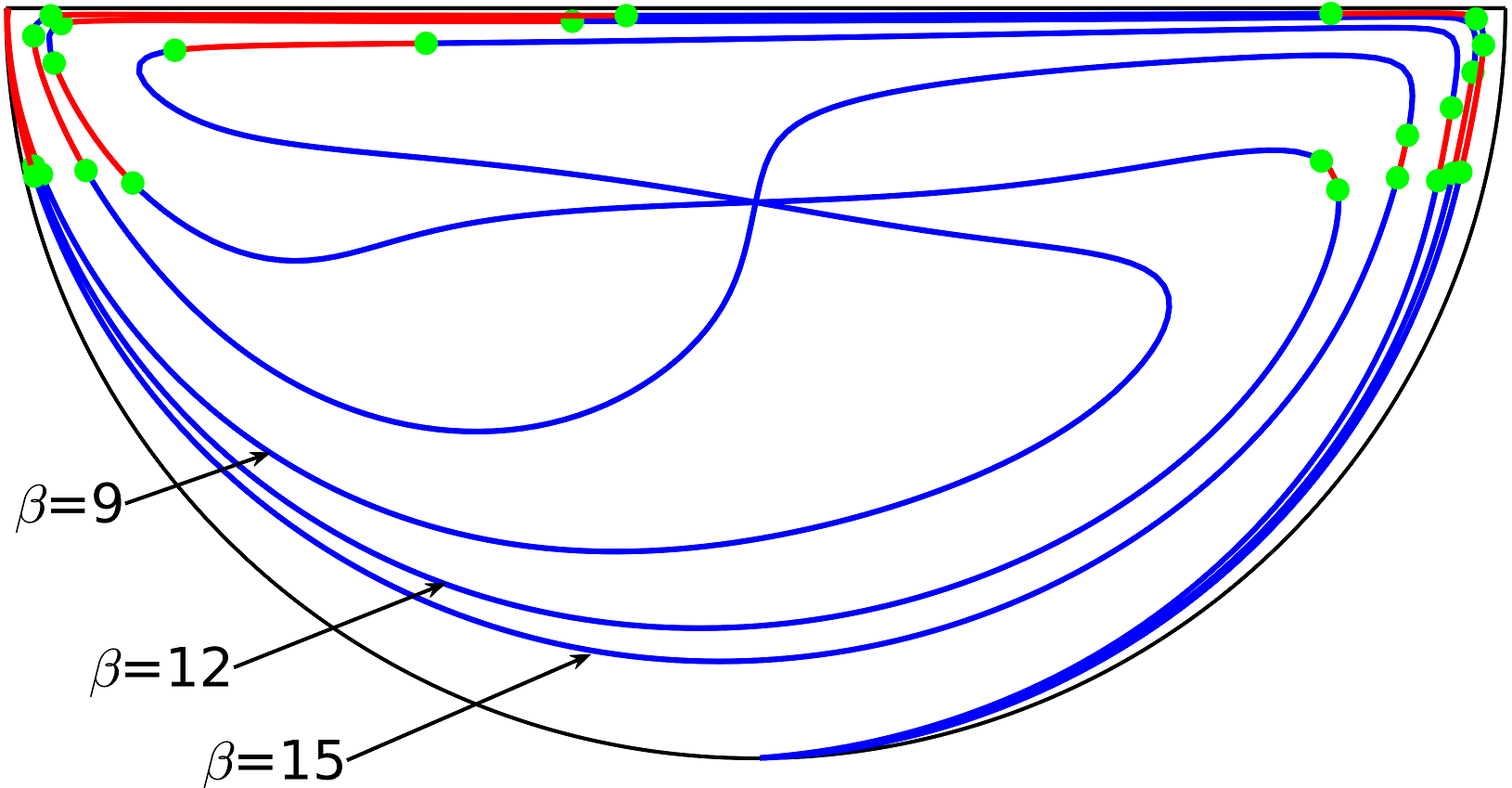} & 
\hspace*{-1cm} $\beta \in [9,12,15]$ 
\end{tabular}
\caption{P1 lines viewed perpendicular to the symmetry plane in
  nomal space coordinates for $\Theta=\pi/4$ and indicated values of
  $\beta$.  Blue (red) are elliptic (hyperbolic) segments; green dots
  are degenerate parabolic points.}
\label{fig:P1_lines}
\end{figure}

Because this main P1 line approaches arbitrarily close to the boundary
with increasing $\beta$, and all P1 lines must lie in the symmetry
plane, it follows there are no other P1 lines attached to the
boundary. A numerical search of the symmetry plane for closed loop P1
lines detected none of size greater than 0.4\% of the hemisphere
diameter.  We conclude that for the Stokes PRHF there is only the
single P1 line that runs from the apex of the hemisphere through the
centroid to an attachment point on the lid or rim.
Figure~\ref{fig:P1_lines} shows for $\Theta = \pi/4$ the sole P1 line
in the symmetry plane for selected values of $\beta$ from 0.1 to 15.
Different values of $\Theta$ do not qualitatively change the picture.
In this figure, lines are coloured by the local character of the line:
blue for elliptic regions and red for hyperbolic regions.  Green dots
are degenerate parabolic points, which play an important role in
organizing transport, as we elaborate in
section~\ref{sec:resonances_on_periodic_lines}.

From figure~\ref{fig:P1_lines}, it is seen that the tangent to a P1
line at the central stagnation point rotates anti-clockwise with
increasing $\beta$.  As $\beta$ increases from $0$, the end of the P1
line which is attached to the lid moves toward the hemisphere boundary
and stops at the rim when $\beta=2 \sin(\Theta/2)$ (see
Appendix~\ref{sec:lid_attach_point}).  Beyond this value of $\beta$,
the P1 attachment point stays at the rim for all $\beta$ values. The
other end of the P1 line is fixed at the bottom of the hemisphere.
Because both the ends of P1 lines are fixed for a given $\beta$ and
the tangent to the P1 line at the central stagnation point rotates
anti-clockwise with increasing $\beta$, the length of P1 line
increases with increasing $\beta$.  As the rotation occurs at the
central stagnation point, the P1 line is pushed towards the boundary.


\subsection{Summary of PRHF Lagrangian Structure}

Due to the single invariant, PRHF consists of nested topological
spheroids (or shells) that correspond to level sets of this invariant,
each of which has a 2D Hamiltonian flow.  Such 2D flows have been
studied extensively, and the range of possible transport structures
and their associated dynamics are well understood
\cite[for example]{Cartwright_chaotic_1996, Malyuga_stokes_2002,
  Speetjens_stokes_2004, Speetjens_inertia_2006,
  Speetjens_merger_2006, Pouransari_inertia_2010,
  Moharana_sphere_2013, Smith_degenerate_2016}.
Because the P1 line is constrained to lie entirely on the symmetry
plane $\boldsymbol{I}_\Theta$ and passes from the hemisphere boundary
to the lid boundary through the centroid of all the shells, it must
pierce each shell at least twice.  Where the P1 line intersects a
shell, the Hamiltonian flow on that shell shares its local deformation
character with the P1 line at that point.

\begin{figure}
\centering 
\includegraphics[width=\columnwidth]{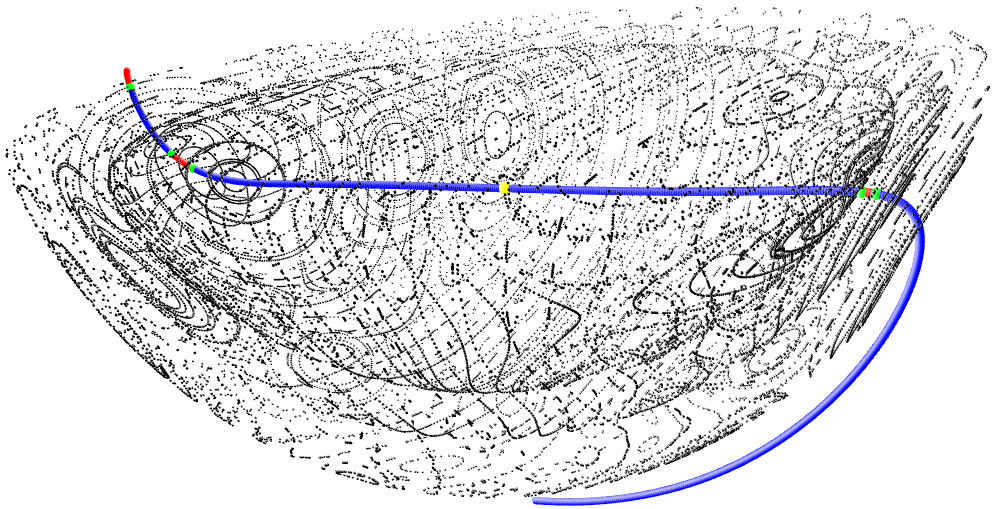}
\caption{P1 line and Poincare maps of flow on three shells.  Oblique
  view.}
\label{fig:poinc_map_shell_3}
\end{figure} 

Figure~\ref{fig:poinc_map_shell_3} shows an example of the P1 line,
coloured by its deformation character as in figure~\ref{fig:P1_lines},
and views of several shells with Poincar{\'e} sections.  The picture
appears complicated with the only organizing principle being the
sharing of deformation character where the P1 line pierces a shell.
In the next section we show how the degenerate points along periodic
lines determine the organizational skeleton of 3D Lagrangian transport
in one-invariant volume-preserving flows.

\section{Resonances On Periodic Lines}
\label{sec:resonances_on_periodic_lines}

In bifurcation theory, the topology of a dynamical system changes as
the parameter of the system goes through a bifurcation (or critical)
point~\citep{Kuznetsov_Bifurcation_1998}.  Analogously, in the PRHF, a
type of bifurcation occurs when moving from one shell to another at
degenerate points on periodic lines.  Degenerate points have zero net
deformation periodically. At such points, in the shell
normal direction the local topology on adjacent shells changes and
this behaviour can be viewed in terms of classical planar bifurcation
theory by establishing an analogy between the PRHF and a 2D system
with a parameter.  Because fluid particles move on nested spheroids,
these spheroids can be considered as the direct product of a two
dimensional phase space (angle variables $\theta_1, \theta_2$) and a
parameter space (action variable $I$, that we also call ``shell
number'').  Conceptually the three-dimensional PRHF can be reduced to
a two-dimensional system with a parameter $I$.

Importantly, deformation at degenerate points is constrained in such a
way that zero net deformation in the neighbourhood of an $n^{th}$
order degenerate point occurs after $n$ periods. Degenerate points are
called resonance points in classical planar bifurcation
theory~\citep{Kuznetsov_Bifurcation_1998}.  An $n^{th}$ order
degenerate point on a P1 line is termed a 1:$n$ resonance point. For
example, a 1:3 resonance point is found on a P1 line and has zero net
deformation in its neighbourhood after 3 periods.  We can categorize
resonant degenerate points as being of type $m$:$p$ if they occur on a
period-$m$ line and have an order $p$, (i.e.\/ zero deformation after
$p$ periods), where $p>m$ and $p =n \times m$, where $n$ is a positive
integer.  Note that not every combination of $m$ and $p$ can occur
because a point cannot simultaneously be a period-$m$ and period-$p$
point, e.g.\/ resonances such as 2:3 can not occur.

The significance of a $m$:$p$ resonance point does {\em not} lie
chiefly in the zero net deformation in the neighbourhood of the
point. We will show that the chief significance is that the
constraints at degenerate points on periodic lines in 3D flows require
the creation and coordination of extended spatial structures of
period-$p$ lines that have specific sequences of elliptic and
hyperbolic segments (each segment separated by another degenerate
point).  These in turn imprint specific Lagrangian structures onto
sizeable 3D volumes of the fluid that extend well beyond the local
neighbourhood of the degenerate point.  It is found that, at
degenerate points, lower order periodic lines and higher order
periodic lines intersect. For example on a P1 line, $n$ period-$n$
lines pass through a 1:$n$ resonance point except for 1:1 and 1:2
resonance points. At a 1:1 resonance, the P1 line is tangent to the
invariant surface, and the degenerate nature of this point results in
a cusp appearing in the Lagrangian structure on this shell.  A
period-2 line passes through a 1:2 resonance point on a P1 line. The
reason only one period-2 line exists at a 1:2 resonance point instead
of two period-2 lines is discussed in
section~\ref{subsec:resonance_1_2}.  An analogy can be made between
resonance points on a P1 line to resonance points on a higher order
periodic line in the following way. A period-$m$ line of map
$\bm{\Psi}$ can be considered as a P1 line of the map $\bm{\Psi}^m$.
This way, resonance points on the period-$m$ line are treated as
resonance points on a period -1 line but with the map $\bm{\Psi}^m$.
In general, on a period-$m$ line, $n$ $(=p/m)$ period-$p$ lines
intersect at a $m$:$p$ resonance point. This can be seen from the
following: consider a 1:3 resonance point on a period-1 line.  There
will be three period-3 lines which intersect with the 1:3 resonance
point. Again, consider a 3:9 resonance point on one of the three
period-3 lines.  There will be 3:9 resonance points similarly on the
other two period-3 lines, totalling three 3:9 resonance points. Three
period-9 lines pass through each of these 3:9 resonance points,
totalling 9 period-9 lines which belong to the three 3:9 resonance
points.  These resonance bifurcations are local bifurcations in that
they occur in the neighbourhood of periodic (or fixed) points.

Because the Lagrangian structure on a shell is completely determined
by periodic line piercings, this hierarchical set of resonance points
and accompanying periodic lines allows {\em all} the complicated
structure exemplified in figure~\ref{fig:poinc_map_shell_3} to be
uncovered in a systematic manner.  This is undertaken in practice by
finding the P1 lines and calculating the local and extended 1:$n$
structures, then finding the P2 lines and calculating the local and
extended 2:$2n$ structures, etc.\/, to whatever degree of detail is
required.

The essentially 2D behaviour on invariant surfaces was noted as
typical of periodic lines in \cite{Malyuga_stokes_2002} and
\cite{Pouransari_inertia_2010} but the consequences of the degenerate
points on global transport were not fully elaborated.  Period-doubling
and period-tripling bifurcations, which are associated with second
order and third-order degenerate points, have also been observed
\citep{Dullin_1999, Mullowney_2005, Smith_degenerate_2016}, but these
studies did not explore the general $m$:$p$ resonance structure.  One
of the novelties of the current paper is to systematically expose the
set of possible structures.


Although resonance points can be identified on a periodic line of any
order and used to calculate corresponding higher periodic lines, the
analysis of resonance bifurcation points in this study is restricted
mainly to P1 lines.  In the rest of this section we first elaborate
the effects of constraints on deformation on transport, then focus on
the P1 line and give examples of the first four 1:$n$ spatially
extended resonance structures.  We also provide an example of a 2:6
resonance to demonstrate resonance points on higher-order periodic
lines.

\subsection{Constraints on Eigenvalues of the Deformation Tensor}
\label{sec:resonances_eigenvalues}

\begin{figure}
\centering 
\includegraphics[width=0.8\columnwidth]{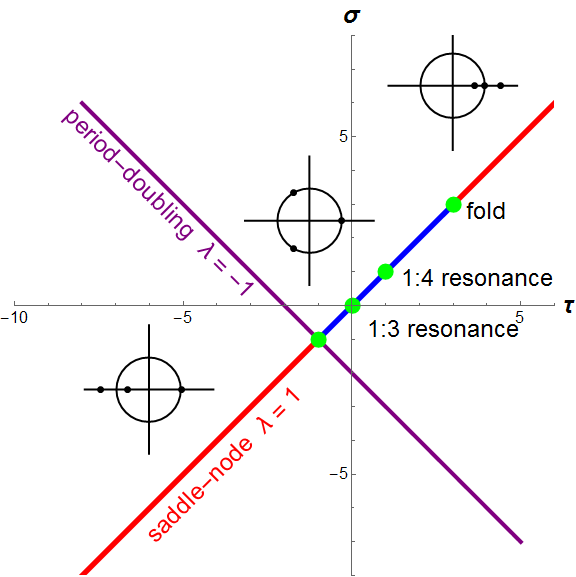}
\caption{Eigenvalues $\lambda$ of the deformation tensor of a
  volume-preserving map plotted in terms of the trace $\tau$ and
  second trace $\sigma$ tensor invariants. As there is no deformation
  along any periodic line, one eigenvalue on every periodic line is
  one, and the other two eigenvalues lie along the diagonal line
  $\sigma = \tau$.  The blue line segment denotes complex conjugate
  pairs---elliptic segments of the periodic line---and the red line
  segments denote real pairs---hyperbolic segments of the periodic
  line.  Green dots locate the four strong resonances.  The inset
  complex planes schematically depict the eigenvalues on the unit
  circle.  Adapted from~\protect\cite{Lomeli_quadratic_1998}.}
\label{fig:eigenvalue_trace}
\end{figure}

Classification of periodic points by the eigenvalues of the
deformation tensor is similar to the classification of critical points
of a flow field \cite{Chong_general_1990}.  Like all $3 \times 3$ matrices, the
deformation tensor $\boldsymbol{F}$ has three matrix invariants: the trace
$\tau \equiv tr\left(\boldsymbol{F}\right)$; second trace
$\sigma \equiv \frac{1}{2}\left(tr\left(\boldsymbol{F}^2\right) -
  tr^2\left(\boldsymbol{F}\right)\right)$;
and, determinant $\det\left(\boldsymbol{F}\right)$.  In terms of the
invariants, the characteristic equation for the eigenvalues
$\lambda_i$ of $\boldsymbol{F}$ is
\begin{equation}
\lambda^3 - \tau \lambda^2 + \sigma \lambda - \det\left(\boldsymbol{F}\right) = 0.
\end{equation}
Volume-preservation constrains the deformation such that
$\det\left(\boldsymbol{F}\right) = \lambda_1 \lambda_2 \lambda_3 = 1$
everywhere in the flow.  An additional constraint appears on periodic
lines.  Because each point on a period-$n$ line \emph{must} return to
the same point after $n$ periods, there cannot be any deformation
tangent to a periodic line, and thus one of the deformation
eigenvalues must equal one on a periodic line.  We define $\lambda_3$
as the eigenvalue associated with the direction tangent to the
periodic line, hence $\lambda_3=1$ always, and $\lambda_1$,
$\lambda_2$ are associated with the directions tangent to the shell.
This turns out to constrain the trace and the second trace to be equal
on a periodic line (see
Appendix~\ref{app:invariants_deformation_tensor}).  Figure~\ref{fig:eigenvalue_trace}, adapted from
Lomeli and Meiss~\cite{Lomeli_quadratic_1998}, is part of the general
stability diagram for a volume-preserving map.  Plotted in the
$(\tau,\sigma)$ parameter space, all eigenvalues of the deformation
tensor on periodic lines correspond to points that lie on the line
$\tau = \sigma$.  In the blue part of the $\tau = \sigma$ line,
deformation near the periodic line is elliptic with rotation in closed
orbits around the periodic line. The corresponding
eigenvalues transverse to the periodic line's tangent direction appear
as complex conjugate pairs of the form
$\left(e^{i \alpha}, e^{-i \alpha}\right)$ and lie on the unit circle
in the complex plane.  In the two red parts of the $\tau = \sigma$
line deformation near the periodic line is hyperbolic with a sheet
stretching material away from the line in one direction and an equal
amount of compression in a sheet towards the line.  The corresponding
eigenvalues transverse to the line's tangent direction are purely real
and inverse to each other, i.e.\/ of the form
$(\lambda, \lambda^{-1})$.


At an $n^{th}$ order degenerate point on the P1 line (a 1:$n$
resonance point), $\boldsymbol{F}^n=I$, which in turn implies
$\lambda_1^n=1$, $\lambda_2^n=1$ and $\lambda_3^n=1$. The eigenvalues
of the deformation tensor at such a point are thus $\lambda_3=1$ and
$\lambda_{1,2}=e^{\pm i (2 \pi /n) } $.  The local rotation $\alpha$
at the $n^{th}$ order degenerate point is
\begin{equation}
\alpha=\frac{2\pi}{n}.
\end{equation}
The trace of the deformation tensor $\boldsymbol{F}$ at an $n^{th}$
order degenerate point ($\bm{x}_{1:n}$) is (via
equation~\ref{eq:traceeqn_degen})
\begin{equation}
\label{eq:traceeqn_degen}
\tau(\bm{x}_{1:n})=\Tr \boldsymbol{F}(\bm{x}_{1:n}) = 1+2\cos (2\pi/n).
\end{equation}
Eigenvalues, the local rotation angle, and the trace of the deformation
tensor for general $m$:$p$ resonance points are given in
table~\ref{tab:degen_pt_eigenvalues}.


\begin{table}
\centering
\caption{ Resonances, their corresponding eigenvalues, and trace
  values on periodic lines; $\lambda_3=1$ for each periodic point}
\label{tab:degen_pt_eigenvalues}
\begin{sideways}
\begin{tabular}{ | M{0.8cm} | M{1.3cm}| M{1.5cm} | M{1.8cm} |M{0.4cm} |M{0.6cm} | M{0.4cm} |  M{1.2cm} |} 
\hline $\alpha$ &
 $\lambda_1$  & $\lambda_2$ &  Trace & \begin{turn}{270}P1 line ($\bm{\Psi}$)\end{turn} & \begin{turn}{270} P2 line ($\bm{\Psi}^2$) \end{turn}& \dots & \begin{turn}{270} P$m$ line ($\bm{\Psi}^m$) \end{turn}\\ 
\hline
0 & 1  & 1 &  3 & 1:1 & 2:2 & \dots & $m$:$m$ \\ 
$\pi$ &  -1 & -1 & -1 & 1:2 & 2:4 & \dots & $m$:$2m$ \\
2$\pi/3$ & -$\frac{1}{2}+\frac{\sqrt{3}}{2} i$ & $-\frac{1}{2}-\frac{\sqrt{3}}{2} i$ &  0  & 1:3 & 2:6 & \dots & $m$:$3m$ \\ 
$\pi/2$ & $i$ & $-i$ &  1 & 1:4 & 2:8 &  \dots & $m$:$4m$ \\
\vdots & \vdots & \vdots & \vdots & \vdots &  \vdots &  \vdots &  \vdots \\
$2 \pi /n$ & $e^{i (2 \pi /n) }$ & $e^{-i (2 \pi /n) }$ &  $1+2\cos(\frac{2 \pi}{n})$ & 1:$n$ & 2:2$n$ &  \dots & $m$:$n \times m$ \\
\hline
\end{tabular}
\end{sideways}
\end{table}

Because the trace values of resonance points are known, we can
identify these points by finding the roots of
$\tau(\boldsymbol{x}) - \tau({\bm{x}_{m:p}})=0$, where $\bm{x}$ is a
point on the periodic line and $\bm{x}_{m:p}$ is a $m$:$p$ resonance
point.  Once the resonance points are found, the corresponding higher
order periodic lines can then be obtained using the method discussed
in appendix~\ref{app:calc_higher_order_periodic_lines_stokes}.

While there are an infinite number of resonances possible,
corresponding to every root of unity, we will restrict our examination
to the first four resonances (shown in
figure~\ref{fig:eigenvalue_trace}), known classically from planar
bifurcation theory as the ``strong'' resonances.  As higher
resonances, $(n \ge 5)$, are expected to have subharmonic bifurcation
solutions only when exceptional conditions hold
\cite{Gelfreich_resonant_2002}, we focus on the cases $(n \le 4)$ that
will generically be encountered.


\subsection{Parameterization of Periodic Lines}

To determine the fluid transport on a shell, it is necessary to first
identify the periodic line piercings on the shell and their stability.
These piercings are periodic points of the 2D flow on that shell.
Even after calculating a periodic line numerically, it is difficult to
infer how many times a periodic line pierces a particular shell.  In
order to more clearly identify this, we adopt the following approach.
First, we define arc-length along a periodic line as the distance
along the line from the bottom attachment point
$(r = 1, \theta =\pi/2,\phi=\pi/2)$.  Each point along the line thus
has a unique value of arc length.

Then, we enumerate shell number at each point along the line
arbitrarily in such way that shell number has a value of zero at the
central fixed point and one on the lid and hemisphere boundaries.
Shell number is a proxy for the action variable.  Although this is not
the only way to number shells, the choice makes no difference to the
arguments that follow.  Finally, the shell number versus arc length
for the periodic line can be plotted, and the number of piercings of
any shell is the number of times this curve crosses that shell number.
In this plot, we also colour code the stability of each point (i.e.\/
elliptic, hyperbolic or degenerate) along the line.  As will be seen
in the following sections, this will allow the number of piercings to
be rapidly determined and will help in understanding global transport
in this flow.

We now consider the structure associated with each of the strong resonances in turn.

\subsection{1:1 Fold}

\begin{figure}
\centering
\begin{tabular}{c}
\includegraphics[width=\columnwidth]{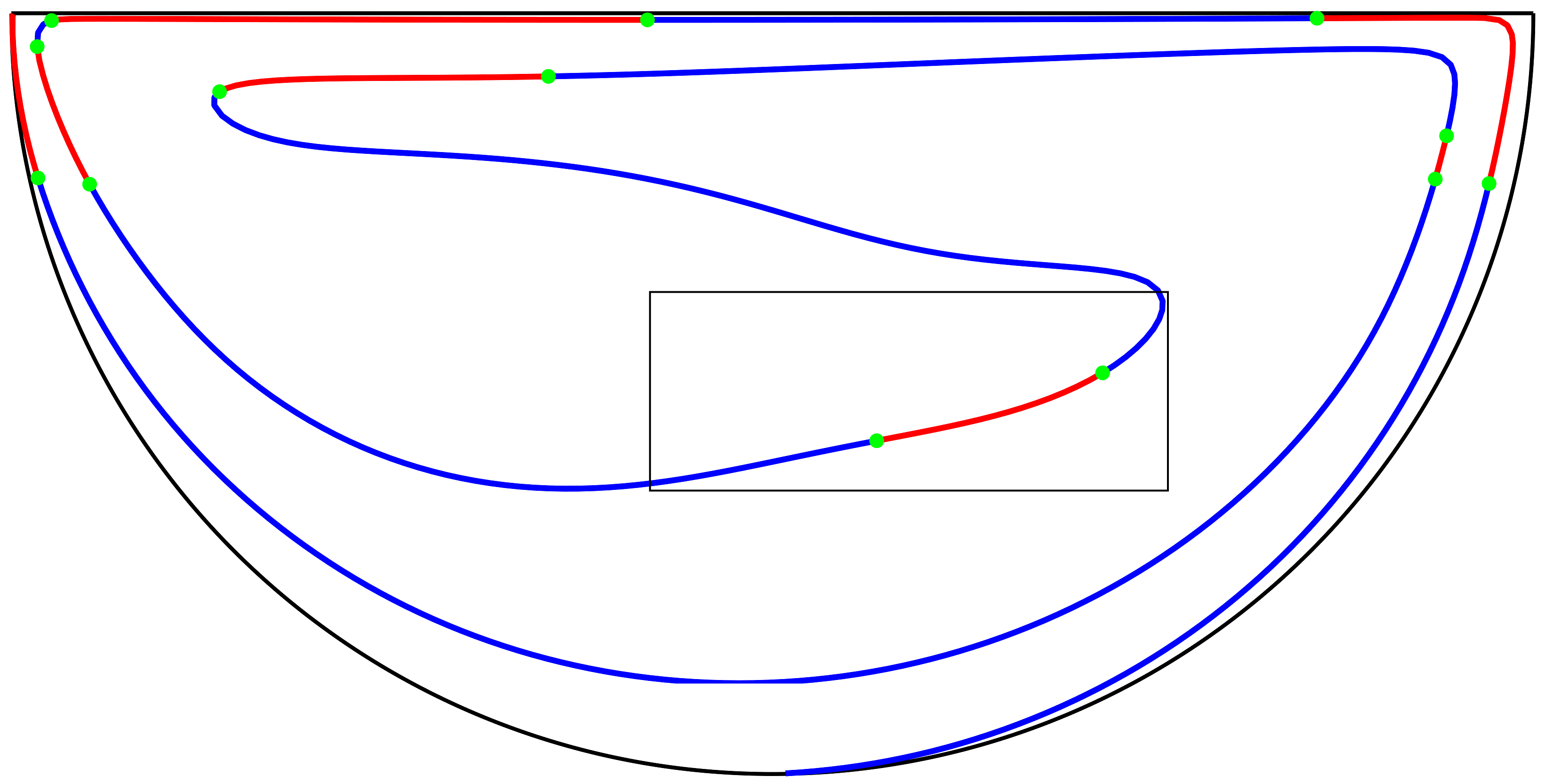} \\
(a) \\
\includegraphics[width=\columnwidth]{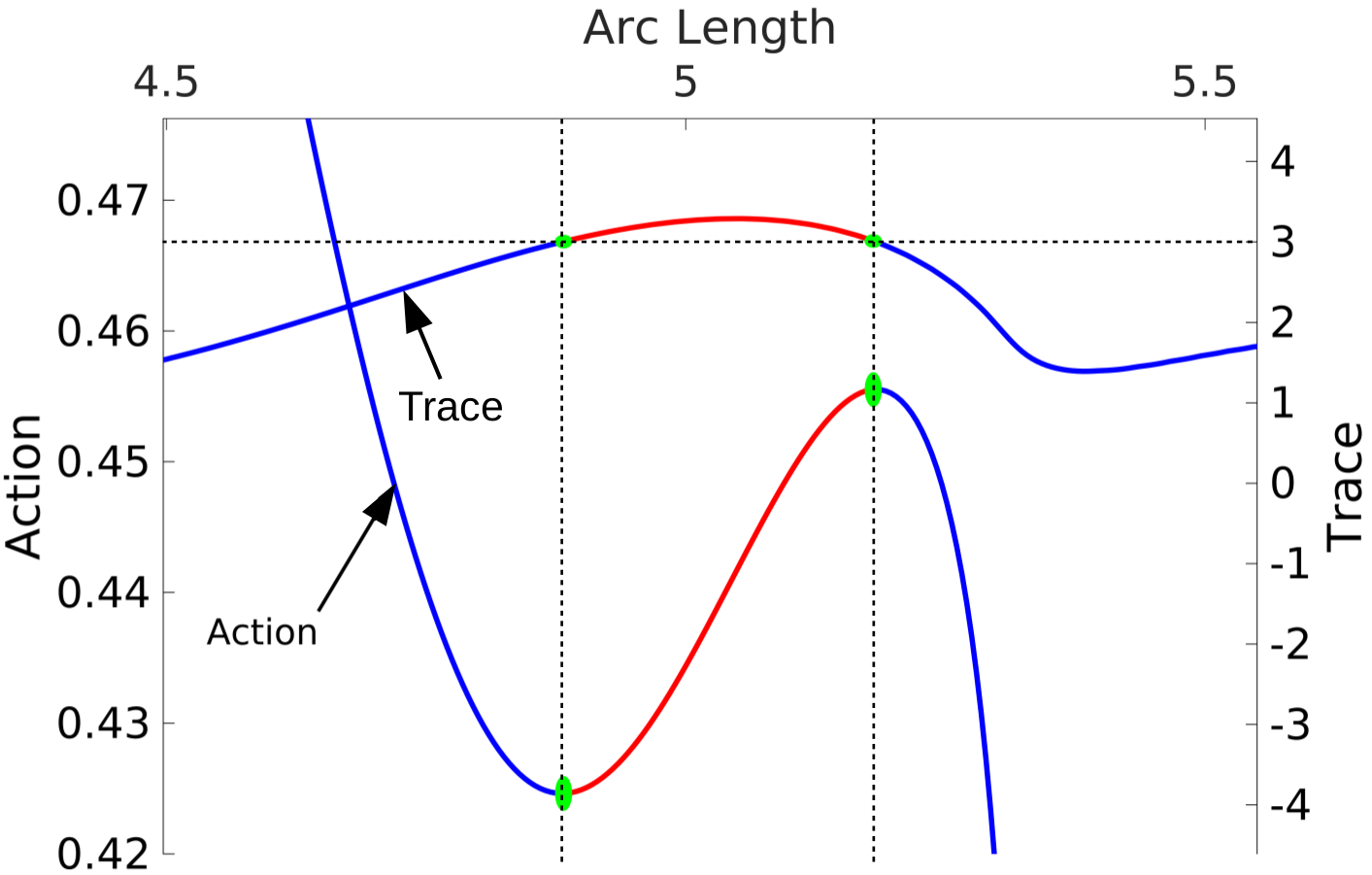}\\
(b)
\end{tabular}
\caption{(a) P1 line in the symmetry plane for
  $(\beta, \Theta) = (16, \pi/4)$.  Red (blue) is hyperbolic
  (elliptic) line segment.  (b) The action and trace values versus Arc
  length of points on the P1 line inside the rectangular box in (a)}
\label{fig:prd1_line_sgmnt_beta_16}
\end{figure}

A 1:1 resonance occurs in the flow at locations where $\tau=3$ on the
periodic line and all three eigenvalues are one.  As an example, the
P1 line for $\Theta=\pi/4$ and $\beta=16$ is shown in
figure~\ref{fig:prd1_line_sgmnt_beta_16}(a).  Inside the box in this
figure, two 1:1 resonance points occur, marked with green dots.
Elliptic segments of the P1 line are marked in blue, and hyperbolic in
red.  As seen in the figure, the P1 line changes its characteristic
from elliptic to hyperbolic (or {\it vice versa}) on opposite sides of
a degenerate point.  Additionally, the eigenvalues $\lambda_1$ and
$\lambda_2$ change from complex to real (or real to complex) through
$\lambda_1=\lambda_2=1$.  In appendix~\ref{app:tangent} we prove that
extrema of P1 lines are always tangent to shells and that the
character of the line must change either side of an extremum.

The action and trace of the deformation tensor on the P1 line segment
inside the box are plotted versus arc length in
figure~\ref{fig:prd1_line_sgmnt_beta_16}(b).  The value of the trace
is 3 at the two degenerate points.  The corresponding points on the
action versus arc length plot have zero slope (i.e. they are local
extrema on this curve).  This in turn means that the P1 line is
tangent to an invariant shell at these points, and with respect to the
action variable (shell number) the line literally folds here.  In the
language of planar catastrophe theory this point is called a fold
bifurcation, whereas in 3D the line is literally folded.

\begin{figure}
\includegraphics[width=\columnwidth]{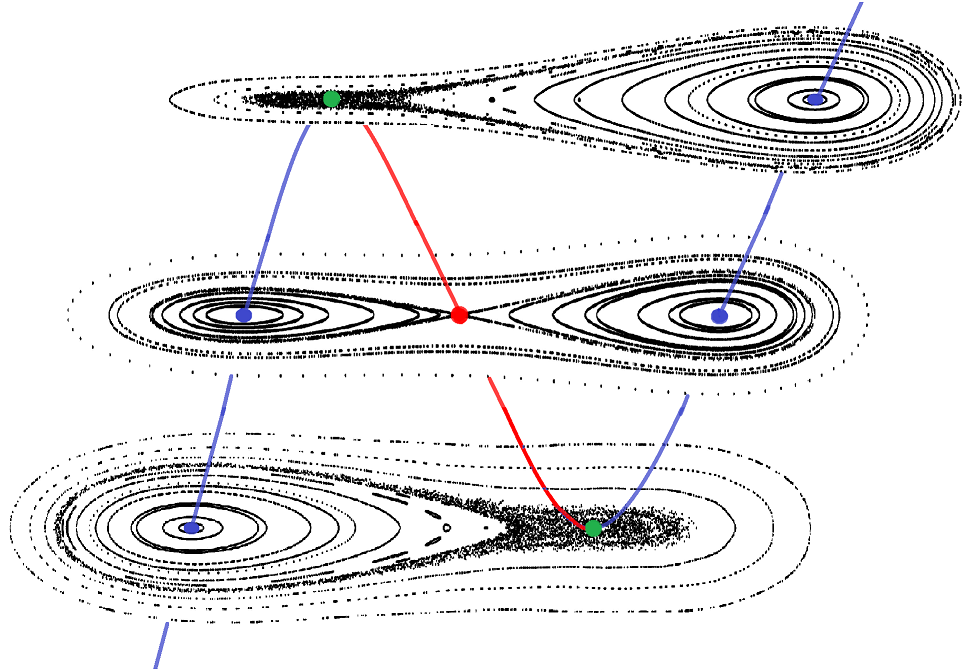} 
\caption{A close up of the P1 line in the rectangular box in
  figure~\ref{fig:prd1_line_sgmnt_beta_16}(a) with local Poincar\'{e}
  sections superposed.  The upper and lower sections are shells with
  degenerate points that coincide with cusps in the sections.  The
  middle section is on a shell half way between.  For clarity the
  shell-normal coordinate is expanded.}
\label{fig:prd1_line_sgmnt_beta_16_PS}
\end{figure}

The local Poincar\'{e} sections on shells containing the 1:1 resonance
points inside the box in figure~\ref{fig:prd1_line_sgmnt_beta_16}(a)
are shown in figure~\ref{fig:prd1_line_sgmnt_beta_16_PS} (with one
shell in between).  The normal coordinate has been stretched in order
to more clearly see the effect of the 1:1 resonances.  As seen, the P1
line is tangent to a shell at a 1:1 resonance point.  The line goes
from elliptic to hyperbolic through the first 1:1 resonance
point and from hyperbolic to elliptic through the second. 1:1
resonance points observed in the PRHF always occur in pairs, and are
connected through a hyperbolic segment.  They always form a
wiggle-like structure in action vs arc length plots as shown in
figure~\ref{fig:prd1_line_sgmnt_beta_16_PS}.  The character of the P1
line at piercing points is imprinted on the Poincar\'{e} sections.
The upper and lower Poincar\'{e} sections contain a degenerate point
and an elliptic point, and the middle section contains a hyperbolic
point and two elliptic points. The stable and unstable manifolds of
the hyperbolic point in the middle section have homoclinic orbits.

\subsection{1:2 Flip}
\label{subsec:resonance_1_2}

At 1:2 resonance points, a period-2 (P2) line intersects the P1 line.
The eigenvalues are $\lambda_1=\lambda_2=-1$ and $\lambda_3=1$, and
the local rotation angle is $\pi$. The net deformation at a 1:2
resonance point is zero after two periods.  The stability of period-1
points on the P1 line changes at the 1:2 resonance point from elliptic
to hyperbolic (or {\it vice versa}), and the eigenvalues
($\lambda_1, \lambda_2$) change from complex to real negative (or real
negative to complex).

\begin{figure}
\centering
\begin{tabular}{c}
\includegraphics[width=\columnwidth]{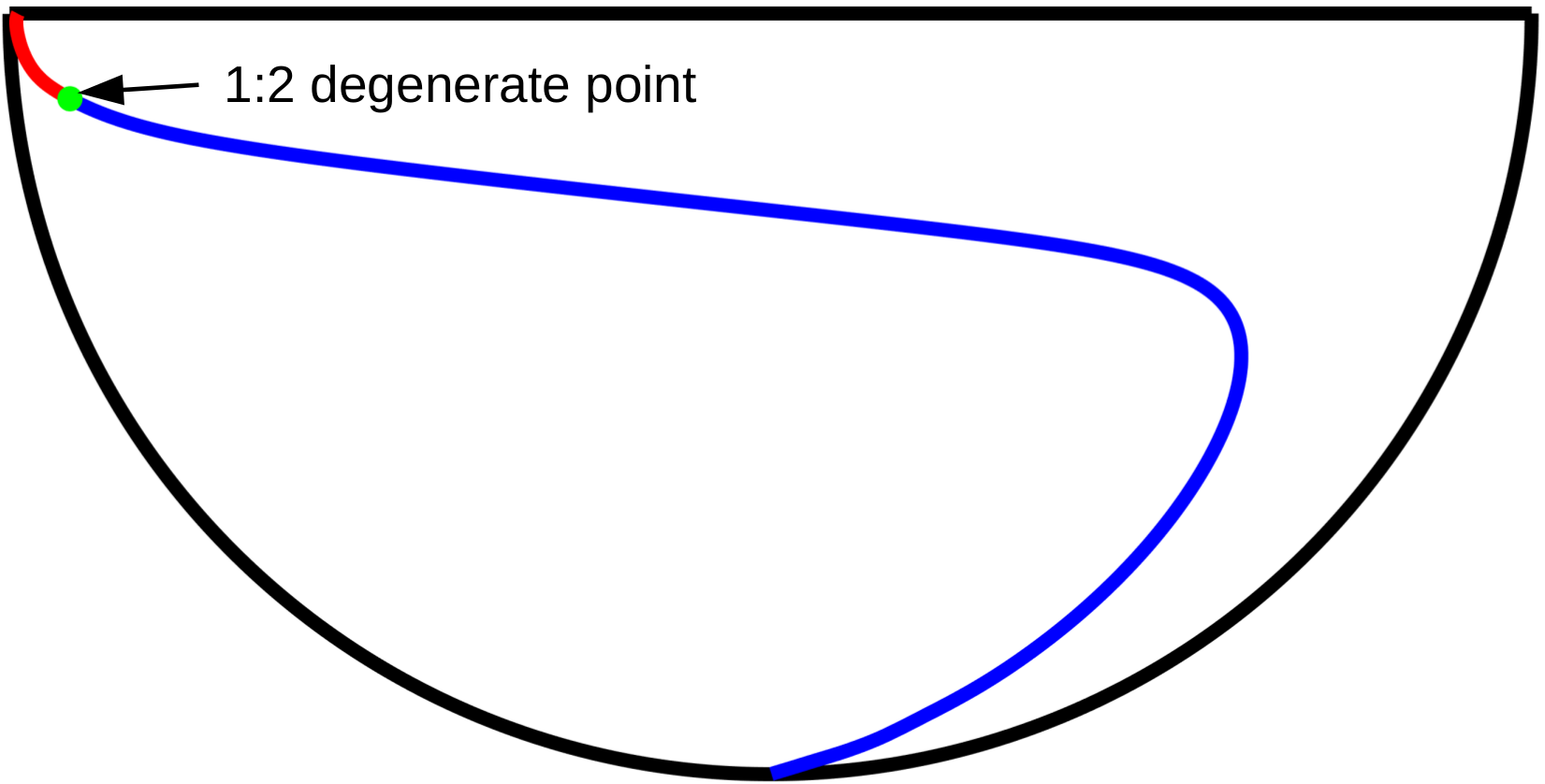} \\
(a) \\
\includegraphics[width=\columnwidth]{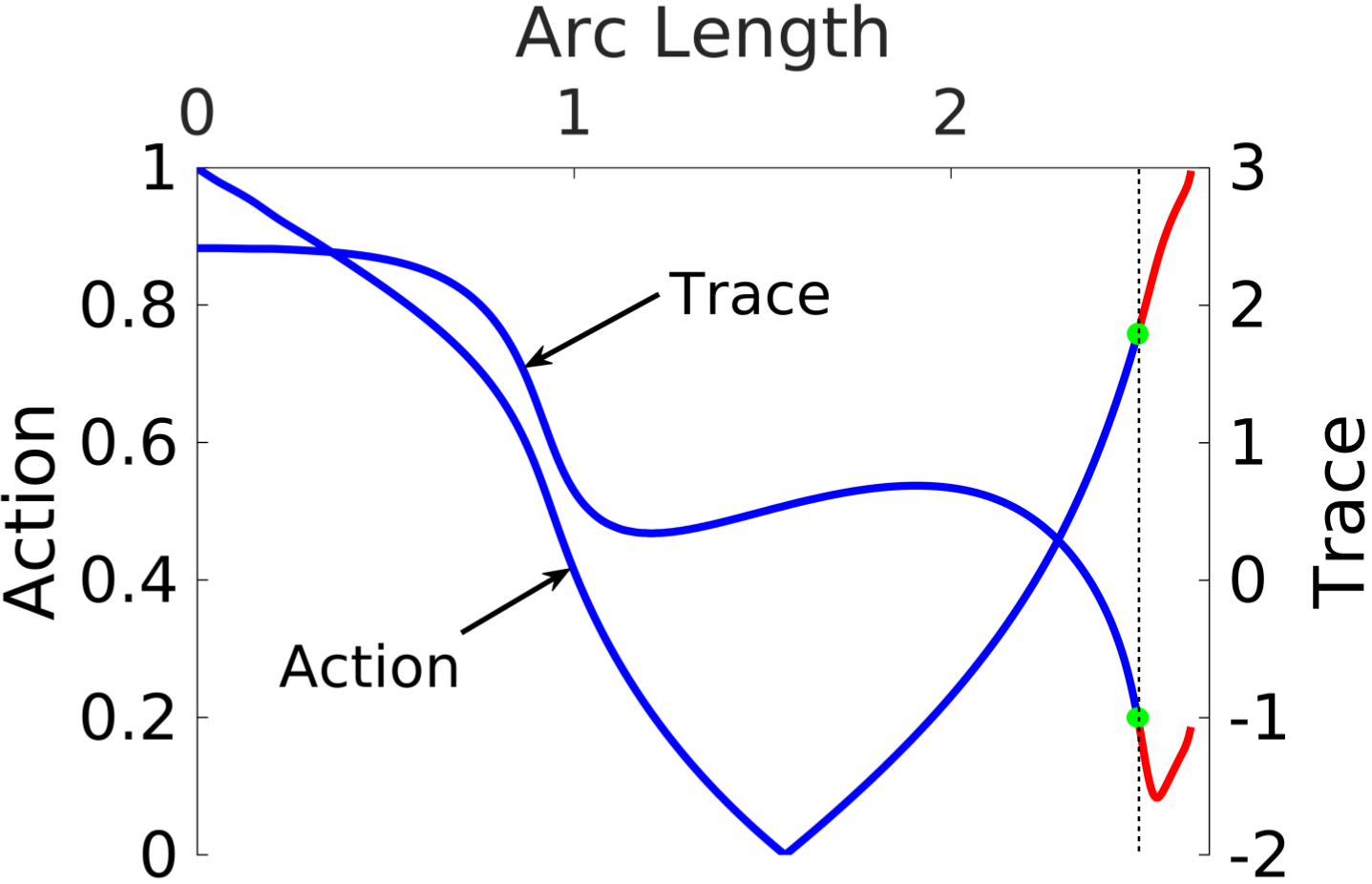} \\
(b)
\end{tabular}
\caption{(a) The P1 line for $\beta=2$ and $\Theta=\pi/4$ on the symmetry plane.  (b) Action and trace versus arc length on P1. }
\label{fig:p1_theta_pi_over4_beta2}
\end{figure} 

\begin{figure}
\centering
\includegraphics[width=\columnwidth]{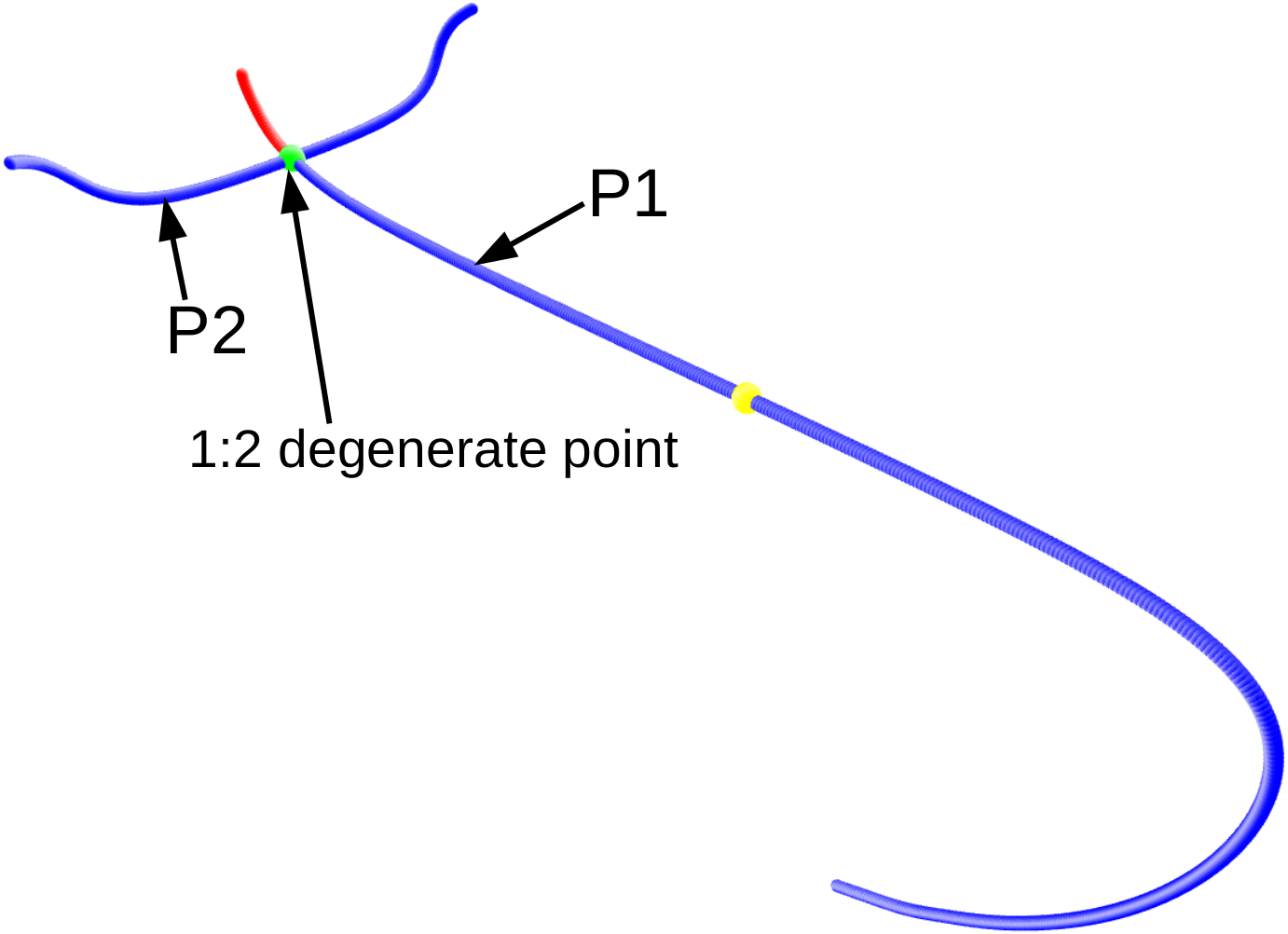}
\caption{P1 and P2 lines for $\beta=2$ and $\Theta=\pi/4$}
\label{fig:p1_p2_lines_3d}
\end{figure}

\begin{figure}
\centering
\includegraphics[width=\columnwidth]{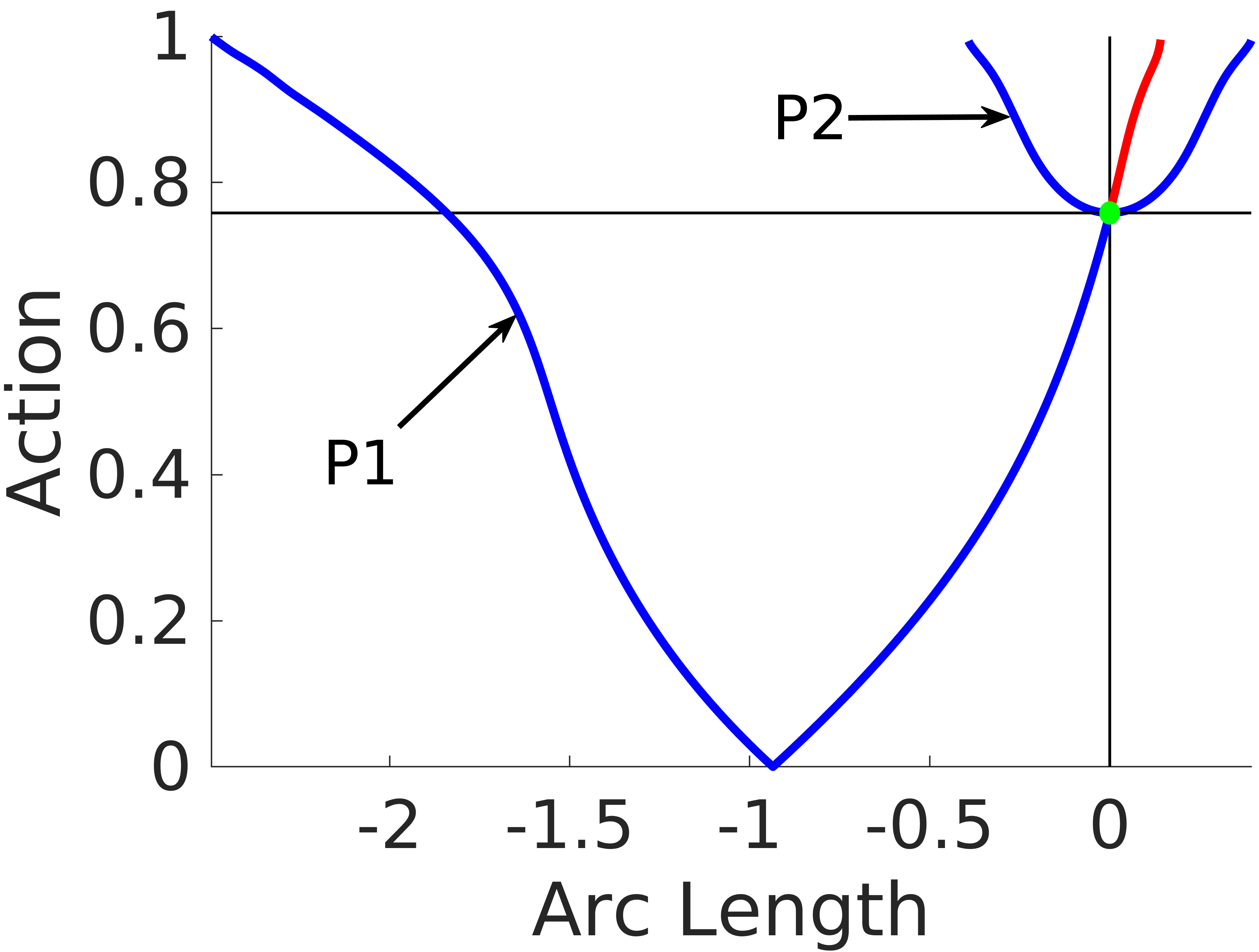}
\caption{Action vs arc length of P1 and P2 lines for $\beta=2$ and $\Theta=\pi/4$}
\label{fig:action_arcln_p1_p2_theta_pi_over4_beta2}
\end{figure}

As an example, the P1 line on the symmetry plane for the case
$\Theta=\pi/8$ and $\beta=2$ is shown in
figure~\ref{fig:p1_theta_pi_over4_beta2}(a).  The action and trace
versus arc length are shown in
figure~\ref{fig:p1_theta_pi_over4_beta2}(b).  In this figure, the
single 1:2 degenerate point is coloured green and has a trace value of
$-1$ (see table~\ref{tab:degen_pt_eigenvalues}).  The change of
stability of P1 at a 1:2 degenerate point is also seen in
figure~\ref{fig:p1_theta_pi_over4_beta2}(b) in the change from blue
(elliptic) to red (hyperbolic).

\begin{figure}
\includegraphics[width=\columnwidth]{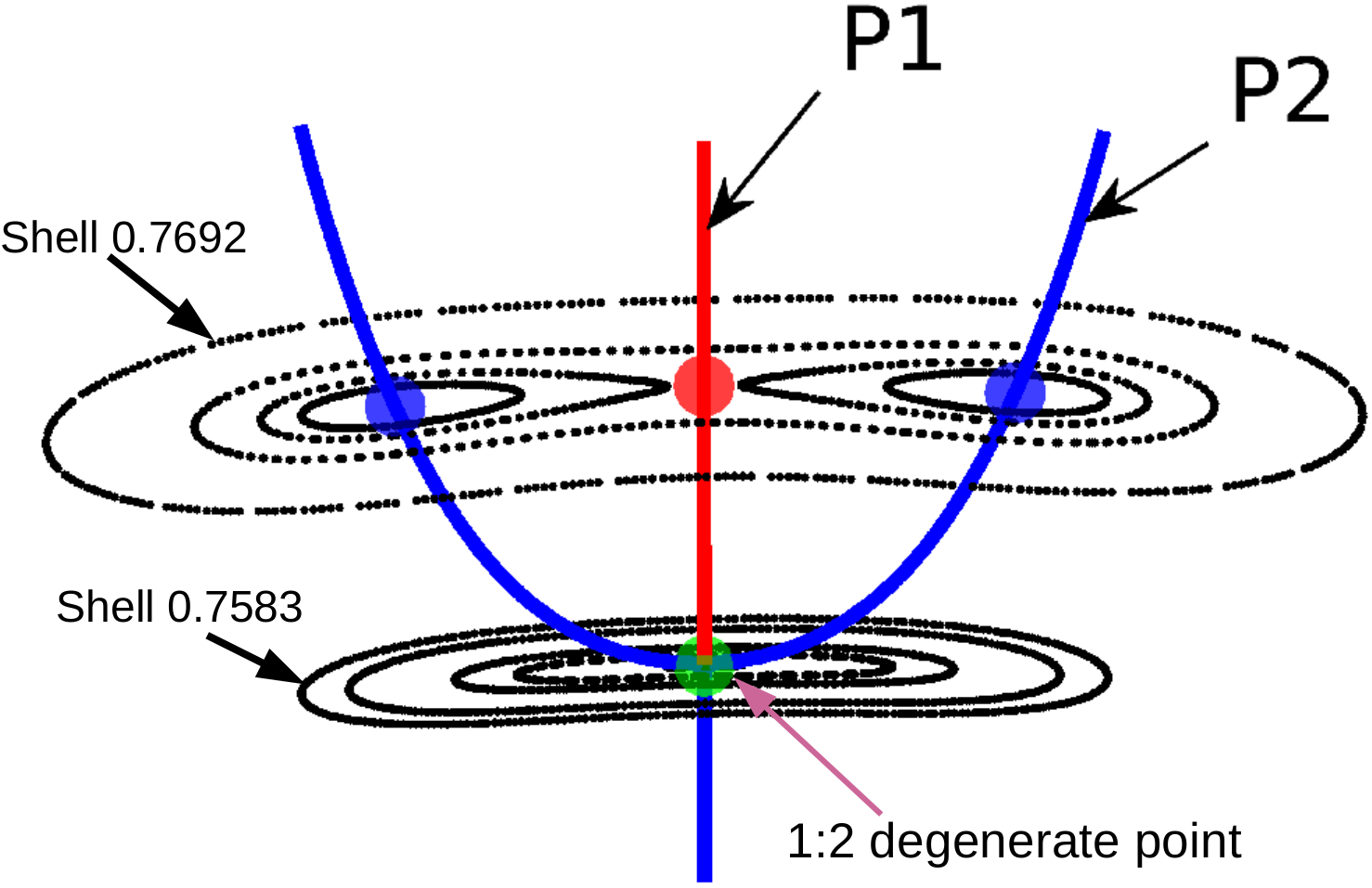}
\caption{3-dimensional 1:2 resonance structure; period-doubling
  bifurcation in space with the action or shell number as the
  bifurcation parameter.  Blue is elliptic, red hyperbolic, and green
  degenerate.  The symmetry plane is normal to the page along the P1
  line.  The period-2 elliptic lines are symmetric about the symmetry
  plane.}
\label{fig:strobo_map_1_2_resonance}
\end{figure}

The period-2 line which goes through the 1:2 resonance point is
computed numerically using the method discussed in
appendix~\ref{app:calc_higher_order_periodic_lines_stokes} and a
perspective view is in figure~\ref{fig:p1_p2_lines_3d}.  The action
value along P1 and P2 is plotted against arc length in
figure~\ref{fig:action_arcln_p1_p2_theta_pi_over4_beta2} where here,
arc length is shifted to have its origin at the 1:2 resonance point.
From figures~\ref{fig:p1_p2_lines_3d} and
\ref{fig:action_arcln_p1_p2_theta_pi_over4_beta2}, we observe that the
period-2 line is symmetric about the symmetry plane and its ends are
attached to opposite sides of the hemisphere boundary.  Local
Poincar\'{e} sections on a shell containing the 1:2 resonance point
and a neighbouring shell are shown in
figure~\ref{fig:strobo_map_1_2_resonance}.  Again, the shell-normal
coordinate is stretched to highlight the key features of the
Lagrangian structures. The lower section (action $I=$ 0.7583) contains
the 1:2 degenerate point and the upper section (action $I=$ 0.7692)
contains one period-1 hyperbolic point and two period-2 elliptic
points. All piercings on sections are shown with solid spheres, blue
for elliptic, red for hyperbolic points, and green for degenerate.
The stable and unstable manifolds of the hyperbolic point on P1 have
homoclinic orbits as seen in
figure~\ref{fig:strobo_map_1_2_resonance}.

The reason a single period-2 line passes through a second order
degenerate point instead of two period-2 lines is the following.
Because eigenvalues of the deformation tensor at degenerate points of
order $n \geq 3$ are complex as shown in
table~\ref{tab:degen_pt_eigenvalues} and the eigenvalues change
continuously along the P1, the P1 line character does not change at
these degenerate points and remains as elliptic. For example, the P1
line character changes at the 1:2 resonance point (see
figure~\ref{fig:strobo_map_1_2_resonance}), where as at 1:3 and 1:4
resonance points the P1 line character does not change (see
figures~\ref{fig:strobo_map_1_3_resonance}
and~\ref{fig:strobo_map_1_4_resonance}). In general, a chain of $n$
period-$n$ elliptic islands appear on an invariant surface in the
neighbourhood of an $n^{th}$ order degenerate point and the centres of
these islands are period-$n$ elliptic points, as shown in the upper
sections of figures~\ref{fig:strobo_map_1_2_resonance},
\ref{fig:strobo_map_1_3_resonance}
and~\ref{fig:strobo_map_1_4_resonance}.  The chain of $n$ elliptic
islands are connected by the manifolds of $n$ period-$n$ hyperbolic
points for $n \geq 3$, forming heteroclinic connections (e.g. see the
upper sections of figures~\ref{fig:strobo_map_1_3_resonance}
and~\ref{fig:strobo_map_1_4_resonance}).  In contrast, at a second
order degenerate point, the two elliptic islands are connected by
manifolds of a hyperbolic period-1 point instead of two period-2
hyperbolic points as shown in the
figure~\ref{fig:strobo_map_1_2_resonance}, forming a homoclinic
connection.  Because of the existence of this hyperbolic period-1
point which is due the P1 line character change at the 1:2 degenerate
point, two additional hyperbolic period-2 points are not possible to
connect the islands.  Hence there is only one period-2 line which
passes through the 1:2 degenerate point.

\subsection{1:3}

\begin{figure}
\centering
\begin{tabular}{c}
\includegraphics[width=\columnwidth]{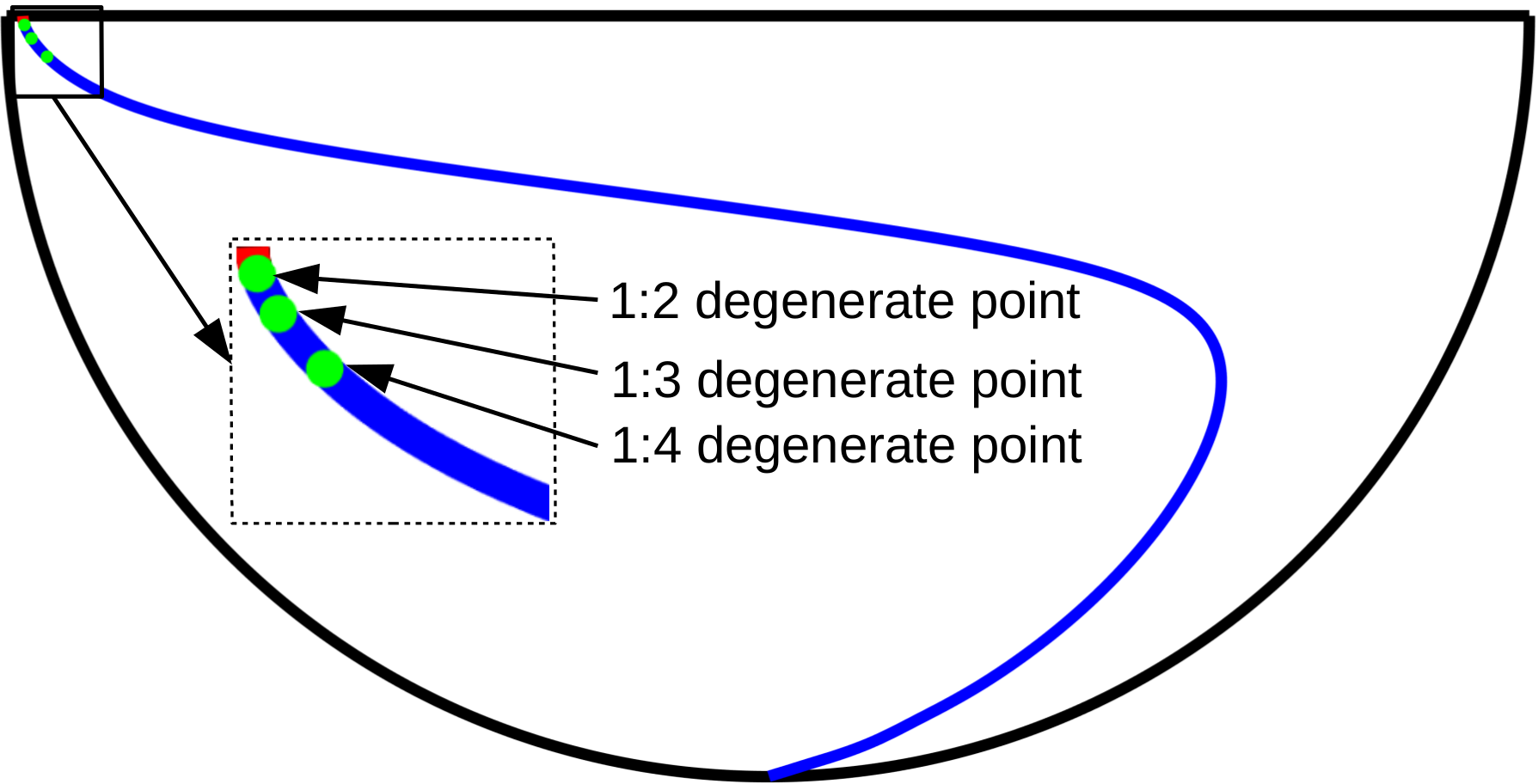} \\
(a) \\
\includegraphics[width=\columnwidth]{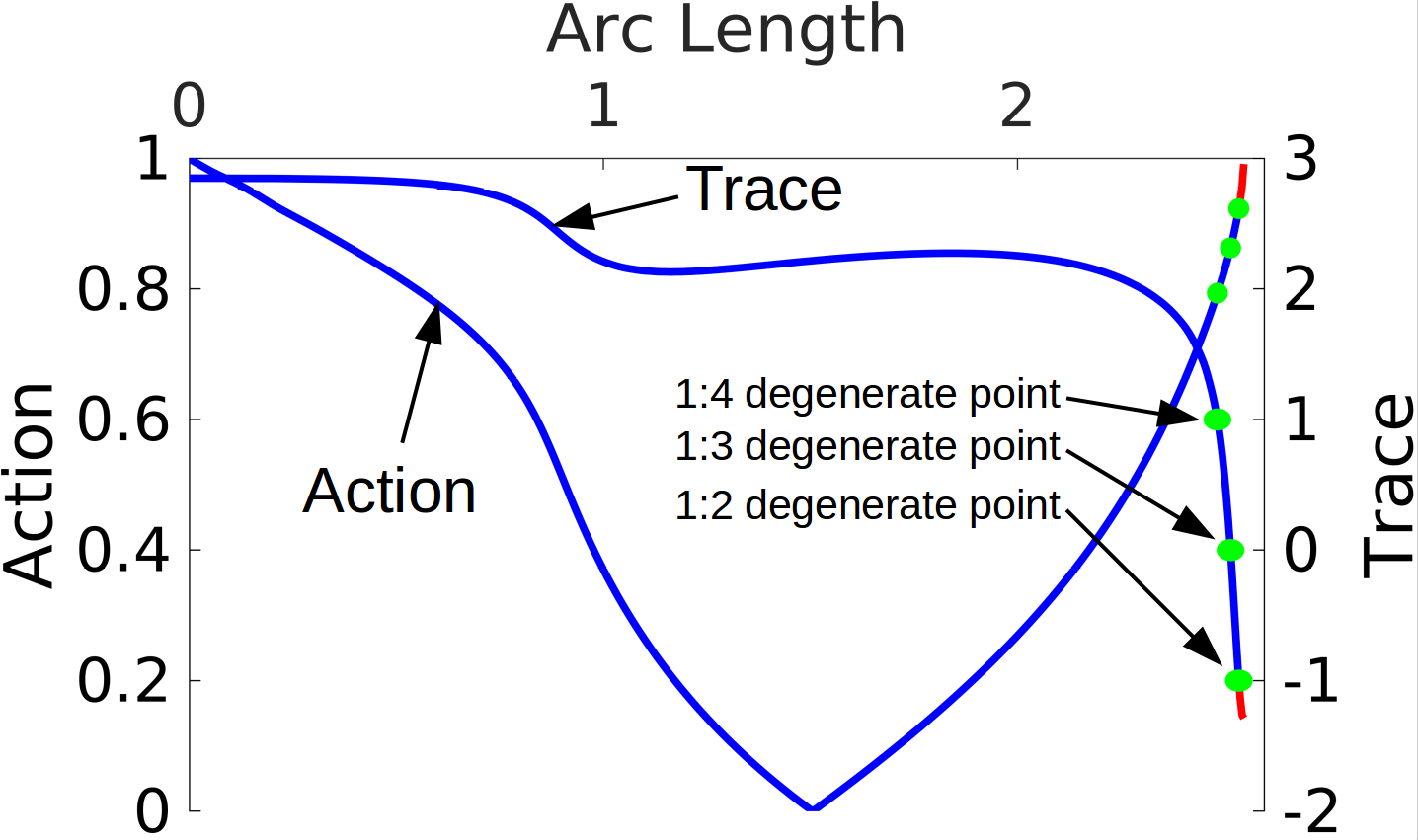} \\
(b)
\end{tabular}
\caption{(a) P1 line for $\beta=1$ and $\Theta=\pi/8$ on the symmetry plane.  (b) Action and trace along P1.}
\label{fig:p1_theta_pi_over8_beta1}
\end{figure}

At a 1:3 resonance point, three period-3 lines (P3$_1$,P3$_2$ and P3$_3$) intersect P1. The local rotation angle is $2\pi/3$ and
the deformation tensor has eigenvalues
$\lambda_{1,2}=e^{(\pm i 2\pi/3)}, \lambda_3=1$.  The net deformation
at the resonance point is zero after 3 periods.  The properties of a
1:3 resonance point are described using an example with $\Theta=\pi/8$
and $\beta=1$.  The P1 line lies in the symmetry plane and is shown in
figure~\ref{fig:p1_theta_pi_over8_beta1}(a), where it is seen in the
inset that 1:2, 1:3 and 1:4 resonance points exist on P1 in close
proximity to each other.  In this section we consider only the 1:3
resonance.  

\begin{figure}
\centering
\includegraphics[width=\columnwidth]{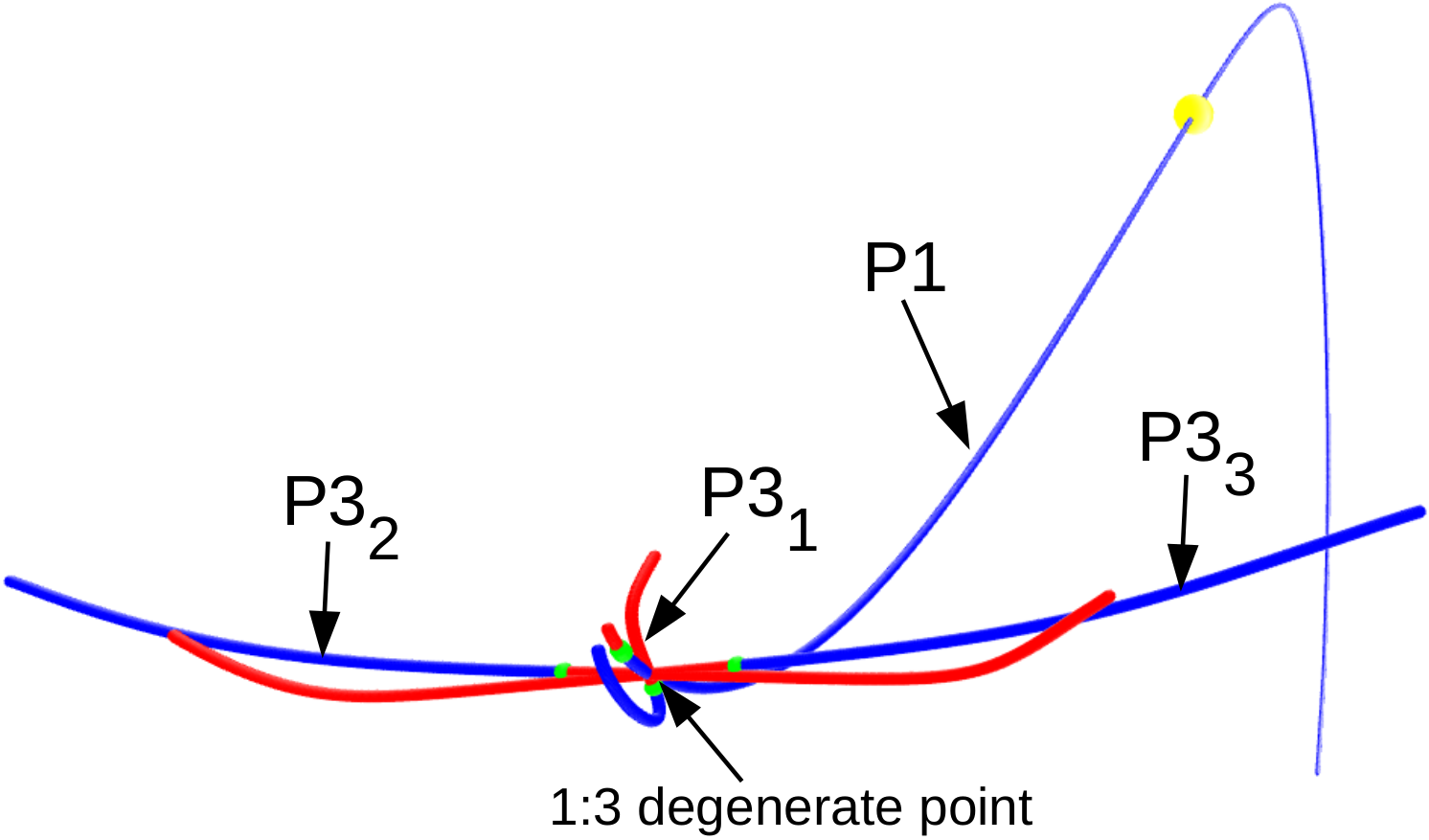}
\caption{P1 and P3 lines for $\beta=1$ and $\Theta=\pi/8$}
\label{fig:p1_p3_lines_3d}
\end{figure}

\begin{figure}
\centering
\begin{tabular}{c}
\includegraphics[width=\columnwidth]{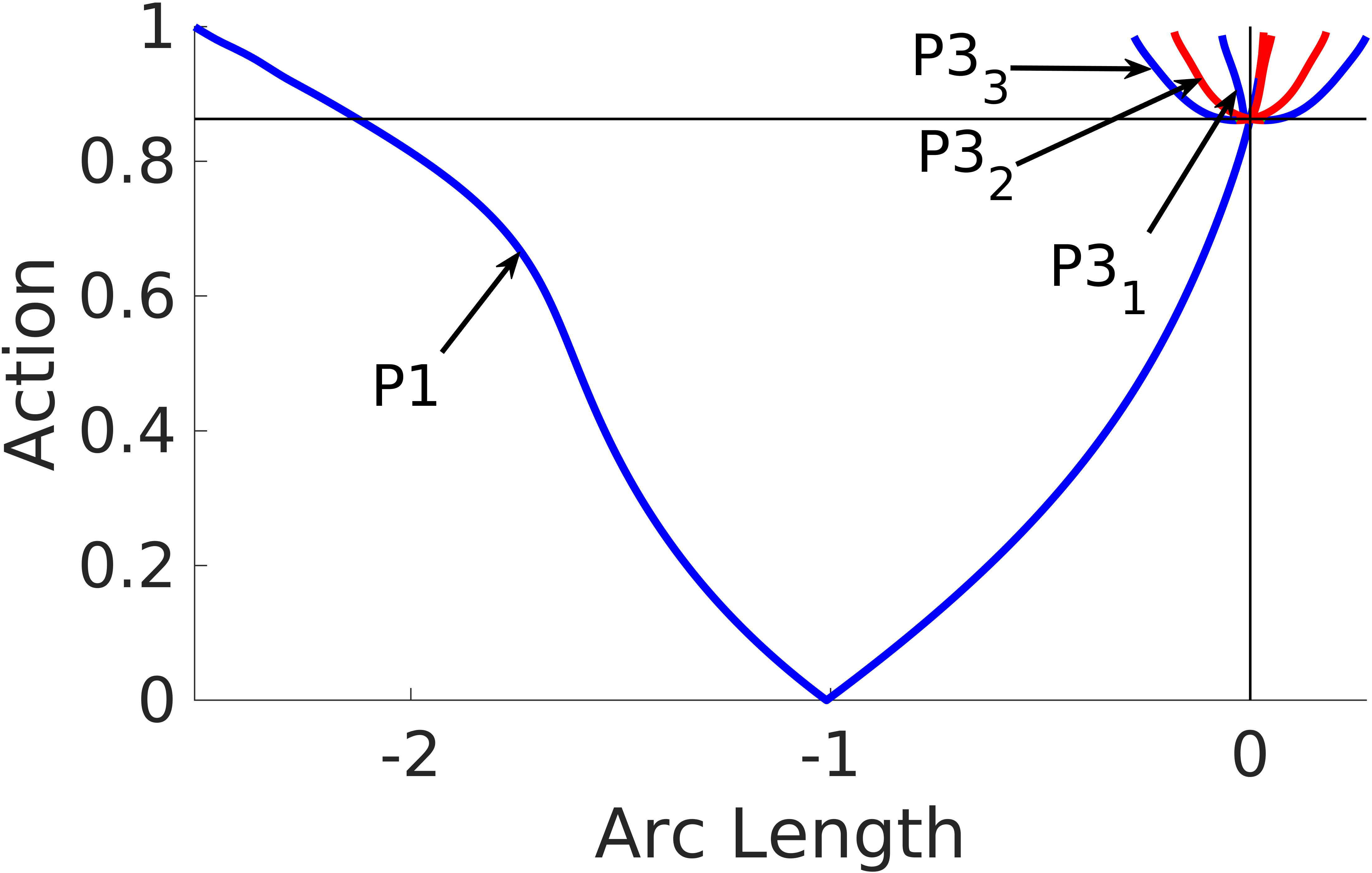} \\
(a) \\
\includegraphics[width=\columnwidth]{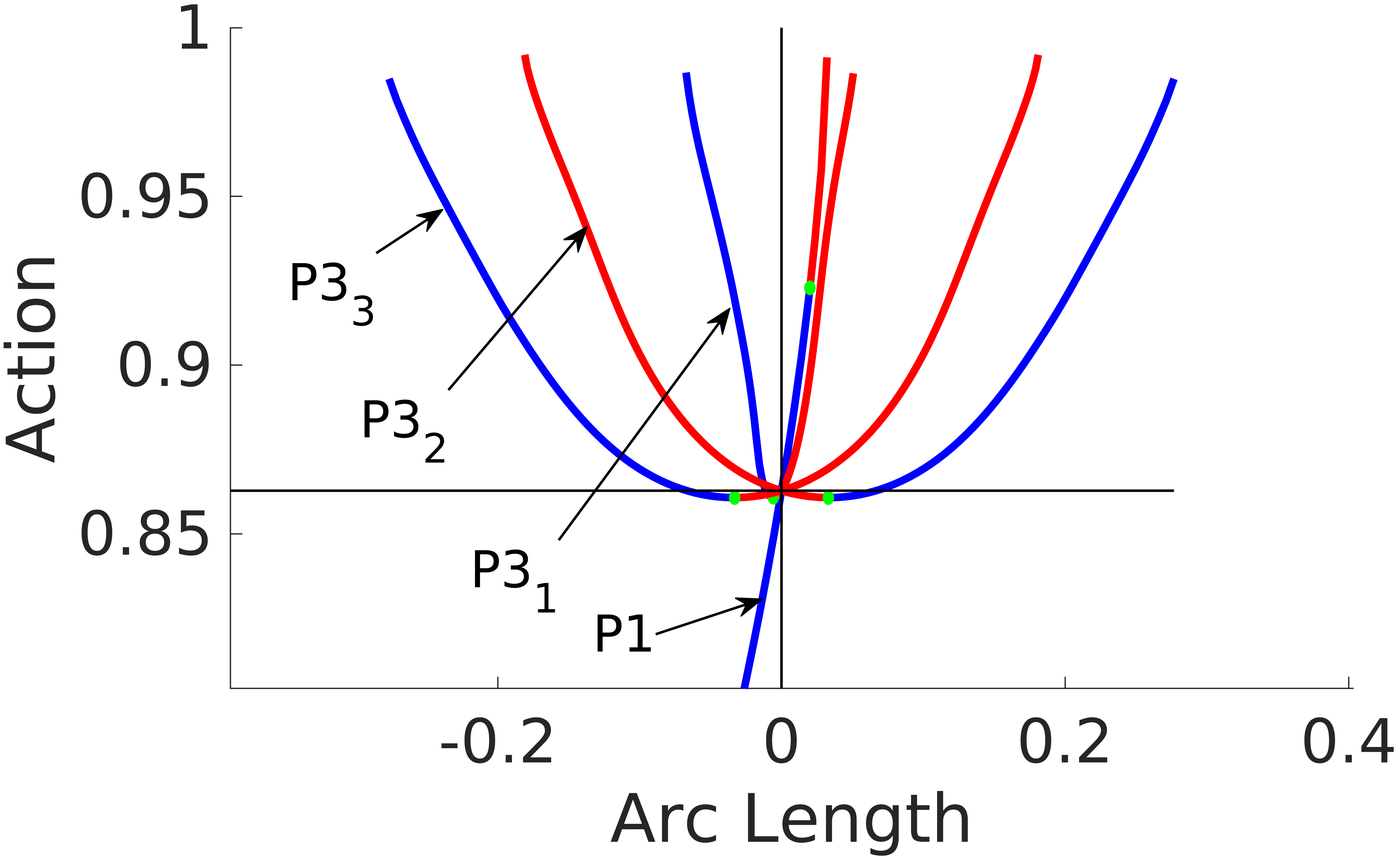} \\
(b)
\end{tabular}
\caption{(a) Action is plotted along P1 and P3 lines for
  $\Theta=\pi/8$ and $\beta=1$ (b) Top right segment of figure (a).}
\label{fig:action_arclength_p1_p3_lines}
\end{figure}

Action and trace values of the deformation tensor on the P1 line are
shown in figure~\ref{fig:p1_theta_pi_over8_beta1}(b) where the trace
value of 0 corresponds to the 1:3 resonance point (see
table~\ref{tab:degen_pt_eigenvalues}).  A perspective view of P1 and
the three period-3 (P3) lines are shown in
figure~\ref{fig:p1_p3_lines_3d}.  Action values on P1 and the three P3
lines are plotted against arc lengths in
figure~\ref{fig:action_arclength_p1_p3_lines}(a) where P3 arc length
is redefined with an origin at the 1:3 resonance point.  A close-up of
the rectangular box in
figure~\ref{fig:action_arclength_p1_p3_lines}(a) is shown in
figure~\ref{fig:action_arclength_p1_p3_lines}(b).  P1 and P3$_1$ lie
in the symmetry plane, and from
figure~\ref{fig:action_arclength_p1_p3_lines}(b) and
figure~\ref{fig:p1_p3_lines_3d}, it is observed that, all period-3
line ends are attached to the hemisphere boundary.  $P3_2$ and $P3_3$
are reflections of each other across the symmetry plane, and can be
defined by $P3_3 = S_{\Theta} P3_2$.

\begin{figure}
\includegraphics[width=\columnwidth]{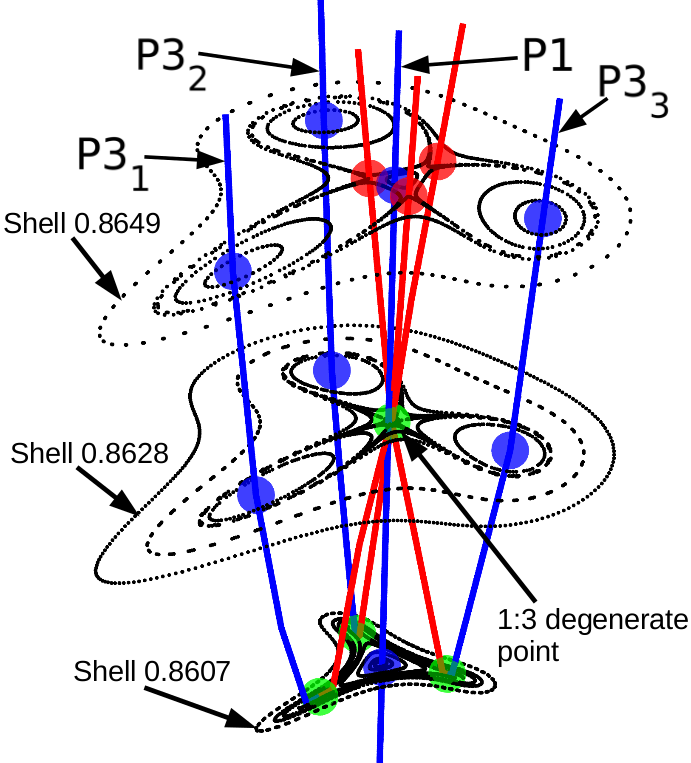}
  \caption{3-dimensional 1:3 resonance structure.  Blue is elliptic,
    red hyperbolic, and green degenerate.  Left is a 3-dimensional
    view of selected shells, exploded along the tangent direction of
    the P1 line.  Right are planar projections of the shells.
    Period tripling birfurcation in space with the action or shell
    number as the bifurcation parameter.}
\label{fig:strobo_map_1_3_resonance}
\end{figure}

The local Poincar\'{e} sections on a shell containing the 1:3
resonance point and two neighbouring shells (one below, one above) are
shown in figure~\ref{fig:strobo_map_1_3_resonance}. The shell-normal
coordinate is stretched for clarity.  Each of the P3 lines has an
extremum on the lower shell (action value 0.8607), i.e.~the three
period-3 lines are all tangent to this same shell.  This is seen in
the tri-cuspid structure with a central period-1 elliptic point.  The
middle section (action value 0.8628) contains three period-3 elliptic
points and the 1:3 resonance point through which P1 and all period-3
lines pass (by definition).  The upper section (action value 0.8649)
contains three period-3 elliptic points, three period-3 hyperbolic
points, and a period-1 elliptic point. All piercings on the
Poincar\'{e} sections are shown with solid spheres. The stable and
unstable manifolds of the period-3 hyperbolic points on the upper
section have heteroclinic connections which create four connected
islands, which act as transport
barriers. \citep{Smith_degenerate_2016} gave a detailed analysis of a
1:3 resonance; a notable feature is that the sense of rotation along
the P1 line reverses either side of the degenerate point.

\subsection{1:4}
\label{subsec:1_4_res}

\begin{figure}
\centering
\includegraphics[width=\columnwidth]{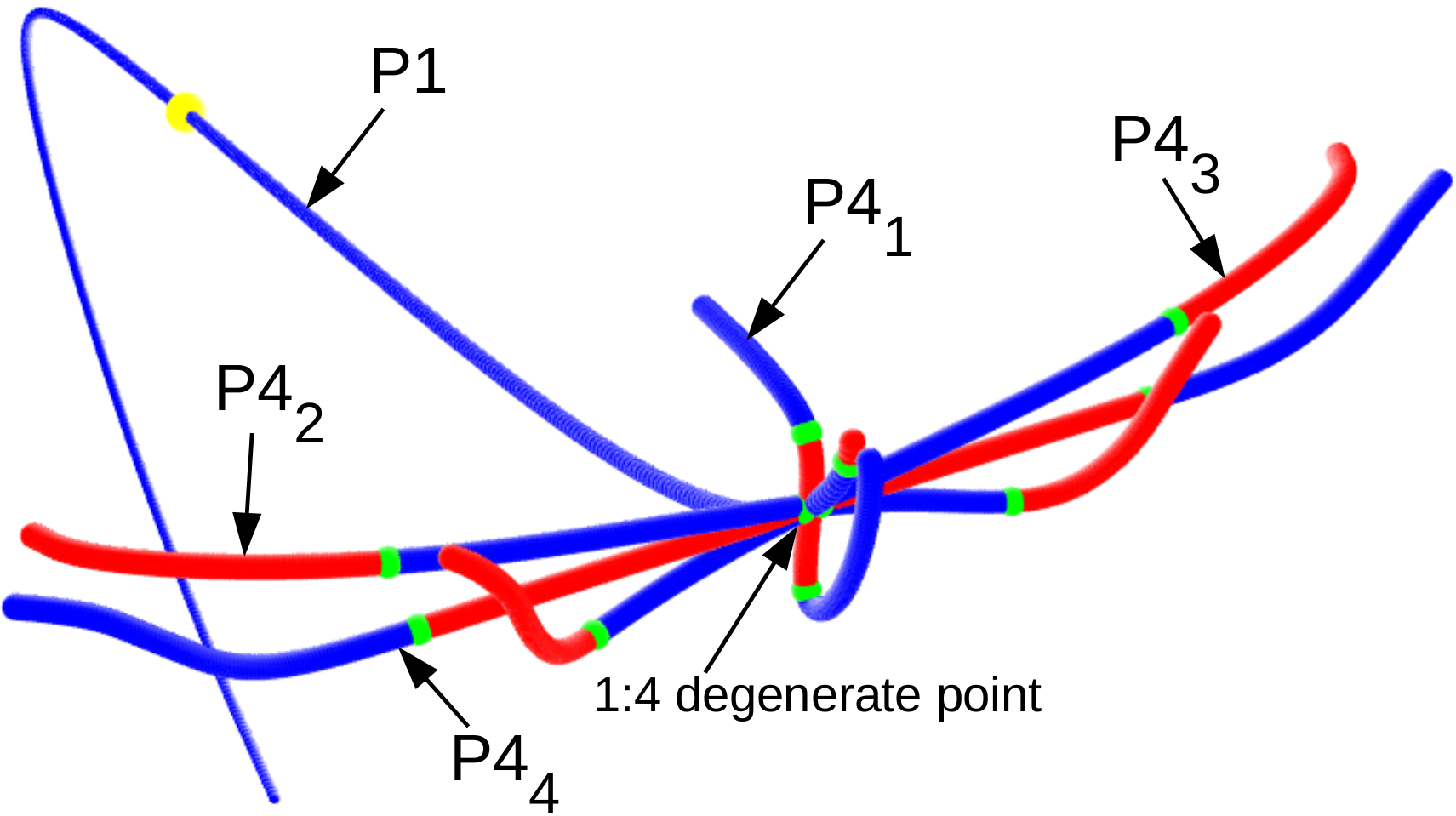}
\caption{P1 and P4 lines for $\beta=1$ and $\Theta=\pi/8$}
\label{fig:p1_p4_lines_3d}
\end{figure}

\begin{figure}
\centering
\begin{tabular}{c}
\includegraphics[width=\columnwidth]{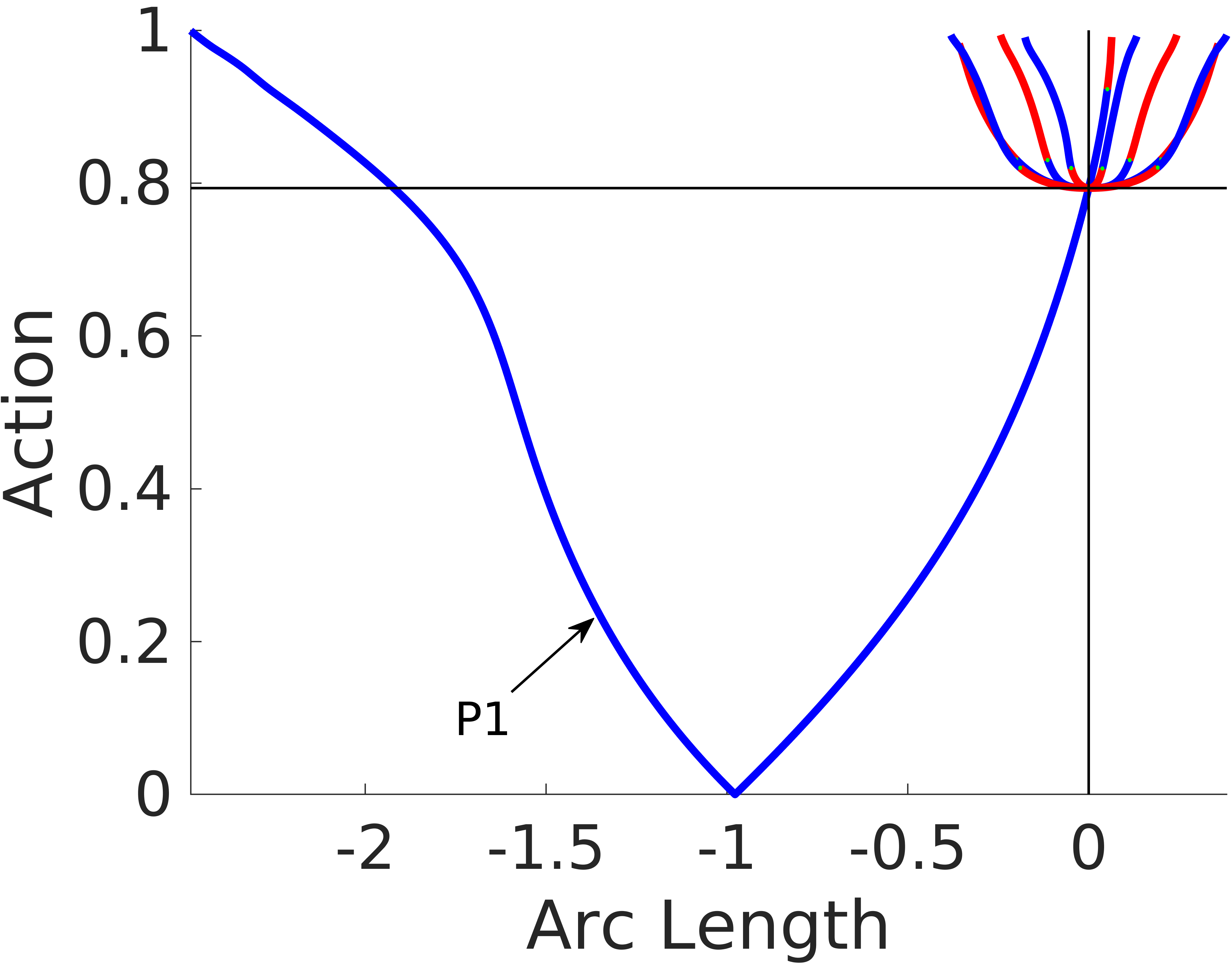} \\
(a) \\
\includegraphics[width=\columnwidth]{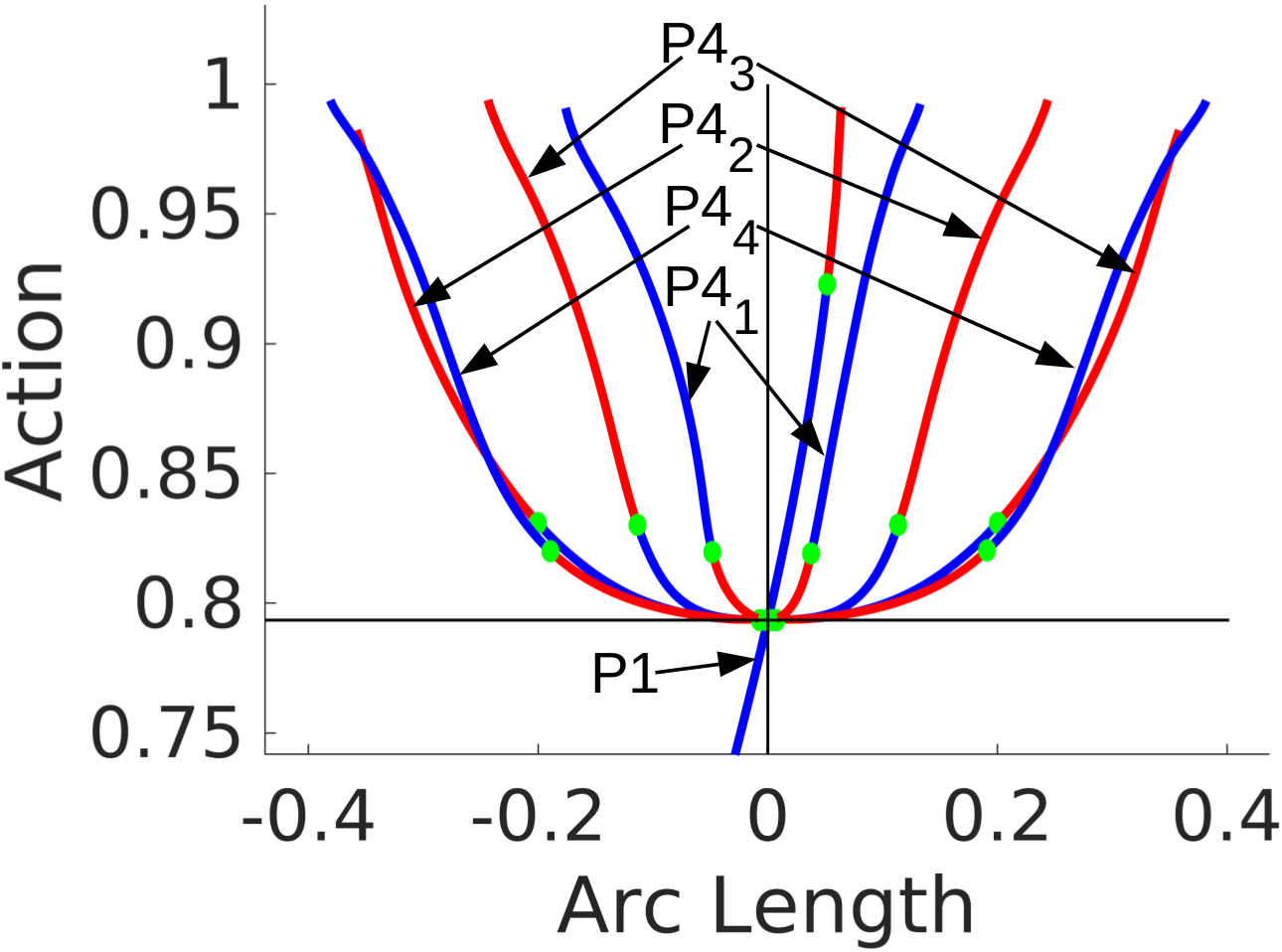} \\
(b)
\end{tabular}
\caption{(a) Action is plotted along P1 and P4 lines for $\Theta=\pi/8$ and $\beta=1$ (b) Top right segment of figure (a)}
\label{fig:action_arclength_p1_p4_lines}
\end{figure}

At a 1:4 resonance point, four period-4 lines intersect P1.  The
eigenvalues of the 1:4 resonance are
$\lambda_{1,2}=e^{ (\pm i 2\pi/4)}, \lambda_3=1$ (and $\Tr=1$). The
net deformation is zero after four periods, and the local rotation
angle is $\pi/2$.  We use the same parameters ($\Theta=\pi/8$ and
$\beta=1$) here as in the previous section to discuss the case in
figure~\ref{fig:p1_theta_pi_over8_beta1}(b) with trace $ = 1$ which is
the 1:4 resonance point.  This point is also shown on the P1 line in
figure~\ref{fig:p1_theta_pi_over8_beta1}(a).  P1 and the four period-4
lines (P4$_1$-P4$_4$) are shown in perspective in
figure~\ref{fig:p1_p4_lines_3d}.  The action value of all lines
plotted against arc length are shown in
figure~\ref{fig:action_arclength_p1_p4_lines}(a), where again, arc
length is redefined with its origin at the 1:4 resonance point.

A close-up of the rectangular box in
figure~\ref{fig:action_arclength_p1_p4_lines}(a) is shown in
figure~\ref{fig:action_arclength_p1_p4_lines}(b).  Perhaps not obvious
from figure~\ref{fig:p1_p4_lines_3d} and
figure~\ref{fig:action_arclength_p1_p4_lines}(b), is that $P4_1$ and
$P4_4$ lie on the symmetry plane, and that $P4_3$ and $P4_2$ are
reflections of each other about the symmetry plane and are related by
$P4_3=S_{\Theta} P4_2$.  All four P4 lines are attached to the
hemisphere boundary at both ends.

\begin{figure}
\includegraphics[width=\columnwidth]{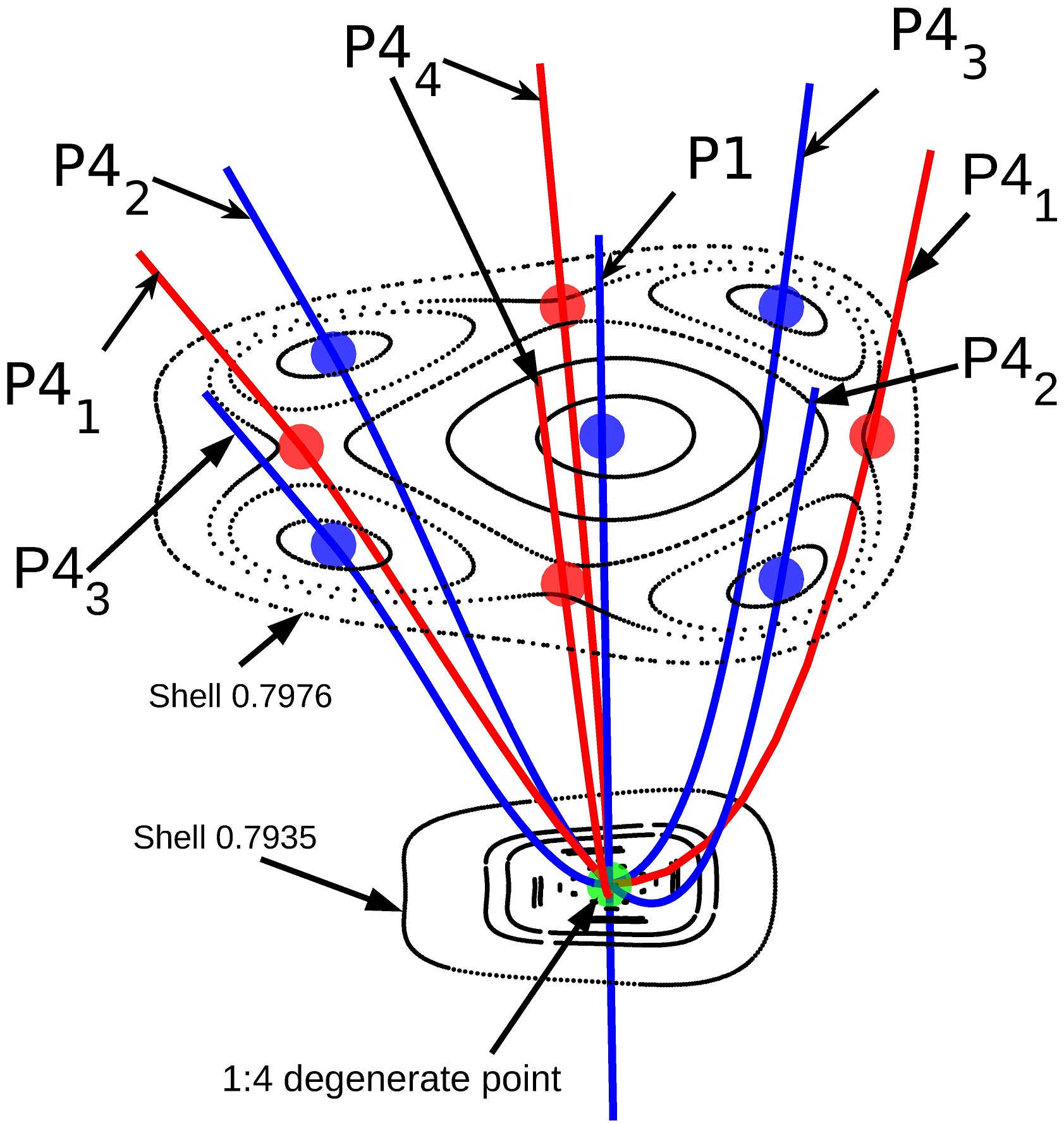}
\caption{3-dimensional 1:4 resonance structure.  Blue is elliptic, red
  hyperbolic, and green degenerate.  A 3-dimensional view of selected
  shells, exploded along the tangent direction of the P1 line.
  Birfurcation in space with the action or shell number as the
  bifurcation parameter.}
\label{fig:strobo_map_1_4_resonance}
\end{figure}

The local Poincar\'{e} sections on the shell containing this 1:4 resonance point and a shell above is shown in figure~\ref{fig:strobo_map_1_4_resonance}. The lower section (action value 0.7935) contains the 1:4 degenerate point and the upper section (action value 0.7976) contains the four period-4 elliptic points, four period-4 hyperbolic points and one P1 elliptic point. All the piercings sites on a Poincar\'{e} section are shown with solid spheres. The stable and unstable manifolds of the period-4 hyperbolic points on upper section have heteroclinic connections. The heteroclinic connections create five islands (four period-4 islands and one period-1 island), and these islands act as transport barriers.

\subsection{Hierarchy of Resonances: a 2:6 example}

\begin{figure}
\centering
\includegraphics[width=1.0\columnwidth]{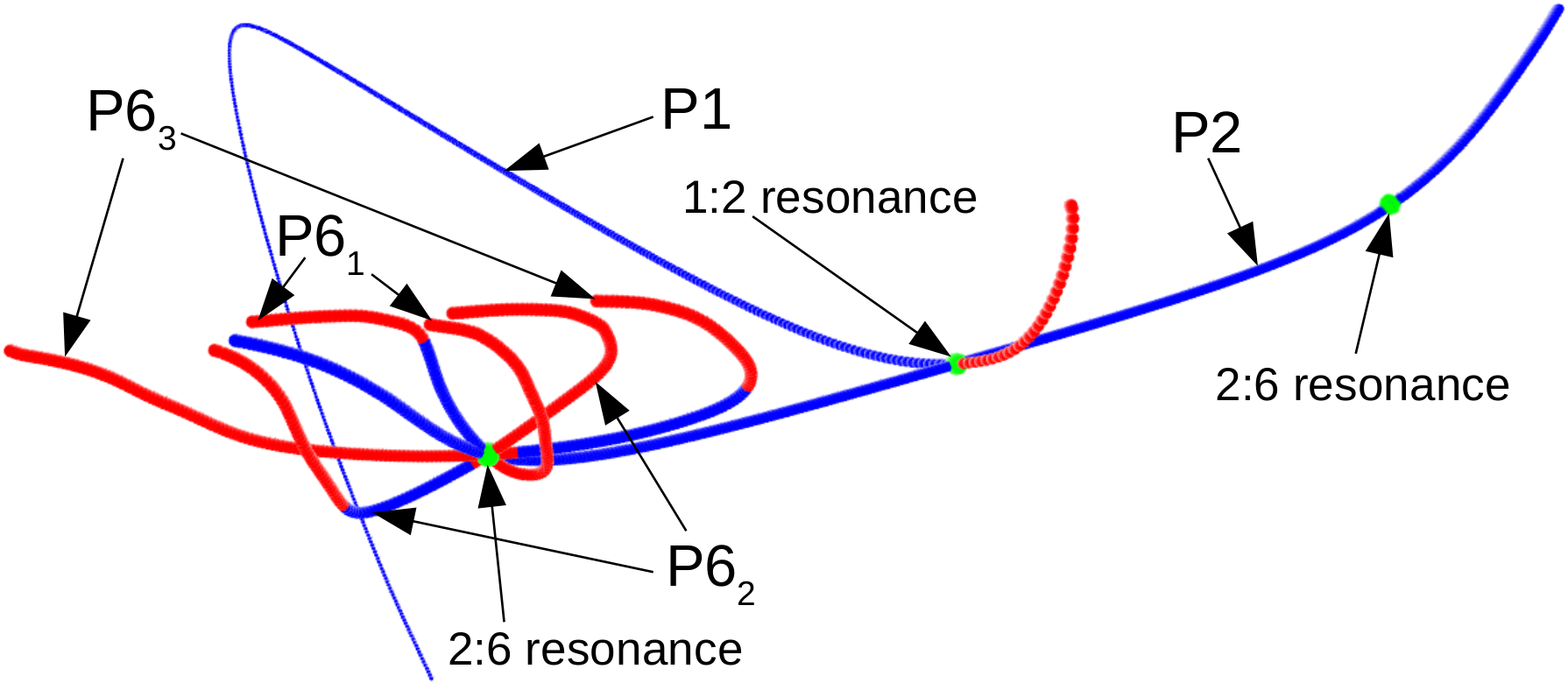}
\caption{A period-1 line (P1), a period-2 line (P2) and three period-6 lines ($P6_1, P6_2$ and $P6_3$) coordinating via 1:2 resonance point and 2:6 resonance points for $\Theta=\pi/4$ and $\beta=2$.}
\label{fig:resonance_2_6_structures}
\end{figure}

One of the main findings of this study is that resonance points,
coordinating lower-order periodic lines and higher-order periodic
lines, act as nodes in the Lagrangian network of periodic lines in one
invariant flows.  These resonance points organise periodic lines,
which then controls fluid transport. Resonances on P1 lines are
discussed so far. To demonstrate the organisation of periodic lines by
resonances on higher-order periodic lines, a 2:6 resonance is
described with the example of $\Theta=\pi/4$ and $\beta=2$, which is
the same example used in section~\ref{subsec:resonance_1_2} to
describe 1:2 resonance. A 1:2 resonance point is identified on the P1
line of $\Theta=\pi/4$ and $\beta=2$, and the period-2 line which
passes through the 1:2 resonance point is calculated. On this period-2
line, 2:6 resonance points are identified, and the corresponding
period-6 lines which pass through them are calculated numerically
using the method discussed in
appendix~\ref{app:calc_higher_order_periodic_lines_stokes}.  The 2:6
resonance points and their associated higher-order periodic lines are
shown in figure~\ref{fig:resonance_2_6_structures}. In this figure, a
period-2 line (P2) intersects the period-1 line (P1) at the 1:2
resonance point and three period-6 lines ($P6_1, P6_2$ and $P6_3$)
intersect the period-2 line at the 2:6 resonance point.  In this
figure, there is a symmetric structure on the other side of the
P2 line (due to the symmetry in the PRHF), however, for clarity only the 2:6
resonance point, not the P6 lines, is shown.

Following this approach, the detailed Lagrangian skeleton of a
one-invariant flow can be calculated systematically, by first
identifying resonance points on P1 lines and calculating their
corresponding higher-order periodic lines, next identifying resonance
points on P2 lines and calculating their corresponding higher-order
periodic lines. This process is continued recursively until enough
Lagrangian structure detail is obtained for any given purpose, e.g.\
those in figure~\ref{fig:poinc_map_shell_3}.

\subsection{Global Bifurcations}

In the examples of resonances discussed so far, higher order periodic lines extend from the resonance points and terminate on the hemisphere boundary.  However it is also possible that a higher order periodic line from one resonance point can connect with another resonance point of the same order within the fluid bulk.  In this way, the line forms a truly global Lagrangian structure. 

\begin{figure}
\centering
\begin{tabular}{c}
\includegraphics[width=\columnwidth]{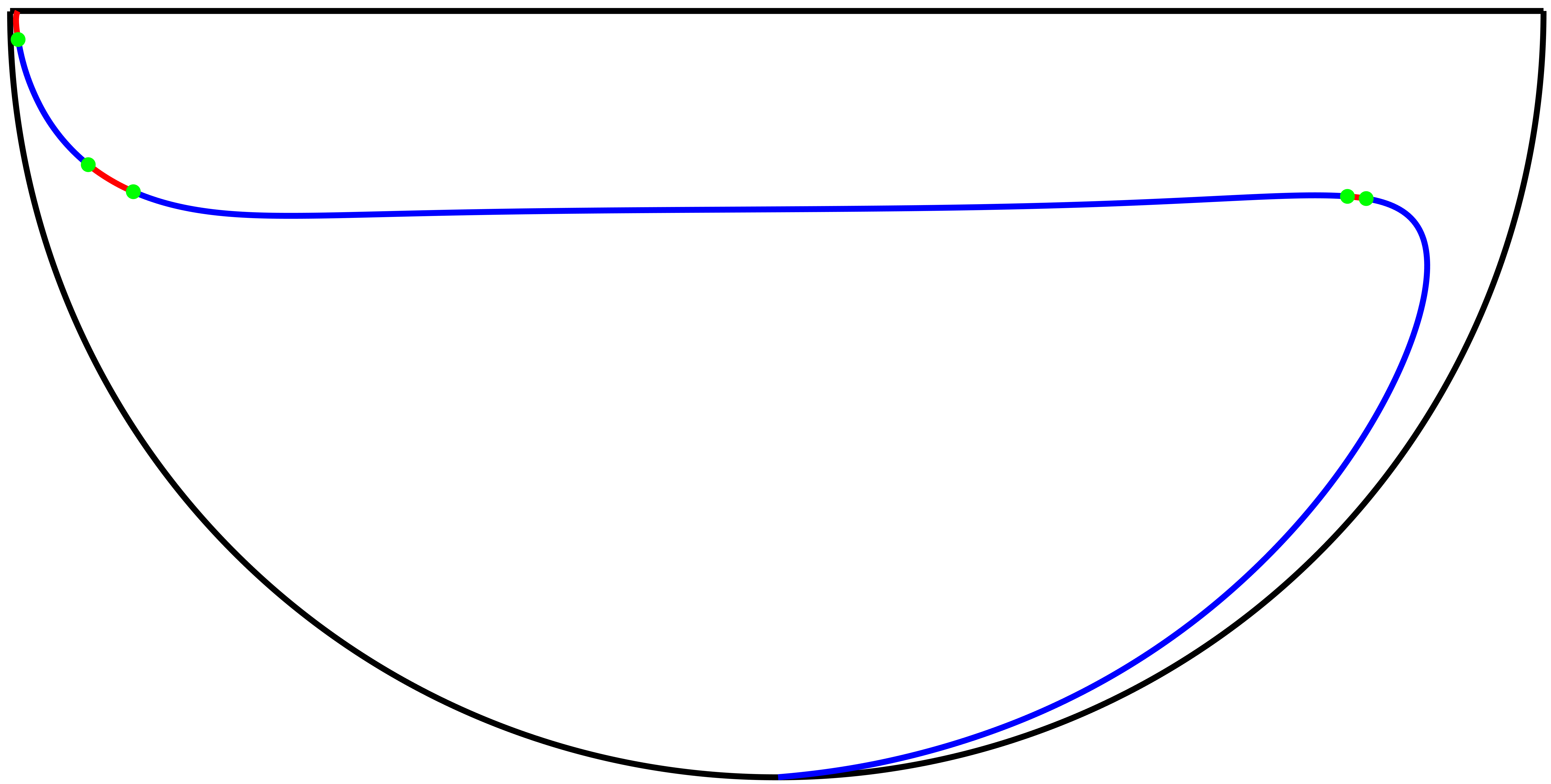} \\
(a) \\
\includegraphics[width=\columnwidth]{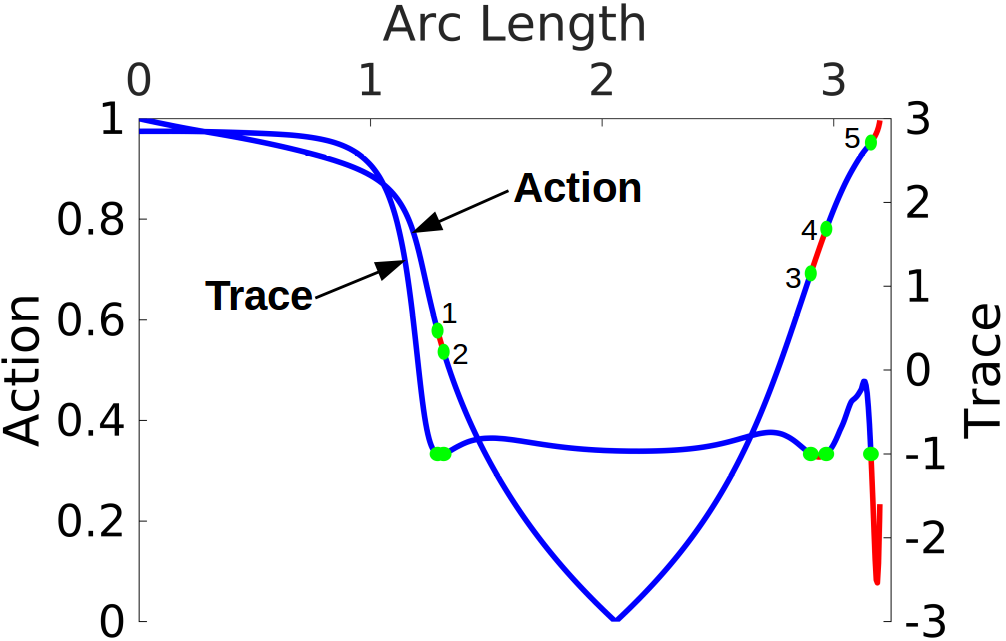} \\
(b)
\end{tabular}
\caption{(a) P1 line for $\beta=4$ and $\Theta=\pi/8$ on symmetry plane.  (b) Action and trace along P1 }
\label{fig:p1_theta_pi_over8_beta4}
\end{figure}

An example where such structures arise occurs in the flow with $\Theta=\pi/8$ and $\beta=4$.  The P1 line is shown in figure~\ref{fig:p1_theta_pi_over8_beta4}(a) and the action and trace as functions of arc length along P1 are shown in figure~\ref{fig:p1_theta_pi_over8_beta4}(b).   Although a little difficult to discern in figure~\ref{fig:p1_theta_pi_over8_beta4}(b), the trace has five points with value -1, corresponding to five 1:2 resonance points along P1. Each of these points lie on different shells and are labelled from 1-5 with increasing arc length. The P2 Line connected to point 5 attaches to the hemisphere wall like the P2 line seen previously in figure~\ref{fig:p1_p2_lines_3d}, whereas the first four 1:2 resonance points have different behaviour.

\begin{figure}
\centering
\includegraphics[width=\columnwidth]{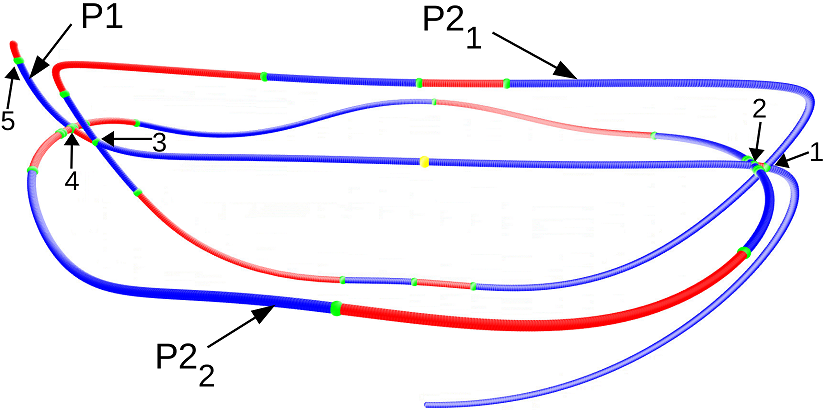}
\caption{$\Theta=\pi/8$ and $\beta=4$, Period-1 line $P1$ and period-2 line $P2_1$ are on symmetry plane, and period-2 line $P2_2$ is symmetric about symmetry plane. Elliptic line segments are coloured blue, hyperbolic line segments are coloured red and degenerate points are coloured green. The first four $1:2$ resonance points on $P_1$ are also shown with numbers $1-4$.} 
\label{fig:global_bifur_1_2_resonance_P1_P2lines_pi_over8_beta4}
\end{figure}

The period-2 lines corresponding to these four points are shown in
figure~\ref{fig:global_bifur_1_2_resonance_P1_P2lines_pi_over8_beta4},
where the period-2 lines extending from resonance points 1 and 3 join
together to form a single closed period-2 line (P2$_1$) that lies in
the symmetry plane, recalling that the P1 line marks the symmetry
plane.  The period-2 lines extending from resonance points 2 and 4
also join together to form a single closed period-2 line (P2$_2$) that
extends in an almost normal direction to the symmetry plane.  These
two closed P2 loops are interlinked.

\begin{figure}
\centering
\includegraphics[width=\columnwidth]{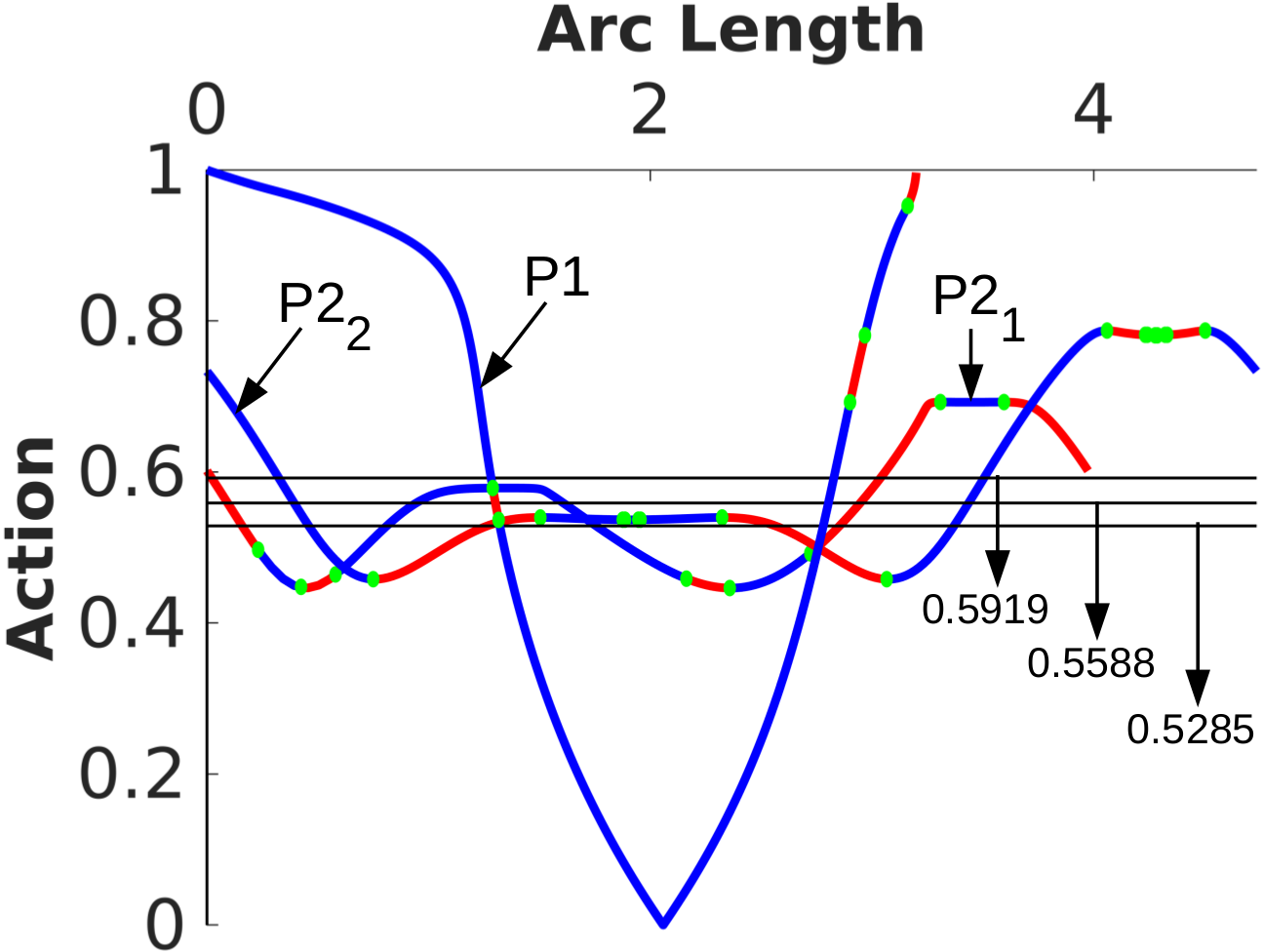}
\caption{Action versus arc length along periodic lines (P1, $P2_1$ and $P2_2$) shown in figure~\ref{fig:global_bifur_1_2_resonance_P1_P2lines_pi_over8_beta4} are shown. Three lines which correspond to action values (0.5285, 0.5588, 0.5919) are also shown.}
\label{fig:p1_p2lines_global_bifur_shlnum_arcln} 
\end{figure}

\begin{figure}
\centering
\begin{tabular}{c}
\includegraphics[width=0.9\columnwidth]{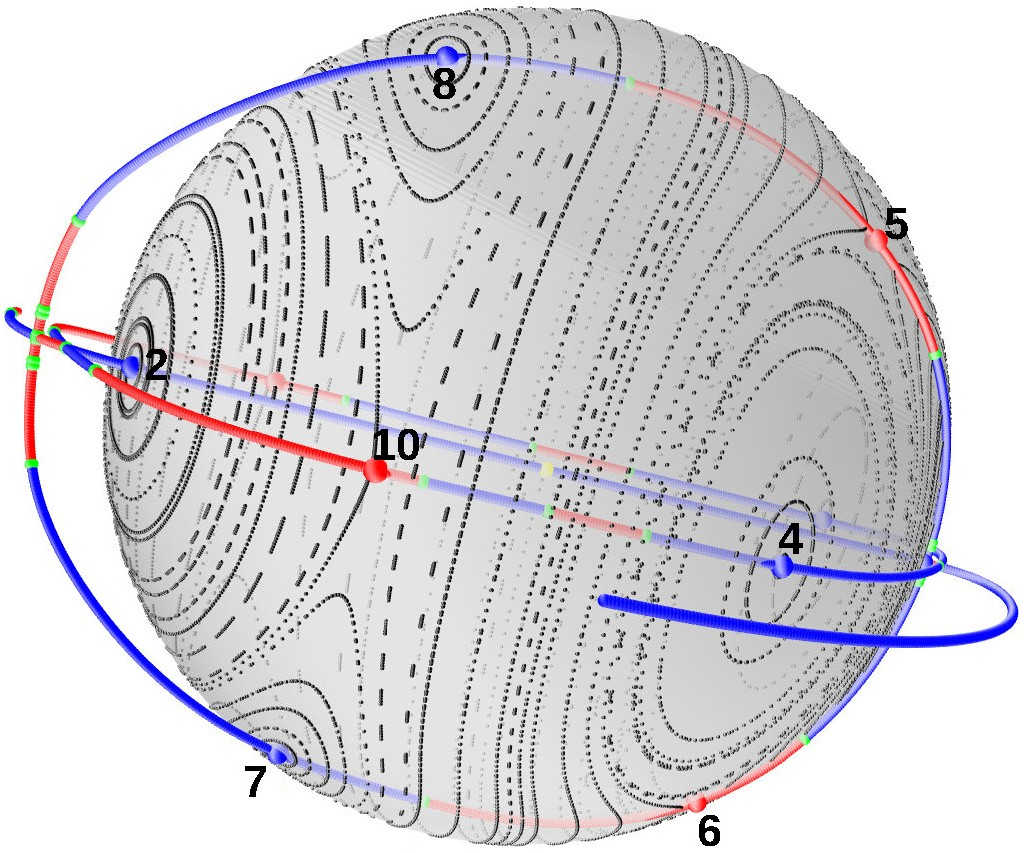} \\
(a) \\
\includegraphics[width=0.9\columnwidth]{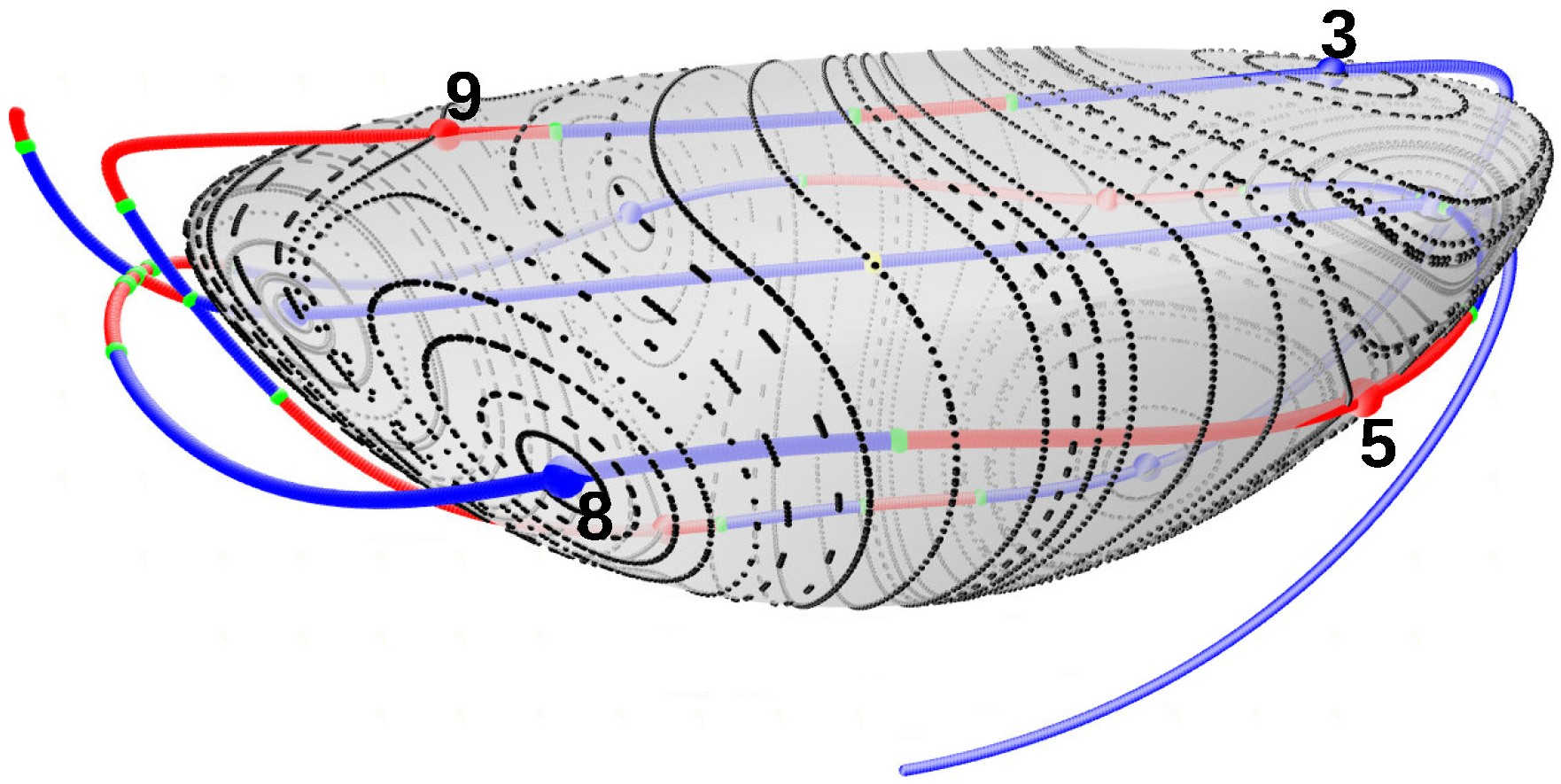} \\
(b) \\
\includegraphics[width=0.9\columnwidth]{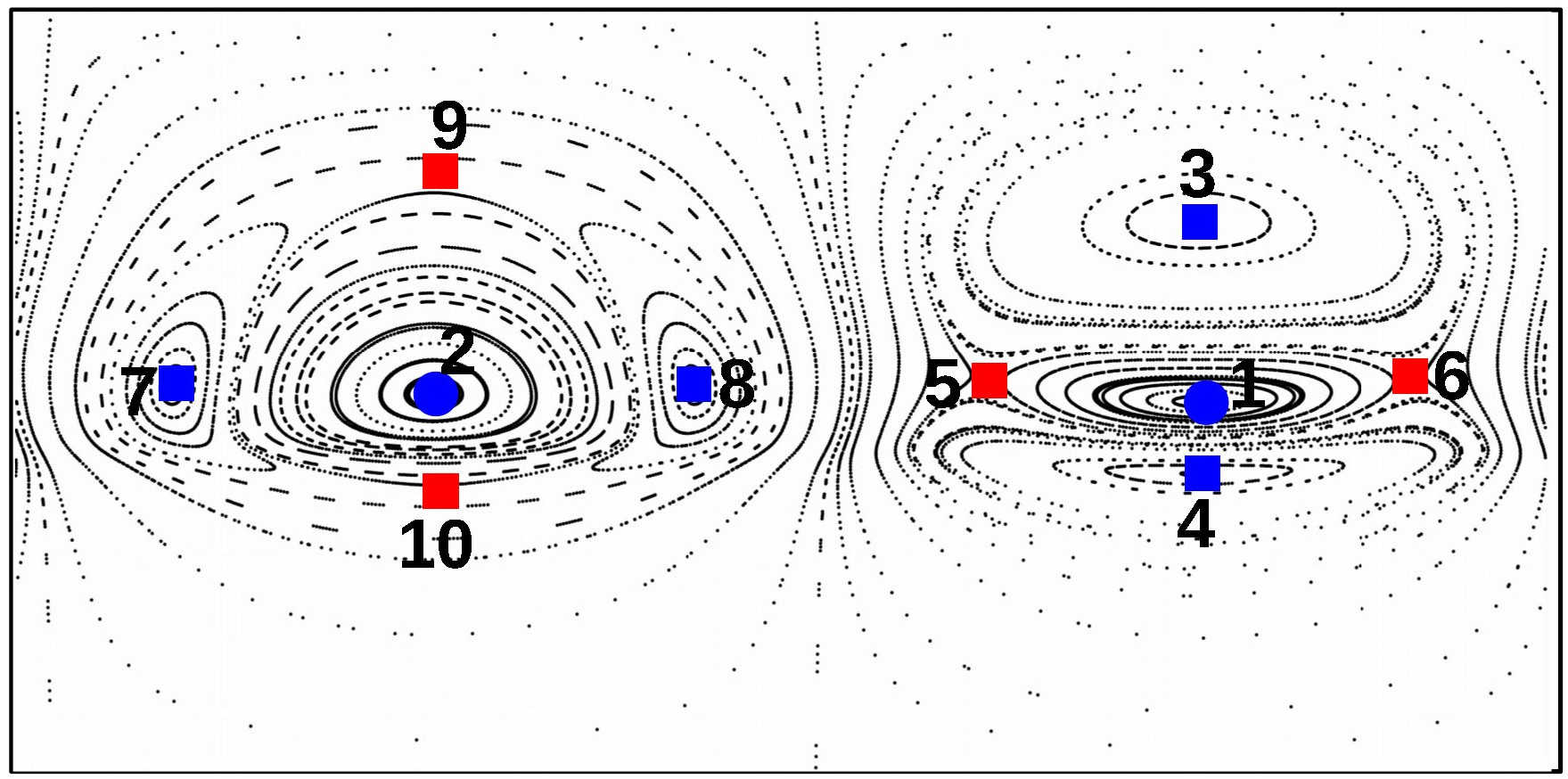} \\
(c)
\end{tabular}
\caption{The periodic lines shown in figure~\ref{fig:global_bifur_1_2_resonance_P1_P2lines_pi_over8_beta4} impart their character on to shell 0.5285; Stroboscopic map on Shell seen from (a) Bottom (b) Side and above (c) Stroboscopic map projected on to a plane.  P1 piercing points are represented by circles and P2 piercing points represented by squares; blue for elliptic and red for hyperbolic. }
\label{fig:res_1_2_global_bif_2_shell_5285}
\end{figure}

\begin{figure}
\centering
\begin{tabular}{c}
\includegraphics[width=0.9\columnwidth]{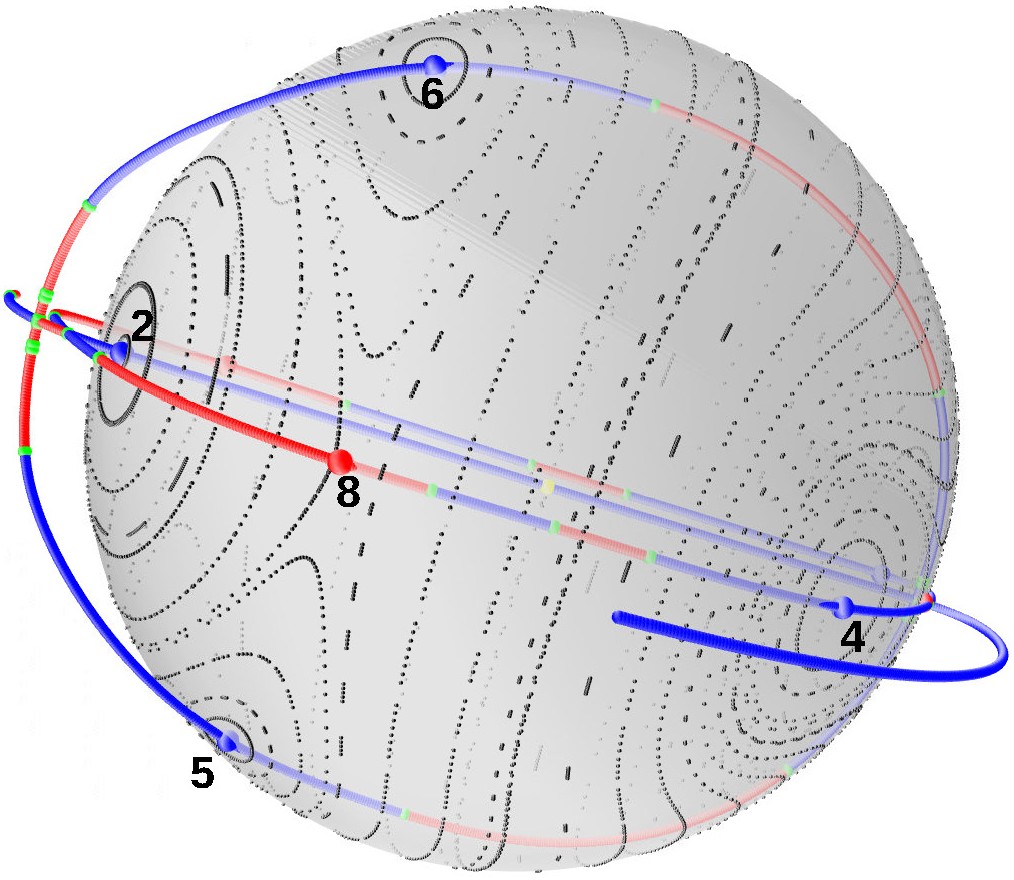} \\
(a) \\
\includegraphics[width=0.9\columnwidth]{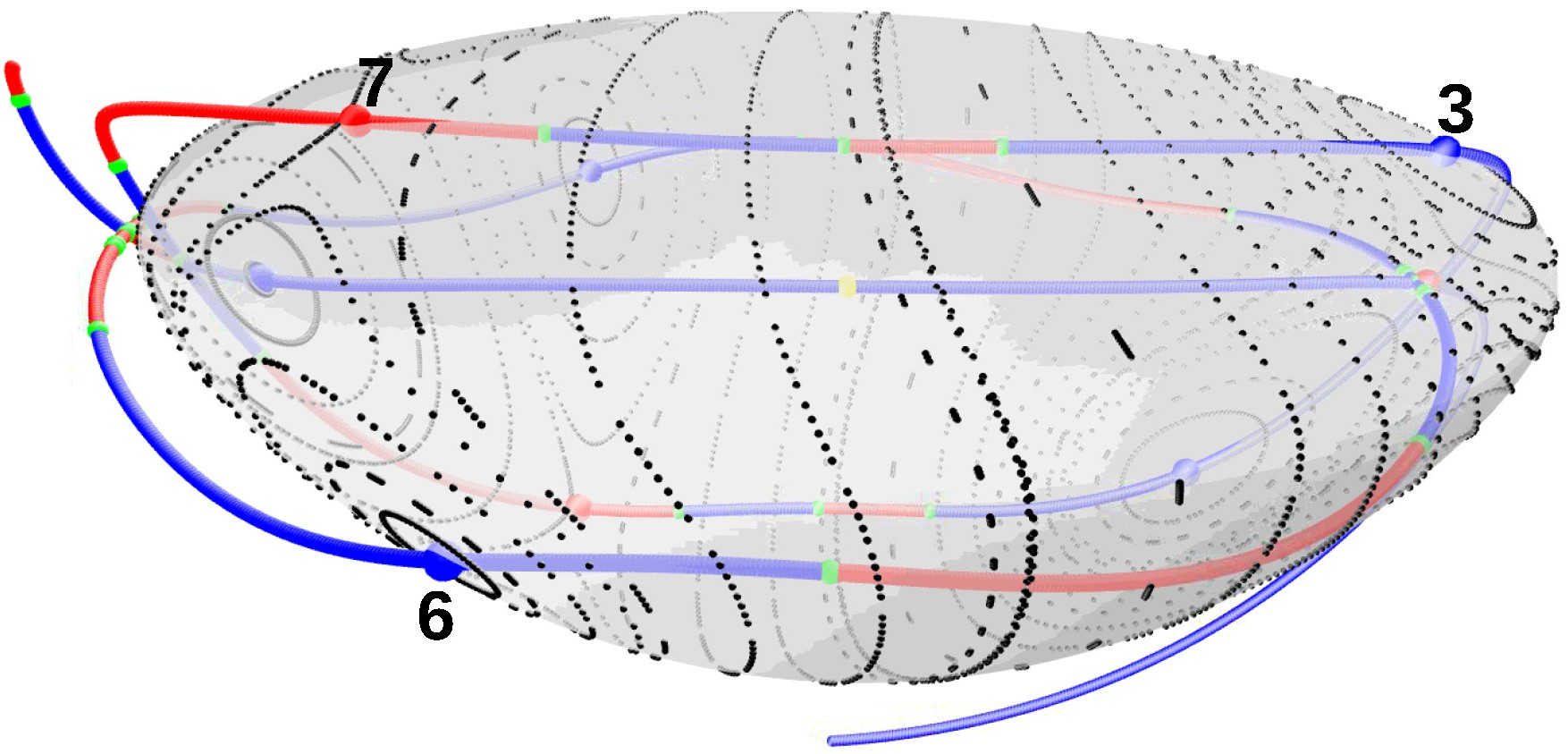} \\
(b) \\
\includegraphics[width=0.9\columnwidth]{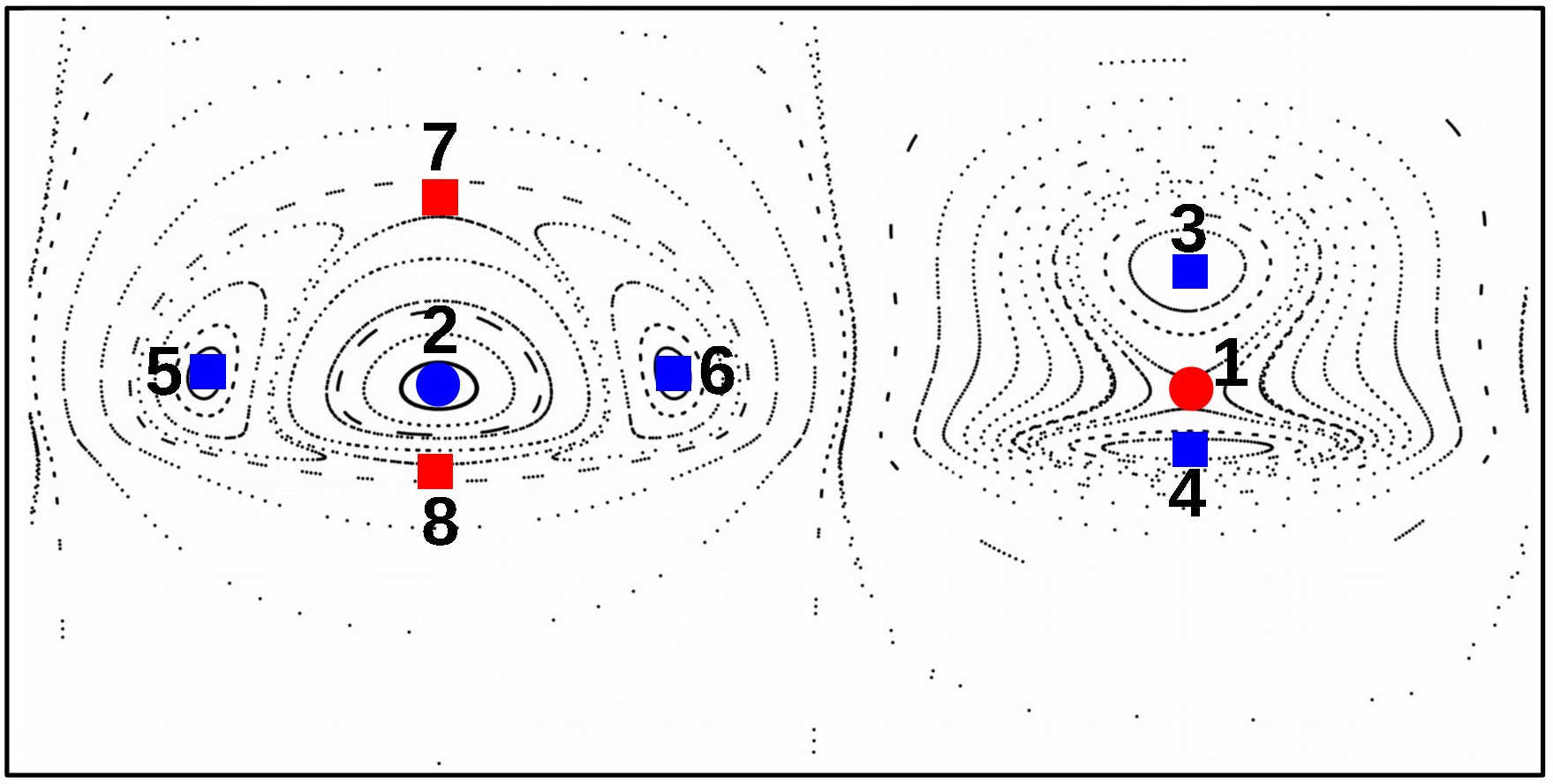} \\
(c)
\end{tabular}
\caption{Poincar\'{e} sections on shell 0.5588.  See caption for figure~\ref{fig:res_1_2_global_bif_2_shell_5285} for details}
\label{fig:res_1_2_global_bif_2_shell_5588}
\end{figure}

\begin{figure}
\centering
\begin{tabular}{c}
\includegraphics[width=0.9\columnwidth]{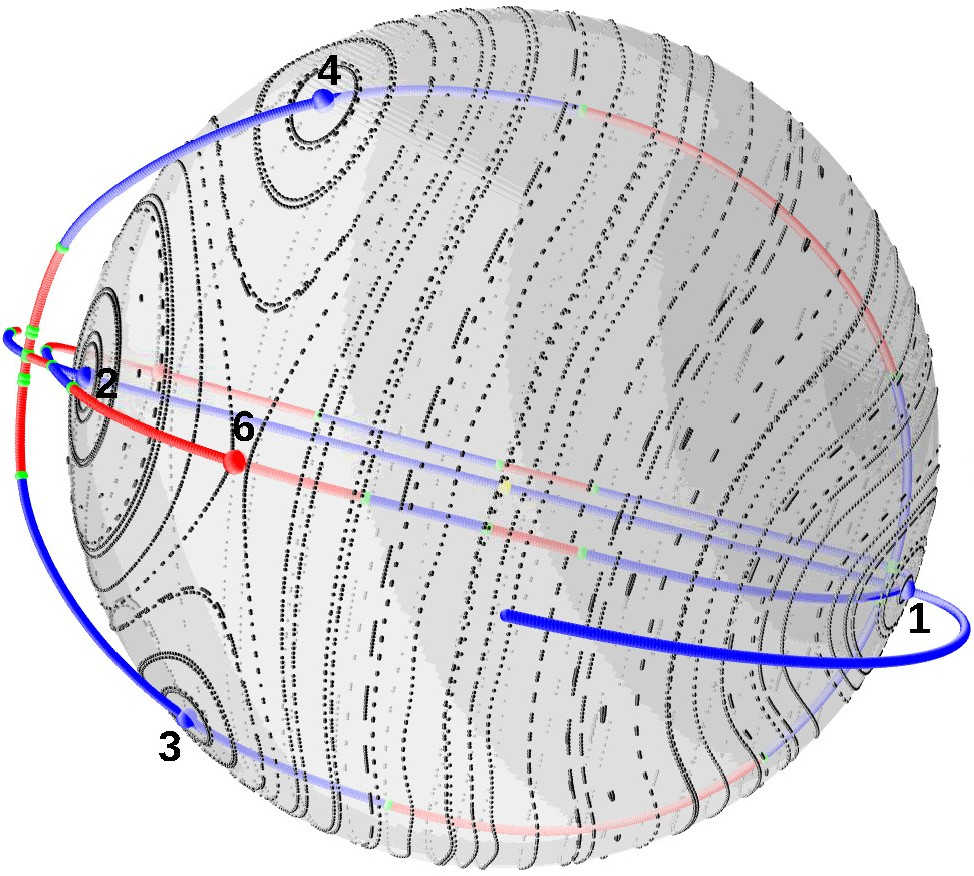} \\
(a) \\
\includegraphics[width=0.9\columnwidth]{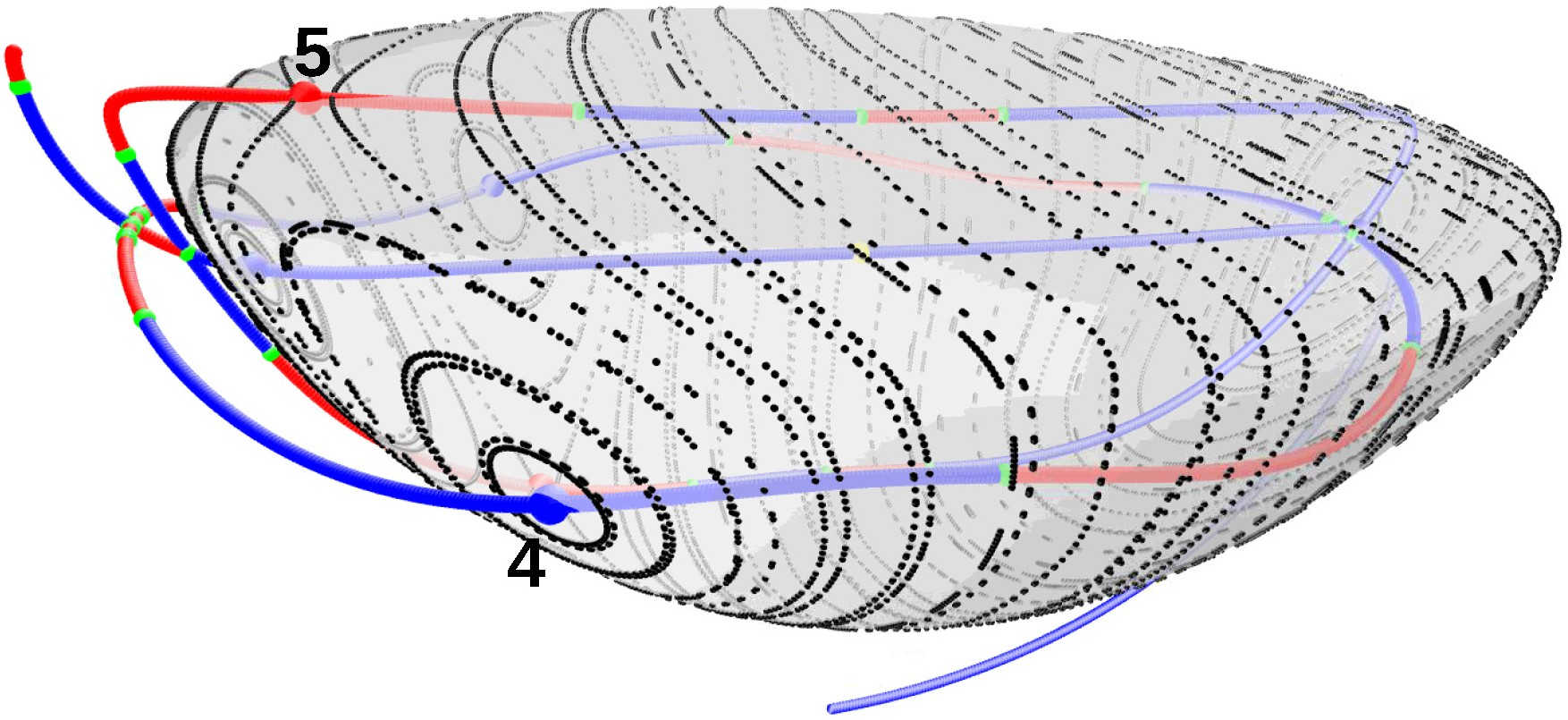} \\
(b) \\
\includegraphics[width=0.9\columnwidth]{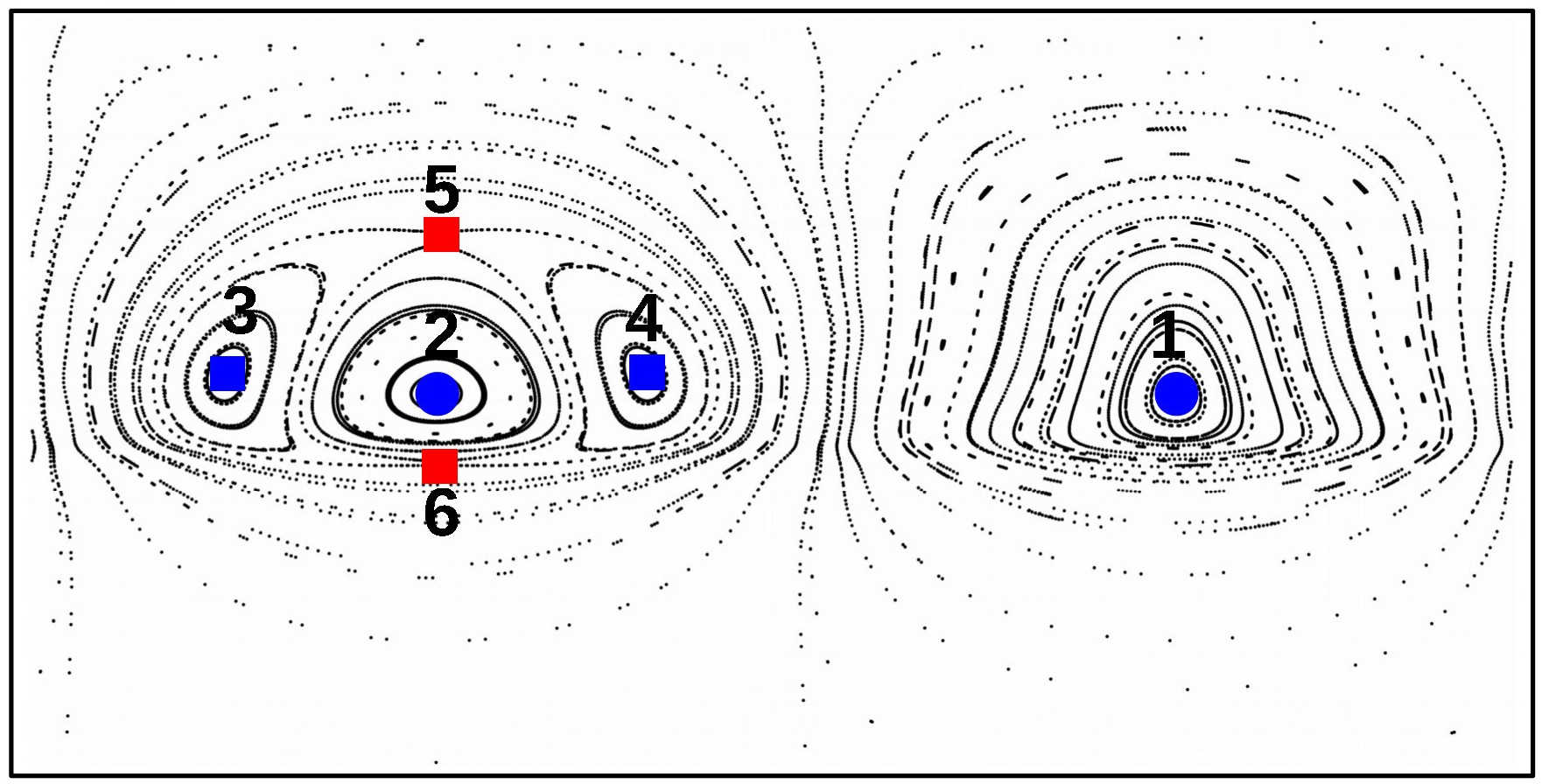} \\
(c)
\end{tabular}
\caption{Poincar\'{e} sections on shell 0.5919.  See caption for figure~\ref{fig:res_1_2_global_bif_2_shell_5285} for details}
\label{fig:res_1_2_global_bif_2_shell_5919}
\end{figure}

The P1 and P2 lines impart their character locally on shells at the piercing points and the looped nature of these closed curves coordinates the behaviour of the 2D surface flow at distant points on a shell. Action values on the P1 and P2 lines are plotted against arc length in figure~\ref{fig:p1_p2lines_global_bifur_shlnum_arcln}. Since the period-2 lines here are closed, arc length values for them are calculated from arbitrary starting points on the lines. To examine influence of the period-2 lines on the Lagrangian behvaiour of the shells, three nearby shells (shell values 0.5285, 0.5588 and 0.5919) are considered (shown as the three horizontal black lines in figure~\ref{fig:p1_p2lines_global_bifur_shlnum_arcln}).\\ \\
\textbf{\underline{Shell 0.5285}}\\
The Poincar\'{e} section on shell 0.5285 is shown in figure~\ref{fig:res_1_2_global_bif_2_shell_5285} and includes P1 and both period-2 lines.  Periodic line piercing points are highlighted with a small solid sphere and are labelled 1 to 10 for reference.  The shells are made translucent so that periodic line segments inside and outside of the shell can be distinguished.   Lines and spheres are coloured blue for elliptic and red for hyperbolic.  Figure~\ref{fig:res_1_2_global_bif_2_shell_5285}(a) presents the view from the bottom of the hemisphere, (b)~presents the view from the side and a little above the shell and (c)~presents a flattened out view of the section projected onto a 2D sheet.   As most clearly seen in figure~\ref{fig:res_1_2_global_bif_2_shell_5285}(c), shell 0.5285 contains four period-2 elliptic points, four period-2 hyperbolic points and two period-1 elliptic points. The stable and unstable manifolds of the four period-2 hyperbolic points have heteroclinic connections.\\ \\
\textbf{\underline{Shell 0.5588}}\\
As shell number increases (i.e.~we move further away from the central stagnation point), two of the hyperbolic piercing points on P2$_2$ disappear (as the corresponding segment of the P2 line no longer pierces this shell which fully contains 1:2 resonance point number 2).  The character of P1 also changes from elliptic to hyperbolic.  Similar plots of the Poincar\'{e} section on shell 0.5288 are shown in figure~\ref{fig:res_1_2_global_bif_2_shell_5588}.  This shell now contains four period-2 elliptic points, two period-2 hyperbolic points, a period-1 elliptic point and a period-1 hyperbolic point. The stable and unstable manifolds of two period-2 hyperbolic points have heteroclinic connections. The stable and unstable manifolds of the period-1 hyperbolic point have homoclinic connections.\\ \\
\textbf{\underline{Shell 0.5919}} \\
As shell number increases further to 0.5919, 1:2 resonance point
number 1 now falls inside this shell and the P2 lines no longer pierce
it in the vicinity of P1. Thus two period-2 elliptic points disappear
and the hyperbolic P1 line becomes elliptic once more.  Similar plots
of the Poincar\'{e} section for shell 0.5919 are shown in
figure~\ref{fig:res_1_2_global_bif_2_shell_5919}.  The shell contains
two period-2 elliptic points, two period-2 hyperbolic points and two
period-1 elliptic points. The stable and unstable manifolds of two
period-2 hyperbolic points have heteroclinic connections.

If shell number was increased still further, 1:2 resonance point 3 would cease to impart its structure locally on the Poincar\'{e} sections and eventually 1:2 resonance point 4 would follow suit.  The most important point to come out of the results presented in this section is that global and connected Lagrangian structures can be born from resonance points along the P1 line.  Although we do not explicitly show it here, similar statements can be made regarding higher order resonances that rise on higher order periodic lines.

\section{Conclusions and Discussion}

In this paper we considered the Periodically Reoriented Hemisphere
Flow (PRHF), an example of an 1-invariant flow which consists of a set
of nested, topologically spherical, space-filling shells on which all
transport occurs. In one invariant flows, isolated periodic points
cannot exist (see Appendix~\ref{app:no_iso_periodic_points}); closed
periodic lines or periodic lines with ends attached to the boundary
are possible. In the Stokes PRHF, a single P1 line exists on the
symmetry plane for all parameter values and forms the basis of the
Lagrangian skeleton of the flow.  This P1 line connects three
non-trivial fixed points: a point at the bottom of the hemisphere; a
central stagnation point which is at the centre of the topological
spheres; and, a point on the lid.

We define shell number as a unique (continuous) variable that acts as
a proxy for the action variable (in an action-angle-angle coordinate
system).  By plotting shell number as a function of distance along P1,
the local Lagrangian topology generated by the periodic lines can be
quickly and simply determined based solely on the number of shell
piercings and their character.  Although demonstrated here for the
PRHF, this same methodology can be used for any similar 1-invariant
system in which the action variable can be quantified in some way.
This then allows a global understanding to be quickly and simply
obtained of the likely types of Lagrangian structure on different
invariant surfaces of the flow.

Using this approach on the PRHF allowed us to show that the P1 line
passes through each shell at least twice.  Wherever P1 pierces a
shell, the stability of the piercing period-1 point is imparted to the
2D transport on that shell in the neighbourhood of the piercing site.
In the limit $\beta \to 0$, the P1 line is a stagnation line that
coincides with the $y$ axis. As $\beta$ increases, the tangent to the
P1 line at the central stagnation point rotates anti-clockwise which
increases the total length of the P1 line (since both the ends of a P1
line are fixed). For very low $\beta$ values, the entire P1 line is
elliptic and pierces each invariant surface only twice. As $\beta$
increases from very low to low values, hyperbolic segments start
appearing on the P1 line, although it still pierces each invariant
surface only twice.

As $\beta$ increases further, the P1 line starts folding inside a
shell it has already pierced.  The tangent point delineates a change
of periodic character on P1 from elliptic to hyperbolic (or {\it vice
  versa}) and an increase (or decrease) by two in the number of
period-1 points on shells just inside and outside of the degenerate
shell.  This leads to the creation of two local extrema (one minimum,
one maximum) which constitute a wiggle in the action vs arc length
plot. Each local extrema in the wiggle correspond to a point at which
the P1 line is tangent to a shell and each is also a first-order
degenerate point which corresponds with a cusp on the shell's 2D
Poincar\'{e} section. A single wiggle accommodates two first order
degenerate points, and the P1 line segment between these degenerate
points is always of hyperbolic type, with both sides of the P1 line
segments being elliptic.

The degenerate points on periodic lines are analogous to fixed point resonances in classical planar bifurcation theory. At an $n^{th}$ order degenerate point on a P1 line (also called a 1-$n$ resonance point), $n$ period-$n$ lines intersect the P1 line except for the cases of $n=1$ and $2$. The resonance points on a periodic line can be easily identified by calculating the trace of the deformation tensor value along the periodic line.  Resonance points have unique trace values as specified in  table~\ref{tab:degen_pt_eigenvalues}.  At a first order degenerate point, the P1 line is tangent to one of the shells.  At a second order degenerate point (a 1:2 resonance), a period-2 line intersects the periodic-1 line.  A higher-order periodic line intersecting with a lower order periodic line at a resonance point can be calculated numerically by identifying the resonance point using the method discussed in appendix~\ref{app:calc_higher_order_periodic_lines_stokes}. The significance of these resonances points is that one action flows can be completely understood by finding the resonance points on the P1 lines first, and then computing the corresponding higher-order periodic lines.  Then   period-2 lines can be computed, resonance points located, then determine the corresponding higher-order periodic lines, and so on. This process is pursued recursively until enough Lagrangian transport structures are uncovered for any given purpose.  

A higher-order periodic line extending from a resonance point may
attach to the boundary, or higher-order periodic lines extending from
different resonance points can sometimes join together to form a
closed periodic line.  These resonance bifurcations are local
bifurcations because they can be found in the neighbourhood of
periodic points (or fixed points).  The Lagrangian structures
emanating from different resonance bifurcation points can sometimes
connect with each other and form truly global transport structures.
With this way of building extended structures from lower order to
higher order periodic lines, truly global Lagrangian transport
structures can be uncovered for any one action flow.


In work still to be published, we introduce inertia to the flow to
break the remaining symmetry and perturb the Stokes resonance
structures.  We will see that the resonance structures identified in
this paper are perturbed into 3D Lagrangian coherent structures that
appear to be generic in the sense that they depend on starting from a
specific Stokes resonance and not on specific values of parameters.
An experiment is under construction that will allow these structures
to be visualised in a flow and will be reported in due course.

\begin{acknowledgments}
  BR was supported by the IITB-Monash Academy.  We thank Michael
  Coffey for making figure~\ref{fig:mobile_attachement}.
\end{acknowledgments}

\bibliography{hemisphere}

\appendix

\section{Computational Details}
\label{sec:method}

\subsection{Velocity Discontinuity}

\begin{figure}
\centering 
\includegraphics[width=\columnwidth]{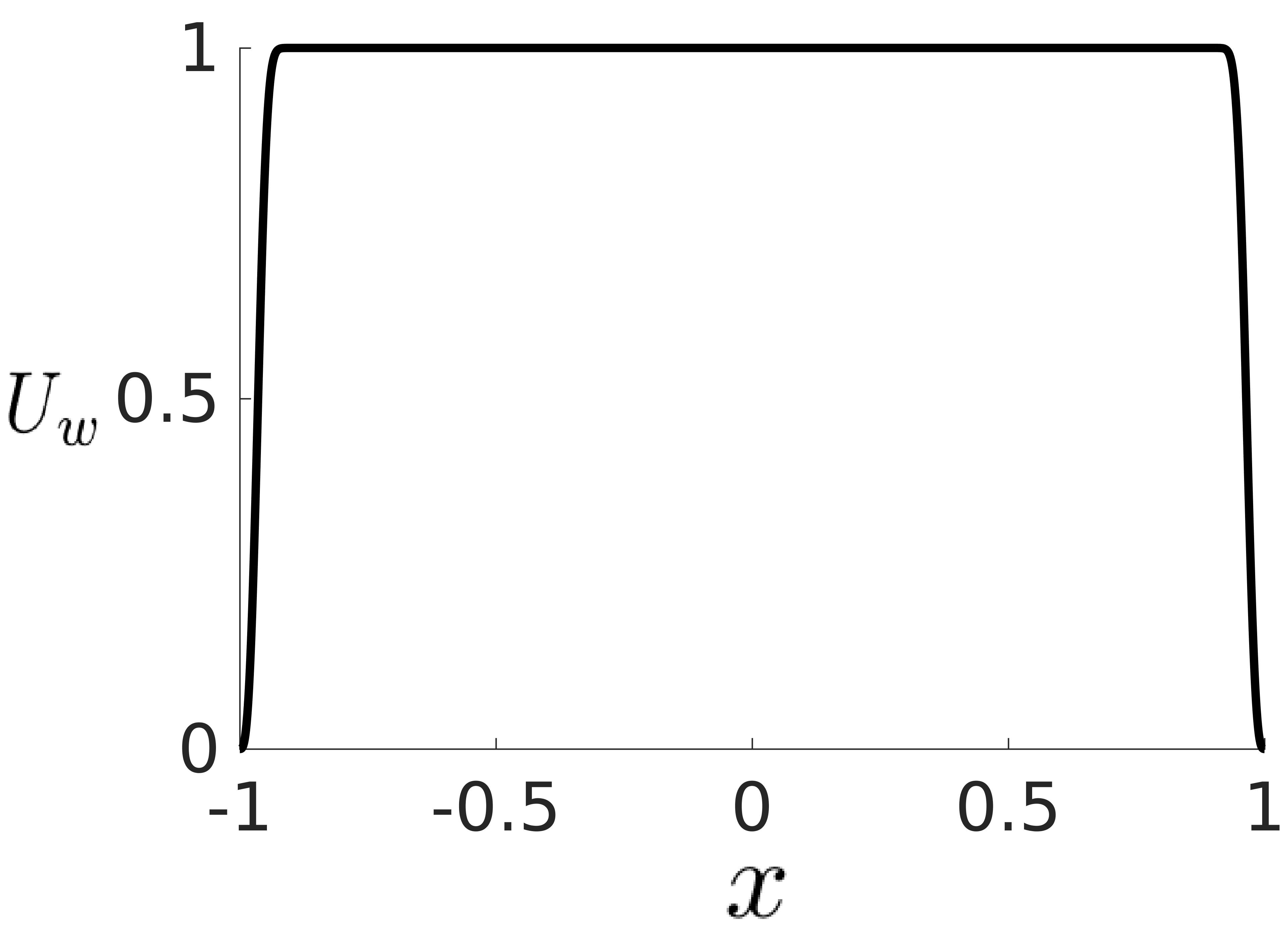}
\caption{Lid Velocity $U_w$ along the $x$ axis used in numerical
  computation to smooth the discontinuous boundary condition along the
  rim.}
\label{fig:lid_velocity}
\end{figure}

There is a velocity discontinuity at the rim of the hemisphere where
the sliding lid meets the stationary hemisphere wall.  Due to the
spectral nature of {\it semtex}, discontinuities present significant
difficulty.  To handle this we replace the discontinuity in velocity with a
continuous function that rapidly ramps the lid velocity from zero to 1
over a small region at the rim.  The lid velocity function we use is
$U_w = 1-e^{-2000(1-x^2)^3}$, which is plotted in
figure~\ref{fig:lid_velocity}.

In order to remove the singular coordinate from the interior of the
domain that arises when $\sin \theta=0$, we preferred to use a
hemispherical domain specified by $r \in [0,1]$, $\theta \in [0,\pi]$,
and $\phi \in [0,\pi]$, where $\theta$ is the polar angle and $\phi$
is the azimuthal angle (instead of $\theta \in [0,\pi/2] $ and
$\phi \in [-\pi,\pi] $). The computational method of Ravu~{\it et
  al.}~\cite{Ravu:2016jcp} fits a 3D spline to the velocity field
obtained from {\it semtex} in a manner that ensures the resulting
analytic velocity field is exactly divergence-free everywhere.  We
found that identification of periodic lines and Lagrangian structures
was critically dependent on this divergence-free interpolation; other
(high-order) interpolation schemes were unable to provide the level of
accuracy required to properly resolve the coherent Lagrangian
structures from a purely numerical data set.

\subsection{Period-1 Lines and their Stability}
\label{app:P1line}

\begin{figure}
\centering 
\includegraphics[width=0.6\columnwidth]{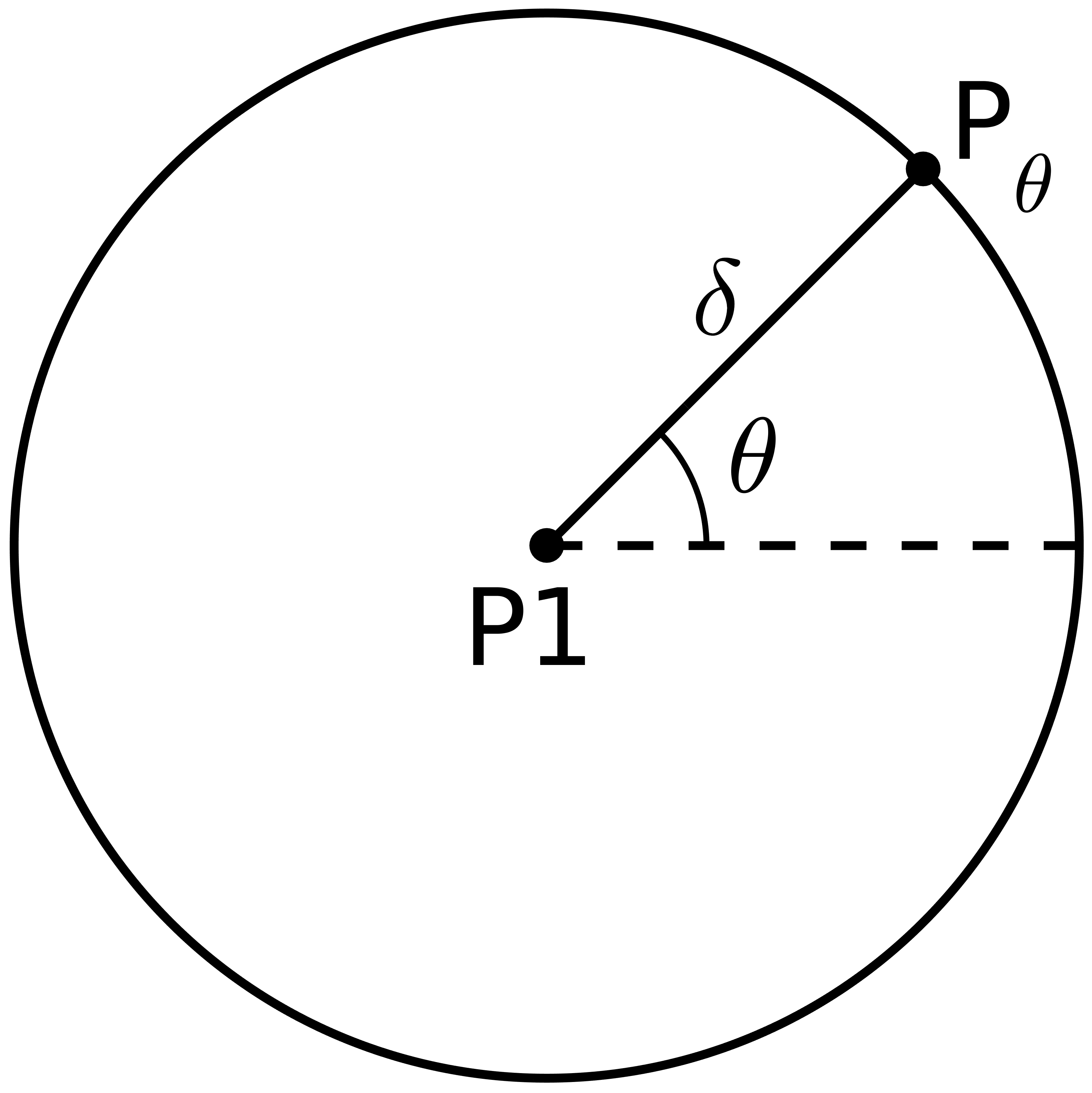}
\caption{Bootstrap method to approximate a periodic line.  At a known
  point P1 on the line the circle, centred on P1 and drawn in the
  symmetry plane $S_{\Theta}$ locates the next point at P$_\theta$.}
\label{fig:circle_sym_plane}
\end{figure}

In order to determine the P1 line, we start at the central fixed
point, $(r_{s},\pi/2,\pi/2)$.  We know from symmetry that this point
is on a line that goes from the origin $(0,\pi/2,\pi/2)$ to the base
of the hemisphere $(1,\pi/2,\pi/2)$.  On this line, $V_{\phi}$ changes
sign and has a value of zero at the central fixed point. We use
bisection to find $r_s \approx 0.2589$.  P1 consists of two
curve segments: the upper part P1$_U$ from the central fixed point to
the attachment point on the lid and the lower part P1$_L$ from the
central fixed point to the attachment point at the apex of the
hemisphere.  We compute P1$_U$ and P1$_L$ separately with the
following algorithm.

We choose a small increment, $\delta$, and create a circle of this
radius centred around a known P1 point on the symmetry plane
(initially the central fixed point).  P1 intersects this circle at two
points, and we choose one, $P_{\theta}$
(figure~\ref{fig:circle_sym_plane}).  As
$\big\{\Phi_{\beta/2} (\boldsymbol{x}_{P1})\big\}_x=0$ for a period-1
point $\boldsymbol{x}_{P1}$ (figure \ref{fig:proj_streamline_xy} and
figure~\ref{fig:proj_streamline_xz}), the $x$--component
(i.e.~$\big\{\Phi_{\beta/2} (P_{\theta})\big\}_x$) changes sign around
the circle and is zero where P1 intersects the circle.  Again, we
use bisection to find the value of $\theta$ at which P1 intersects the
circle.  The value of $P_\theta$ becomes the starting point for the
next iteration of the algorithm.

We determine the stability of P1 by expanding $\boldsymbol{\Psi}_T(\boldsymbol{x})$ at
a period-1 point $\boldsymbol{x}_{P1}$ as
\begin{equation}
\boldsymbol{\Psi}_T(\boldsymbol{x}_{P1}+d\boldsymbol{x})= \boldsymbol{x}_{P1}+ \frac{\partial \boldsymbol{\Psi}_T}{\partial \boldsymbol{x}} \bigg|_{\boldsymbol{x}=\boldsymbol{x}_{P1}} d\boldsymbol{x} + \mathcal{O}({d\boldsymbol{x}}^2),
\end{equation}
where $\boldsymbol{F}=\frac{\partial \boldsymbol{\Psi}_T}{\partial \boldsymbol{x}}$
is the deformation tensor that we compute numerically using central
differences.  The two eigenvalues of $\boldsymbol{F}$ not equal to one determine
the stability of the period-1 point.  If they are real, the point is
hyperbolic; if they are complex, the point is elliptic.
  
\subsection{Calculation of higher order periodic lines of PRHF in the Stokes regime}
\label{app:calc_higher_order_periodic_lines_stokes}

We show in this section how to calculate the period-$n$ lines which go
though an $n^{th}$ order degenerate point on the P1 line.  To compute
a period-$n$ line requires an initial period-$n$ point, which can be
any point on that line.  So, a period-$n$ line is computed here in two
steps: 1) calculate any one period-$n$ point on the period-$n$ line,
2) calculate the full period-$n$ line using the initial period-$n$
point.  In section~\ref{sec:resonances_on_periodic_lines}, the central
role that degenerate points play in higher order periodicities are
discussed.  Without going into detail here, these will provide a
starting guess for higher order points.  For example, $n$ period-$n$
lines intersect at an $n^{th}$ order degenerate point on a P1 line.
This $n^{th}$ order degenerate point cannot be used as an initial
period-$n$ point for all the $n$ period-$n$ lines to calculate them
because this point belongs to all the lines.  To calculate these $n$
period-$n$ lines, a separate initial period-n point for every period-n
line is needed (i.e. total $n$ initial period-$n$ points).  The
$n^{th}$ order degenerate point is used to calculate $n$ initial
period-$n$ points which are needed for the calculation of $n$
period-$n$ lines.  The initial $n$ period-$n$ points are calculated
first as described in
section~\ref{subsubsec:steps_to_generate_starting_points_higher_Pn_line},
then the corresponding $n$ period-$n$ lines are calculated as
described in section~\ref{subsubsec:steps_to_compute_next_pn}.

\subsubsection{Finding $n$ initial period-$n$ points}
\label{subsubsec:steps_to_generate_starting_points_higher_Pn_line}
\begin{itemize}
\item Identify an $n^{th}$ order degenerate point on the P1 line as described in section~\ref{sec:resonances_eigenvalues}.
\item Create a spherical grid of points which are almost equidistant from each other in a small sphere of radius $0.1$ centred around the degenerate point. Because $n$ period-$n$ lines intersect at this degenerate point, this sphere contains segments of all the $n$ period-$n$ lines.
\item Calculate the magnitude of the displacement function $\bm{G}^n(\bm{x})$ at each grid point via
\begin{equation}
\label{eq:pn_fun_root}
\bm{G}^n(\bm{x})=\bm{\Psi}^n (\bm{x})-\bm{x}.
\end{equation}.
\item Arrange the grid points as an array in ascending order of $||\bm{G}^n(\bm{x})||$, since smaller values of  $||\bm{G}^n(\bm{x})||$ are closer to period-$n$ lines, such points are good initial guesses for Broyden's method which is used to find $n$ initial period-$n$ points.
\item The grid points which were sorted according to the magnitude of $\bm{G}^n(\bm{x})$ are then used one by one as initial guess for Broyden's method and initial period-$n$ points are calculated.
\item Continue finding period-$n$ points from the grid points until at least one separate initial period-$n$ point for every period-$n$ line is found.
\item From the computed period-$n$ points, select one period-$n$ point for each period-$n$ line and use them as initial points to calculate $n$ period-$n$ lines as described in the next section~\ref{subsubsec:steps_to_compute_next_pn}.
\end{itemize}

\subsubsection{Calculation of a period-$n$ line using a known initial periodic point on that line}
\label{subsubsec:steps_to_compute_next_pn}
Assume $Pn$ is a period-$n$ line to be computed, and $\bm{x}_k$ is a known initial period-$n$ point on that line. $Pn$ is divided at the point $\bm{x}_k$ into two parts: $Pn_L$ and $Pn_R$. These two parts are calculated separately in the same way as described below.

\begin{enumerate}
\item Compute the deformation tensor $\frac{\partial \bm{\Psi}^n}{\partial \bm{x}}\bigr\rvert_{\bm{x}=\bm{x}_{k}}$ at $\bm{x}_k$. Because the point $\bm{x}_{k}$ belongs to the period-$n$ line, one of its eigenvalues of the deformation tensor (i.e. $\frac{\partial \bm{\Psi}^n}{\partial \bm{x}}\bigr\rvert_{\bm{x}=\bm{x}_{k}}$) will be one (say $\lambda_3=1$), and the eigenvector $\bm{e}_3$ corresponding to this eigenvalue is tangent to the period-$n$ line.
\item Compute eigenvector $\bm{e}_3$.

\item Choose $\bm{x}'_{k+1}=\bm{x}_k \pm \epsilon \bm{e}_3$ where the
  sign is chosen to ensure $\bm{x}'_{k+1}$ continues segment
  $\bm{x}_{k-1} \to \bm{x}_k$. $\epsilon=5 \times 10^{-3}$ is used.
  Note: In the first iteration, to calculate $\bm{x}'_{k+1}$, a minus
  sign is used in the calculation of $Pn_L$ and a plus sign is used in
  the calculation of $Pn_R$.  The signs can be interchanged in the
  first iteration, but the important thing is different signs must be
  used in the calculation of $Pn_L$ and $Pn_R$.  This makes the search
  for periodic points of $Pn_L$ and $Pn_R$ line segments happen in
  opposite directions from the initial periodic point.

\item Create a sphere centred at $\bm{x}'_{k+1}$ with radius
  $\epsilon/5$ and search for the root of
  equation~\ref{eq:pn_fun_root} inside the sphere using Broyden's
  method to find $\bm{x}_{k+1}$ which is the next period-$n$ point on
  the line after $\bm{x}_{k}$. If the root is not found within the
  desired convergence value ($10^{-9}$), increase the radius of the
  sphere and do the same process again to find $\bm{x}_{k+1}$; this
  process, which will always find the root, is continued until the
  root is found within the desired convergence value.
  
\item Continue steps (1) to (4) recursively to calculate period-$n$
  points along the line until the line touches the boundary or the
  line reaches the initial point if the periodic line is closed,
  because a periodic line can be closed or its ends are attached to
  the boundary (see Appendix~\ref{app:no_iso_periodic_points}).
\end{enumerate}



\section{Non-existence of isolated periodic points}
\label{app:no_iso_periodic_points}

In principle, in a one-invariant flow a coordinate transformation can
be used to map fluid particle positions $(x,y,z)$ to
$\bm{\xi}(\theta_1,\theta_2,I)$, where $I$ is the action variable and
$(\theta_{1},\theta_2)$ are angle variables.
\begin{align}
\bm{\xi} &= \begin{bmatrix}
           \theta_1\\
           \theta_2 \\
           I
         \end{bmatrix}
\end{align}
When the system parameters $\Theta$ and $\beta$ are fixed, a map can be defined as
\begin{equation}
\label{eq:action_angle_2d_map}
\bm{\xi} \mapsto \bm{f}(\bm{\xi}) ,
\end{equation}
where $\bm{f}$ is continuous in $(\theta_1,\theta_2,I)$. $\bm{\xi}_0$, $\bm{\xi}_1$, $\bm{\xi}_2$, $\ldots$, $\bm{\xi}_k$ is a fluid particle orbit for $k$ periods, where 
\begin{equation}
\bm{\xi}_n=\bm{f}(\bm{\xi}_{n-1}), \quad n=1,2,3, \ldots, k.
\end{equation}
The period-$n$ fixed point equation, by definition, is
\begin{equation}
\label{eq:fixed_point_2d_map}
\bm{f}^n(\bm{\xi})-\bm{\xi}=0 .
\end{equation}
Since $\bm{f}$ and $\bm{\xi}$ are three-dimensional vectors, equation~\ref{eq:fixed_point_2d_map} can be split component wise into three equations as
\begin{subequations}
\label{eq:fixed_point_2d_map_seperate}
\begin{align}
   f_1^n(\theta_1,\theta_2,I)- \theta_1  &= 0 ,\\ 
    f_2^n(\theta_1,\theta_2,I) - \theta_2  &=  0 ,\\
    f_3^n(\theta_1,\theta_2,I) - I  &=  0 \Rightarrow I-I=0 .
\end{align}
\end{subequations}
Because the action variable of a fluid particle does not change in the
Stokes flow, equation~\ref{eq:fixed_point_2d_map_seperate}c is not an
independent equation (i.e.\ this equation is satisfied at every point
in the domain).  Equations~\ref{eq:fixed_point_2d_map_seperate}a and
\ref{eq:fixed_point_2d_map_seperate}b represent two-dimensional
surfaces that are either closed or attached to the boundary, and the
intersection of these two surfaces is the solution of
equation~\ref{eq:fixed_point_2d_map_seperate} which is the same as
equation~\ref{eq:fixed_point_2d_map}.  If the two surfaces intersect
(i.e.\ if the solution exists for some order of period $n$), the
intersection must be a line and cannot be a point.  So
equation~\ref{eq:fixed_point_2d_map} cannot have point solutions.
Hence, isolated period-1 points cannot exist in the PRHF in the Stokes
limit.  Periodic lines with their ends attached to the boundary or
closed periodic lines are possible.

\section{P1 Points Lie in the Symmetry Plane}
\label{app:symmetry}

Consider a P1 point $\boldsymbol{x}_{P1}$.  We prove that this point
must lie on the symmetry plane $\boldsymbol{I}_\Theta$ by the
following steps.

\begin{figure}
\centering 
\begin{subfigure}[b]{0.5\textwidth}
\includegraphics[width=0.8\columnwidth]{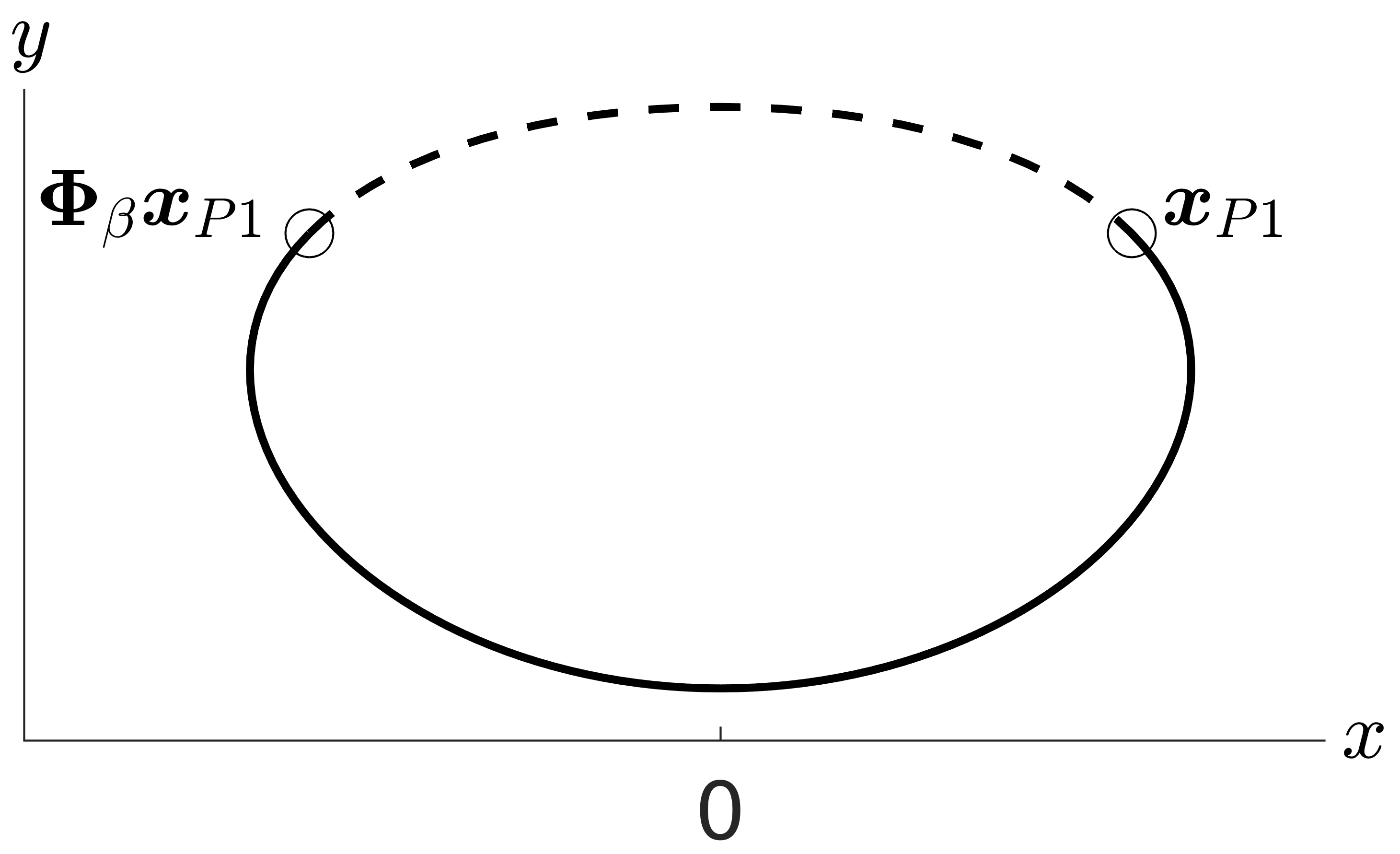}
   \caption{}
   \label{fig:proj_streamline_xy} 
\end{subfigure}

\begin{subfigure}[b]{0.5\textwidth}
\includegraphics[width=0.8\columnwidth]{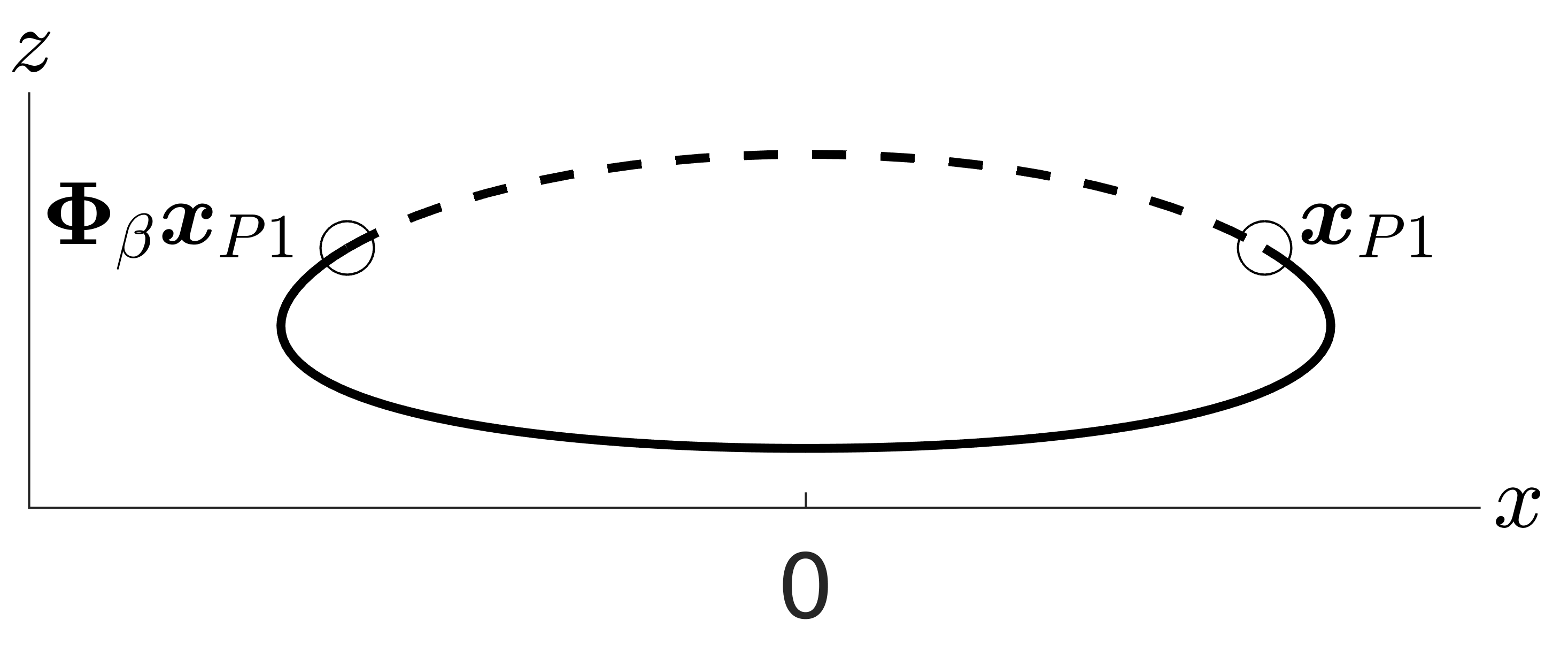}
   \caption{}
   \label{fig:proj_streamline_xz} 
\end{subfigure}
\caption{Projection of a stream line of the base flow: (a) $x$--$y$ projection and (b) $x$--$z$ projection. }
\end{figure}

From the definition of a period-1 point
\begin{equation}
\label{eq:p1_pnt_eqn}
\boldsymbol{x}_{P1}  = \boldsymbol{\Psi}_T \boldsymbol{x}_{P1} 
           = R_{-\Theta} \boldsymbol{\Phi}_{\beta} \boldsymbol{x}_{P1}.
\end{equation}
Applying $R_{\Theta}$ to both sides of equation~\ref{eq:p1_pnt_eqn},
we find that
\begin{equation}
\label{eq:cond_p1}
R_{\Theta}\boldsymbol{x}_{P1} = \boldsymbol{\Phi}_{\beta} \boldsymbol{x}_{P1}.
\end{equation}
Now consider separately how each coordinate of $\boldsymbol{x}_{P1}$
is modified by application of the map $\boldsymbol{\Phi}_{\beta}$.
Because the point $\boldsymbol{\Phi}_{\beta} \boldsymbol{x}_{P1}$ must
be rotated to complete one full period of the reoriented flow map
(equation~\ref{eq:p1_pnt_eqn}) then
$\boldsymbol{\Phi}_{\beta} \boldsymbol{x}_{P1} \neq
\boldsymbol{x}_{P1}$.  Because rotation about the y-axis does not
change the $y$--coordinate of a point, application of $R_{\Theta}$ to
$ \boldsymbol{x}_{P1}$ results in
\begin{equation}
\label{eq:rtheta_nochange_y}
{(R_{\Theta} \boldsymbol{x}_{P1})}_y={(\boldsymbol{x}_{P1})}_y,
\end{equation}
or from equation~\ref{eq:cond_p1} 
\begin{equation}
\label{eq:streamline_ylevel}
{(\boldsymbol{\Phi}_{\beta} \boldsymbol{x}_{P1})}_y={(\boldsymbol{x}_{P1})}_y,
\end{equation}
i.e.~the $y$--coordinate of a period-1 point is not changed by
application of the base flow (figure~\ref{fig:proj_streamline_xy}).
Because streamlines of the base flow are symmetric about $x=0$ and the
$y$ components of $\boldsymbol{x}_{P1}$ and
$\boldsymbol{\Phi}_{\beta} \boldsymbol{x}_{P1}$ are the same, we see
from figures~\ref{fig:proj_streamline_xy} and
\ref{fig:proj_streamline_xz} that it must also be true that
\begin{align}
\label{eq:xz_cond}
\begin{split}
{(\boldsymbol{\Phi}_{\beta} \boldsymbol{x}_{P1})}_z={(\boldsymbol{x}_{P1})}_z  \\
{(\boldsymbol{\Phi}_{\beta} \boldsymbol{x}_{P1})}_x=-{(\boldsymbol{x}_{P1})}_x.
\end{split}
\end{align}
Equations~\ref{eq:streamline_ylevel} and \ref{eq:xz_cond} can then be
written without reference to coordinate directions as
\begin{equation}
\label{eq:baseflow_map_becomes}
\boldsymbol{\Phi}_{\beta} \boldsymbol{x}_{P1}=S_x \boldsymbol{x}_{P1}.
\end{equation}
By substituting equation~\ref{eq:baseflow_map_becomes} into
equation~\ref{eq:cond_p1}, we get
\begin{align}
\label{eq:p1_cond2}
\begin{split}
R_{\Theta} \boldsymbol{x}_{P1} & =S_x \boldsymbol{x}_{P1} \\
\boldsymbol{x}_{P1} & = R_{-\Theta} S_x \boldsymbol{x}_{P1},
\end{split}
\end{align}
and by replacing $R_{-\Theta} S_x$ by $S_{\Theta}$
(equation~\ref{eq:sym_plane}) obtain
\begin{equation}
\boldsymbol{x}_{P1}=S_{\Theta} \boldsymbol{x}_{P1}.
\end{equation}
Hence $\boldsymbol{x}_{P1} \in \boldsymbol{I}_{\Theta}$, i.e.\ $P1$ points must
lie on the symmetry plane $\boldsymbol{I}_{\Theta}$.

\section{Invariants of deformation tensor}
\label{app:invariants_deformation_tensor}

As well as of using eigenvalues of deformation tensor $\bm{F}$ to
determine stability, invariants of $\bm{F}$ (the trace $\tau$, the
second trace $\sigma$ and the determinant $det(\bm{F})$, which are
discussed in section~\ref{sec:resonances_eigenvalues}) can also be
used. The characteristic equation for eigenvalues $\lambda$ of a
deformation tensor $\bm{F}$ is given by
\begin{equation}
\label{eq:char_eqn_def_ten}
p(\lambda)=\lambda^3-(\Tr \bm{F}) \lambda^2 + \frac{1}{2}\big((\Tr \bm{F})^2-\Tr \bm{F}^2 \big) \lambda-det(\bm{F}) ,
\end{equation}
where
\begin{equation}
\Tr \bm{F}=\tau=\lambda_1+\lambda_2+\lambda_3
\end{equation}
and the term $\frac{1}{2}\big((\Tr \bm{F})^2-\Tr \bm{F}^2 \big)$ in equation~\ref{eq:char_eqn_def_ten} is the second trace $\sigma$ .
\begin{equation}
\label{eq:sigma_def}
\sigma=\frac{1}{2}\big((\Tr \bm{F})^2-\Tr \bm{F}^2 \big)=\lambda_1\lambda_2+\lambda_2\lambda_3+\lambda_3\lambda_1 
\end{equation}
Because $det(\bm{F})=1$ for volume preserving maps, the characteristic equation of the Deformation Tensor $\bm{F}$ can be written as,
\begin{equation}
\label{eq:inv_def_ten_eqn}
p(\lambda)=\lambda^3-\tau \lambda^2 + \sigma \lambda-1 .
\end{equation}
Because $\lambda_3=1$ (by definition from section~\ref{sec:resonances_eigenvalues}) and $det(\bm{F})=1$, we can write,
\begin{equation}
\label{eq:lamda1_lamda2}
\lambda_2=1/\lambda_1 .
\end{equation}
Substituting equation~\ref{eq:lamda1_lamda2} in equation~\ref{eq:sigma_def}, we get
\begin{align}
\label{eq:tau_same_sigma}
\begin{split}
\sigma & = \lambda_1\lambda_2+\lambda_2\lambda_3+\lambda_3\lambda_1 \\
  & = 1+1/\lambda_1+\lambda_1 \\
  & = \tau .
\end{split} 
\end{align}
Since $\tau=\sigma$ and $det(\bm{F})=1$, the stability of a period-1 point depends only on one invariant, which is $\tau$ of the deformation tensor.

The characteristic equation~\ref{eq:inv_def_ten_eqn} can be written as,
\begin{align}
\label{eq:char_eqn_deften_stokes}
\begin{split}
p(\lambda) & = \lambda^3-\tau (\lambda^2 -\lambda)-1 \\
& = (\lambda-1) (\lambda^2+(1-\tau)\lambda+1) .
\end{split} .
\end{align}
The solutions of this equation are $1$ and $\frac{1}{2} \big( (\tau-1)\pm \sqrt{\tau^2-2\tau-3}) \big)$. We can characterize the nature of a period-1 point of a period-1 line line by calculating the trace of the deformation tensor evaluated at that point. 
If $\tau \in (-1,3)$, the term $\tau^2-2\tau-3$ is negative, the period-1 point is elliptic. If $\tau > 3$, then the period-1 point is hyperbolic, if $\tau < -1$, then the period-1 point is hyperbolic.

\section{Surviving Invariant and Hamiltonian Flow}
\label{app:invariants}

Here we show that the rotation of the base flow leaves one invariant
and that the motion on each spheroidal shell is Hamiltonian.  The
proof is more general than the hemisphere, being true for any
container generated by any figure of revolution of a convex line
terminating on a plane that is the moving lid of the container.

Although the following proof uses a different coordinate system for
the hemisphere to that shown in figure~\ref{fig:ldhs_schematic}, this
difference is purely cosmetic.  Note 1: Here we treat $\theta$ as the
azimuthal angle and $\phi$ as the polar angle.  In the rest of the
paper we treated $\theta$ as the polar angle and $\phi$ as the
azimuthal angle.  Note 2: In our computations of PRHF, the lid
coincides with the $x$--$z$ plane and translates in the
$\hat{\mathbf{x}}$ direction.  Only in this appendix does the lid
coincide with the $x$--$y$ plane and translate in the
$\hat{\mathbf{x}}$.

Consider Stokes flow inside a container driven by a solid boundary
coincident with the $x$--$y$ plane and uniformly translating in the
$\hat{\mathbf{x}}$ direction at a steady speed $U_w$.  The equations
of motion for the fluid inside the domain are given by
equations~\ref{eq:Stokes} and \ref{eq:incompressibility}.  Any
container having the $x$-$y$ plane as the sliding lid and obtained via
a surface of revolution about the $z$-axis of a curve in the $x$-$y$
plane is left-right ($z$--direction) and fore-aft ($x$--direction)
symmetric.  If the curve is convex, so that the container is
left-right and fore-aft symmetric, then the velocity field components
$\mathbf{U} = (U_r \; U_\phi \; U_\theta)$ have the form
\begin{subequations}
\label{eq:basic_symmetries}
\begin{equation}
U_r      = u_r(r,\phi)  \cos \theta,
\end{equation}
\begin{equation}
U_\phi   = u_\phi(r,\phi)  \cos \theta,
\end{equation}
\begin{equation}
U_\theta = u_\theta(r,\phi)  \sin \theta
\end{equation}
\end{subequations}
in spherical coordinates.  The most general container to which
this analysis applies starts with a curve in the $x$-$z$ plane
connecting the $z$ axis to the $x$ axis.  This curve then generates a
surface of revolution about the $z$-axis that is the container shape
with the $x$-$y$ plane the sliding lid of the container.  The doubly
symmetric container and symmetric driving plane, assumed without loss
of generality to move in the $x$ direction, produce a Stokes flow with
a reflection symmetry across the $x$-axis and a time-reversal symmetry
across the $y$-axis.  

Having two symmetries, the flow has two invariants of the motion
$F_{1,2}$, both given by
\begin{equation}
\label{eq:invariant_basic}
\mathbf{U} \bcdot \bnabla F_{1,2} = 0;
\end{equation}
moreover, equations~\ref{eq:basic_symmetries} imply that the invariant
surfaces have the form
\begin{equation}
\label{eq:F_form}
F_{1,2}(r,\phi,\theta) = f_{1,2}(r,\phi) g(\theta).
\end{equation}

Following Malyuga et al.~\cite{Malyuga_stokes_2002}, putting
equations~\ref{eq:F_form} and \ref{eq:basic_symmetries} into
equation~\ref{eq:invariant_basic} gives
\begin{equation}
\label{eq:master_gradient}
u_r\cos\theta \frac{\partial f}{\partial r} 
+ \frac{u_\phi\cos\theta}{r}g \frac{\partial f}{\partial \phi}
+ \frac{u_\theta\sin\theta}{r\sin \phi}f \frac{\partial g}{\partial \theta} = 0,
\end{equation}
where $f$ is either of the 2-dimensional invariants.  Equation~\ref{eq:master_gradient} is 
separable, and the equation for $g$ is
\begin{equation}
\label{eq:g_eq}
\sin\theta \frac{dg}{d\theta} - \lambda g \cos \theta = 0,
\end{equation}
where $\lambda$ is the separation constant.  For $\lambda = 0$ $g$ is
a constant that, without loss of generality, can be set to one.  For
$\lambda = 1$, $g = \sin \theta$.  Therefore the two invariants are
\begin{eqnarray}
\label{eq:invariants}
F_1 &=& f_1(r,\phi),\\
F_2 &=& f_2(r,\phi) \sin \theta.
\end{eqnarray}
The equation for $f$ is
\begin{equation}
\label{eq:f_eq}
u_r \frac{\partial f}{\partial r} + \frac{u_\phi}{r}\frac{\partial f}{\partial \phi}
+ \frac{\lambda u_\theta}{r\sin\phi} f = 0.
\end{equation}
From equation \ref{eq:invariant_basic}, the two invariant surfaces
intersect at stream lines of the flow, and the equations for the two
invariant surfaces are
\begin{eqnarray}
\label{eq:invariant_eigenvalue}
\lambda = 0 \:\:\: \mathbf{u}^\prime \bcdot \bnabla^\prime f_1 &=& 0,\\
\lambda = 1 \:\:\: \mathbf{u}^\prime \bcdot \bnabla^\prime f_2 &=& 
                                                  -\frac{u_\theta}{r\sin\phi}f_2,
\end{eqnarray}
where $\mathbf{u}^\prime = (u_r, \; u_\phi)$ and $\bnabla^\prime
= (\partial/\partial r, r^{-1}\partial/\partial \phi)$.

\begin{figure}
\centering
  \includegraphics[width=0.8\columnwidth]{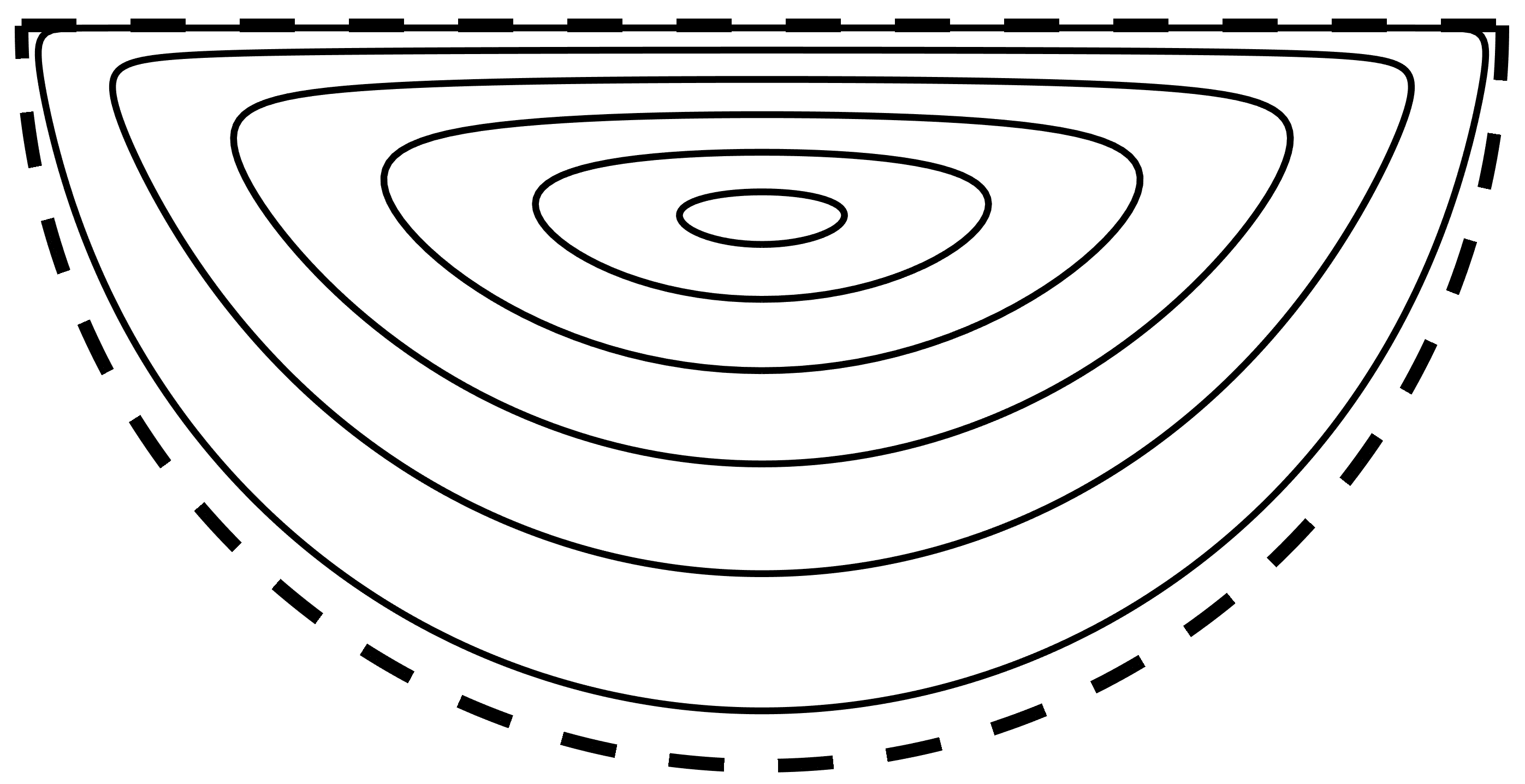}
  \caption{Flow cross-section of streamlines in the $x$--$y$ plane.
    Dashed line is the hemisphere and lid boundary. The lid moves left
    to right in the $+x$ direction.  }
\label{fig:cross_section_stokes}
\end{figure}

From $\bnabla \bcdot \mathbf{U} = 0$ the azimuthal velocity required
to satisfy continuity is
\begin{equation}
\label{eq:continuity}
\left(\frac{2}{r} + \frac{\partial}{\partial r}\right) u_r +
\left(\frac{1}{r} + \frac{\partial}{\partial \phi} + \frac{\cot \phi}{r}\right) u_\phi 
= \frac{-u_\theta}{r \sin\phi}.
\end{equation}
Combining equations~\ref{eq:continuity} and
\ref{eq:invariant_eigenvalue} shows that
\begin{subequations}
\label{eq:streamfunction_1}
\begin{equation}
\frac{1}{r}\frac{\partial f_1}{\partial \phi} = \frac{u_r r^2 \sin\phi}{f_2}
\end{equation}
\begin{equation}
 -\frac{\partial f_1}{\partial r}  = \frac{u_\phi r^2 \sin\phi}{f_2}
\end{equation}
\begin{equation}
\bnabla \bcdot \left(\frac{\mathbf{u}^\prime r^2 \sin\phi}{f_2}\right) = 0,
\end{equation}
\end{subequations}
or more compactly by defining $\mathbf{u}^\ast \equiv
\mathbf{u}^\prime r^2 \sin\phi / f_2$ that
\begin{subequations}
\label{eq:streamfunction_2}
\begin{equation}
\frac{1}{r}\frac{\partial f_1}{\partial \phi} = u_r^\ast(r,\phi)
\end{equation}
\begin{equation}
           -\frac{\partial f_1}{\partial r}   = u_\phi^\ast(r,\phi)
\end{equation}
\begin{equation}
                \bnabla \bcdot \mathbf{u}^\ast = 0.
\end{equation}
\end{subequations}
The Hamiltonian form of equations~\ref{eq:streamfunction_2} reveals
the invariant $f_1$ as a streamfunction in the $r-\phi$ plane for the
rescaled solenoidal vector field $\mathbf{u}^\ast$.  A cross-section
of $f_1$ is shown in figure~\ref{fig:cross_section_stokes}.  For
periodic reorientation $F_2$ is destroyed, but $F_1$, being
rotationally symmetric, is preserved.  In other words, the Hamiltonian
structure is locally preserved.  In the action-angle formalism, $F_1$
is the conserved action, and the periodically reoriented lid-driven
cavity flow in a doubly-symmetric container is an action-angle-angle
dynamical system.

The form of equations~\ref{eq:streamfunction_1}, which are essentially
the same as found for a cylindrical container by Malyuga et
al.~\cite{Malyuga_stokes_2002}, suggests a more general perspective is
possible, which is that all periodically reoriented Stokes flows in
containers of revolution with a moving lid can be put into a Darboux
standard form.  Notice that equations~\ref{eq:streamfunction_1} can be
put into the noncanonical Hamiltonian-Poisson form
\begin{equation}
\label{eq:Darboux}
\frac{d\boldsymbol{x}}{dt} = \frac{f_2}{J} F_c \cdot \nabla H,
\end{equation}
with $\boldsymbol{x} = (r,\phi)^T$, $H = f_1$ the Hamiltonian, $J$ the
Jacobian of the coordinate system, and
\begin{equation}
F_C = 
\begin{pmatrix}
  0 & 1 \\
 -1 & 0
 \end{pmatrix}
\end{equation}
is the canonical skew-symmetric matrix.  A Poisson manifold, which we
specify here for 3D only, is a mixture of canonical and noncanonical
coordinates $(q,p,z)$ with $z$ a Casimir invariant that commutes with
the Hamiltonian: $\{z,H\} = 0$, with $\{ \cdot \}$ the Poisson
bracket.  $(f_2/J)$, then, is the inverse Jacobian for the squashed
hemisphere coordinate system.  It is easier to see the structure if we
``blow up'' the hemisphere into a topologically equivalent actual
sphere.  Then in equation~\ref{eq:Darboux} $f_2 \rightarrow 1$ and
$F_c$ becomes through the Darboux
theorm~\cite{Littlejohn_guiding_1979} the 3D singular Poisson operator
\begin{equation}
\label{eq:Poisson}
{\cal J} = 
\begin{pmatrix}
  0 & 1 & 0 \\
 -1 & 0 & 0 \\
  0 & 0 & 0
 \end{pmatrix}.
\end{equation}
Now with $\boldsymbol{x} = (\theta,\phi,r)^T$
equation~\ref{eq:Darboux} becomes
\begin{equation}
\label{eq:HamiltonPoisson}
\frac{d\boldsymbol{x}}{d\tau} = {\cal J} \cdot \nabla H,
\end{equation}
where the Jacobian has been absorbed into a reparameterization of
time.  The dynamical system \ref{eq:HamiltonPoisson} is Hamiltonian on
level sets of the Casimir function $r = \text{ constant}$.  The
Casimir invariant---here the radial coordinate \emph{after} conformal
transformation to a sphere---is a topological constraint to a foliation
of Hamiltonian leaves.  In the squashed hemisphere coordinates, what
we have called the action coordinate or shell number is the Casimir
function and ${\cal J}$ is a skew-symmetric matrix of generalized
structure functions associated with the conformal transform of sphere
to hemisphere.  We do not calculate the structure functions, but all
lid-driven cavities of Stokes flow are reducible to the
Hamiltonian-Poisson form of equation~\ref{eq:HamiltonPoisson} with
appropriate structure functions.

\section{Lid Attachment Point for $\beta \le 1$}
\label{sec:lid_attach_point}

\begin{figure}
\centering
\begin{tikzpicture}[
         every node/.style={anchor=south west,inner sep=0pt},
         x = 1mm, y = 1mm
     ]
    \node (fig1) at (0,0)
    {\includegraphics[width=\columnwidth]{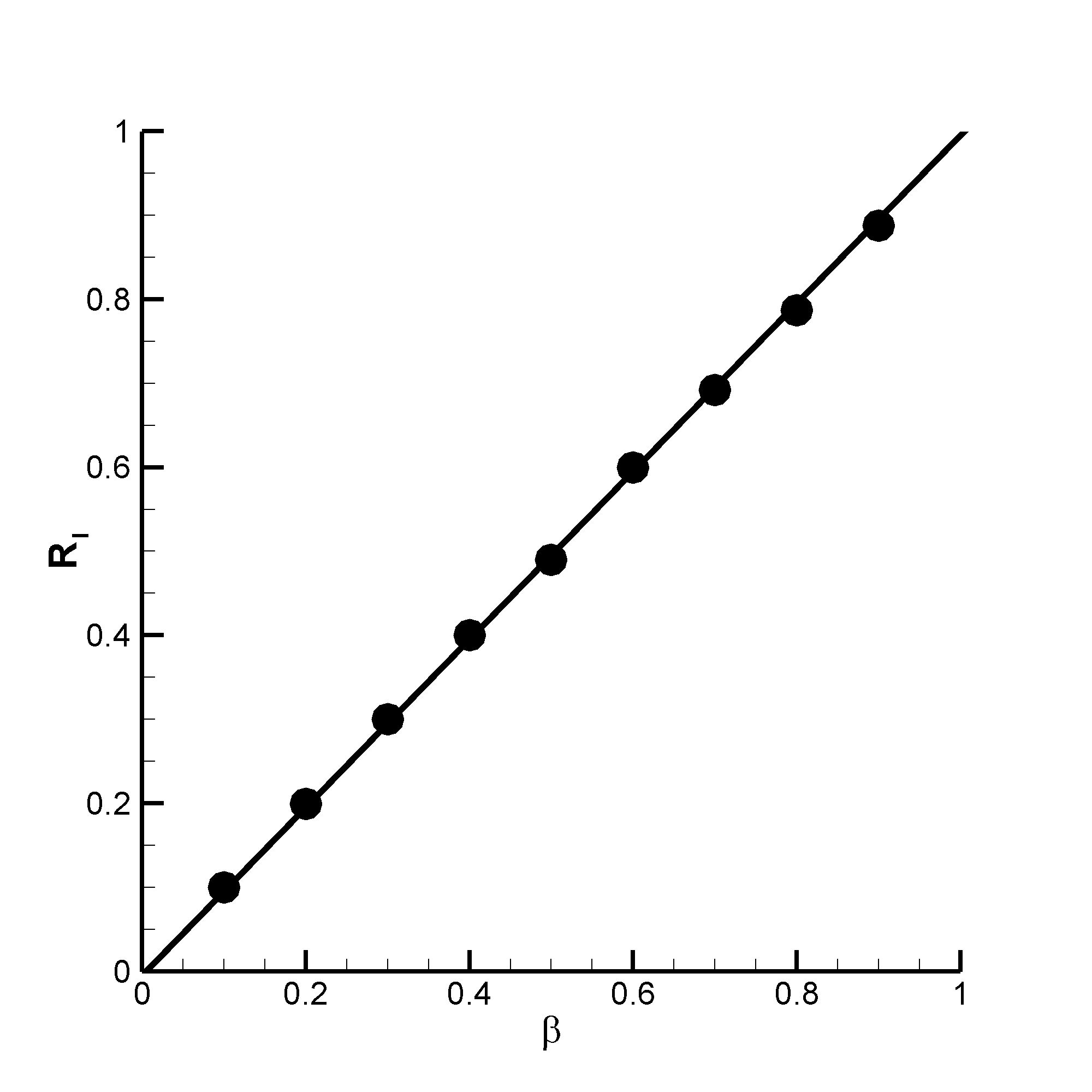}};
    \node (fig2) at (13,42)
    {\includegraphics[width=0.45\columnwidth]{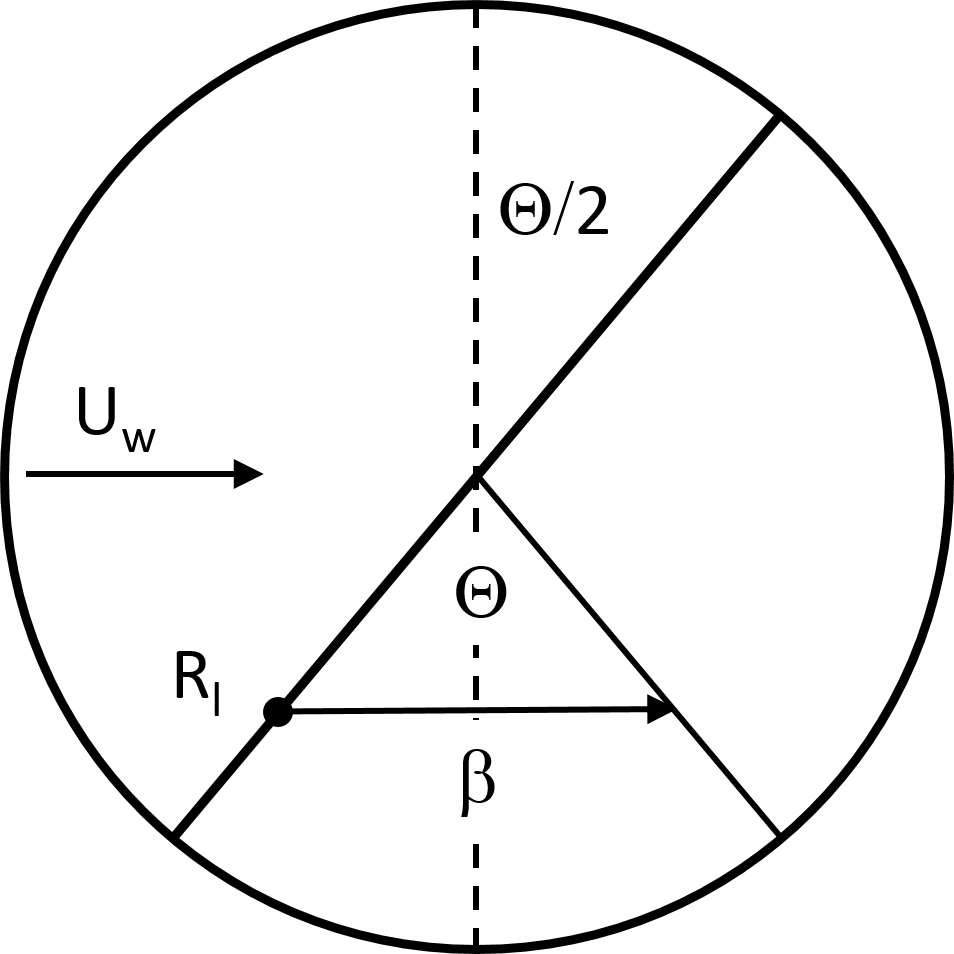}};
\end{tikzpicture}
\caption{The attachment point on the lid starts at the origin for
  $\beta = 0$ and moves to the rim for increasing $\beta \le 1$.
  Plotted is the lid attachment point's distance $R_l$ from the
  origin to the rim as a function of $\beta$ for $\Theta = \pi/3$.
  Dots are computations; solid line is
  equation~\protect\ref{eq:lid_point_location}.  Inset is geometry to
  determine location of the P1 attachment point.}
\label{fig:mobile_attachement} 
\end{figure}

With reference to the inset of figure~\ref{fig:mobile_attachement} the
P1 attachment point on the lid is the point that begins on the
symmetry plane a distance $R_l$ from the center of the lid along the
symmetry plane and is displaced an amount $\beta$ such that subsequent
rotation by $-\Theta$ returns the point to its initial position.  A
trigonometric relations gives
\begin{equation}
\label{eq:lid_point_location}
R_l = \frac{\beta}{2\sin\left(\frac{\Theta}{2}\right)}
\end{equation}
for any $\Theta$ and $\beta \le 1$.
Figure~\ref{fig:mobile_attachement} shows
equation~\ref{eq:lid_point_location} as a solid line and numerically
determined locations of the attachment point as dots.

\section{Extrema of Parameterised Periodic Lines}
\label{app:tangent}

\begin{figure}
  \centering
\includegraphics[width=\columnwidth]{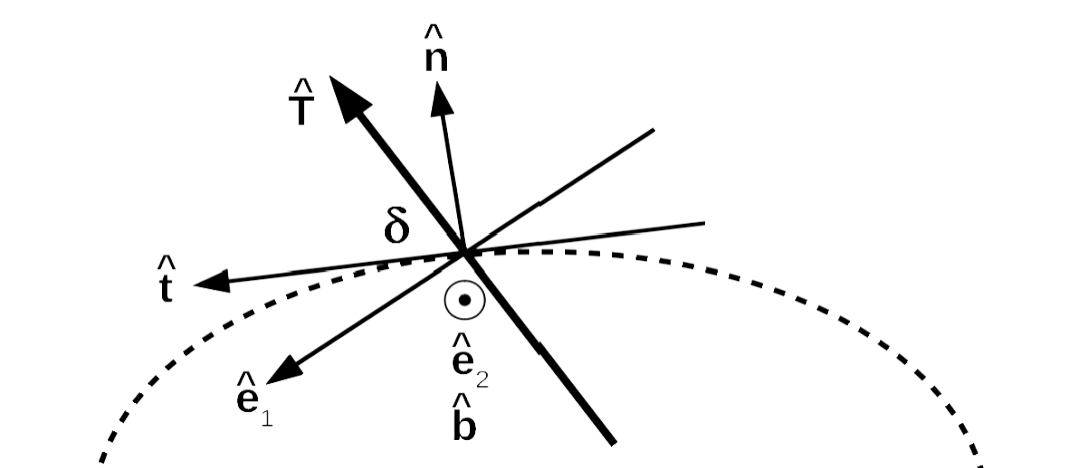}
\caption{Shell (dashed line) pierced by periodic line with $\boldsymbol{\hat{T}}$
  the tangent null direction of the periodic line.
  $\boldsymbol{\hat{e}}_1$ and $\boldsymbol{\hat{e}}_2$ are the other
  eigendirections of the deformation tensor at the piercing point, and
  $\boldsymbol{\hat{n}}$, $\boldsymbol{\hat{t}}$, and
  $\boldsymbol{\hat{b}}$ are respectively the normal, tangent, and
  binormal unit vectors of the shell at the piercing point.  $\boldsymbol{\hat{b}}$
  is normal to the page and $\delta$
  is the angle between the normal plane of the shell and the
  deformation plane perpendicular to $\boldsymbol{\hat{T}}$. }
\label{fig:extrema}
\end{figure}

Figure~\ref{fig:extrema} depicts a generic neighbourhood around a point
at which P1 pierces a shell.  At the piercing point the shell is
described by its normal, tangent and binormal unit vectors,
$(\boldsymbol{\hat{n}}, \boldsymbol{\hat{t}}, \boldsymbol{\hat{b}})$,
with $(\boldsymbol{\hat{t}}, \boldsymbol{\hat{b}})$ defining the
tangent plane of the shell.  The eigendirections of the deformation
tensor at the piercing point are designated
$(\boldsymbol{\hat{T}}, \boldsymbol{\hat{e}}_1,
\boldsymbol{\hat{e}}_2)$
in the figure.  $\boldsymbol{\hat{T}}$ is tangent to the line and is a
null direction of deformation because the point is part of P1 and the
invariance precludes normal (off-shell) transport.  The other
eigendirections of the deformation tensor at the piercing point,
$(\boldsymbol{\hat{e}}_1, \boldsymbol{\hat{e}}_2)$, define a
deformation plane perpendicular to $\boldsymbol{\hat{T}}$.  Without
loss of generality we can make a local coordinate transformation at
the shell and rotate the shell's tangent vector $\boldsymbol{\hat{t}}$
so that the periodic line's tangent $\boldsymbol{\hat{T}}$ is in the
plane $(\boldsymbol{\hat{n}} , \boldsymbol{\hat{t}})$.  This also has
the effect of making $\boldsymbol{\hat{e}}_2$ and
$\boldsymbol{\hat{b}}$ coincide, as in the figure.  The angle
$\delta$ is the angle between the P1 line and the tangent plane at the
point of piercing.  As transport normal to the shell is forbidden,
deformation in the tangent plane of the shell is given by the
projection of the deformation described by
$(\boldsymbol{\hat{e}}_1, \boldsymbol{\hat{e}}_2)$ onto
$(\boldsymbol{\hat{t}}, \boldsymbol{\hat{b}})$.  Since
$\boldsymbol{\hat{e}}_2$ and $\boldsymbol{\hat{b}}$ are already
co-linear, this projection is $\boldsymbol{\hat{e}}_1 \sin\delta$.
The normal form for the deformation tensor in the neighbourhood of the
piercing point becomes
\begin{equation}
  \begin{pmatrix}
    1 & \begin{matrix} \hspace*{0.5cm} 0 & \hspace*{0.5cm} 0\end{matrix}  \\
    \begin{matrix}  0 \\ 0 \end {matrix} &
    \begin{pmatrix}
          \lambda_1 \sin\delta & 0 \\
           0                   & \lambda_2
      \end{pmatrix} \\    
  \end{pmatrix}.
 \end{equation}
 Note that exchanging rows in the submatrix giving deformation in the
 tangent plane of the shell in the neighbourhood of the piercing point
 puts the submatrix in the form of a general linear flow
 \cite{Metcalfe_chaos_2010}, whose eigenvalues $\sigma$ are
 \begin{equation}
   \label{eq:prove_degeneracy}
   \sigma = \pm \sqrt{\lambda_1 \lambda_2 \sin\delta}.
 \end{equation}
Because streamlines cannot cross except at points with zero velocity, 
planar flow topology is restricted to combinations of elliptic, shear, or
hyperbolic flows.  The general linear flow encompasses elliptic,
hyperbolic, and shear flows in one velocity gradient matrix, identical
with the submatrix, with a single parameter that ranges from $-1$ to
$1$: negative (positive) values of the parameter give elliptic
(hyperbolic) streamlines; a zero value gives a shear flow
\cite{Metcalfe_chaos_2010}.
 
The key point about equation~\ref{eq:prove_degeneracy} is that when a
periodic line is tangent to a shell, the tangent vectors
$\boldsymbol{\hat{T}}$ and $\boldsymbol{\hat{t}}$ of the line and
plane coincide.  Thus $\delta$ goes to zero, forcing the deformation
eigenvalues to zero, indicating a degenerate point.  On either side of
degenerate (tangent) points on P1, the shell number is either lower or
higher than at the tangent point, indicating these points represent
local extrema on the plot of shell number versus arc length.  Thus
local extrema in shell number versus arc length necessarily indicate
the location of a degenerate point on P1.  Moreover, as can be seen
from equation~\ref{eq:prove_degeneracy} as $\delta$ goes from just
above zero to just below, the character of the line must change.  We
cannot definitely determine the direction of the change because that
depends on the sign of $\lambda_2$, but the geometry of the periodic
line at the degenerate point forces a change of character (this is
clearly seen in the example shown in
figure~\ref{fig:prd1_line_sgmnt_beta_16}b).  If it is elliptic
(hyperbolic) on one side, it must be hyperbolic (elliptic) on the
other.  Note that we do not say that degenerate points at extrema are
the only degenerate points that exist in the flow; the idiosyncratic
dynamics of a particular flow can (and does) produce other degenerate
points on periodic lines.  In principle P1 could be tangent to a
shell, but not be a local extrema, i.e.\/ it could be an inflexion
point on the shell number {\it vs} arc length plot.  In such cases,
$\delta$ does not change sign, instead it goes to zero then goes
positive (or negative) again.  Thus inflexion in P1 means a degenerate
point but no change of character either side of the degenerate point.


\end{document}